\documentclass{cernrep}
\usepackage{graphicx,epsfig,amsmath,amssymb,wrapfig}

\begin{document}

\pagestyle{myheadings}
\thispagestyle{empty}

\title{Physics Beyond the Standard Model: Supersymmetry}

\author{
{\bf M.M.~Nojiri$^{1,2}$},
{\bf T.~Plehn$^{3}$},
{\bf G.~Polesello$^{4}$} (conveners \& editors), \\[2mm]
M.~Alexander$^{3}$, 
B.C.~Allanach$^{5}$,
A.J.~Barr$^{6}$,
K.~Benakli$^{7}$,
F.~Boudjema$^{8}$, 
A.~Freitas$^{9}$,
C.~Gwenlan$^{10}$, 
S.~J\"ager$^{11}$,
S.~Kraml$^{12}$,
S.~Kreiss$^{3}$,
R.~Lafaye$^{13}$, 
C.G.~Lester$^{14}$,
N.~Kauer$^{15}$,
C.~Milst\'ene$^{16}$,
C.~Moura$^{7}$,
G.S.~Muanza$^{17}$,
A.R.~Raklev$^{5,14}$,
M.~Rauch$^{3}$, 
M.~Schmitt$^{18}$,
S.~Sekmen$^{19}$,
P.~Skands$^{11,20}$,
P.~Slavich$^{8,11}$, 
A.~Sopczak$^{21}$, 
M.~Spannowsky$^{22}$,
D.R.~Tovey$^{23}$,
E.~Turlay$^{24}$,
C.~F.~Uhlemann$^{15}$,
A.M.~Weber$^{25}$,
P.~Zalewski$^{26}$, and
D.~Zerwas$^{24}$
}

\institute{
\begin{small}
\vspace*{5mm}
$^{1}$ KEK Theory Group and Graduate University for Advanced Study (SOUKENDAI),
       Tsukuba, Japan;
$^{2}$ Institute for the Physics and Mathematics of the Universe (IPMU), 
       University of Tokyo, Kashiwa City, Japan; 
$^{3}$ SUPA, School of Physics, University of Edinburgh, 
       Scotland; 
$^{4}$ INFN, Sezione di Pavia, Pavia, 
       Italy;
$^{5}$ DAMTP, CMS, University of Cambridge, 
       UK; 
$^{6}$ Department of Physics, University of Oxford,
       UK; 
$^{7}$ LPTHE, Universit\'e Pierre et Marie Curie --- Paris VI  et 
              Universit\'e Denis Diderot --- Paris VII, France; 
$^{8}$ LAPTH, Annecy-le-Vieux, 
       France;
$^{9}$ Zurich University,
       Switzerland; 
$^{10}$ Department of Physics and Astronomy, UCL, London,
        UK; 
$^{11}$ Theory Group, Physics Department, CERN, 
        Geneva, Switzerland; 
$^{12}$ LPSC, UJF Grenoble 1, CNRS/IN2P3, INPG, 
        Grenoble, France;
$^{13}$ LAPP, Universit\'e Savoie, IN2P3/CNRS, Annecy, 
        France;
$^{14}$ Cavendish Laboratory, University of Cambridge,
        UK; 
$^{15}$ Institut f\"ur Theoretische Physik, Universit\"at W\"urzburg,
        Germany; 
$^{16}$ Fermilab, Batavia,
        U.S.; 
$^{17}$ IPN Lyon, Villeurbanne, 
        France; 
$^{18}$ Northwestern University, Evanston,
        U.S.; 
$^{19}$ Department of Physics, Middle East Technical University, 
        Ankara, Turkey; 
$^{20}$ Theory Group, Fermilab, Batavia
        U.S.; 
$^{21}$ Department of Physics and Astromony, Lancaster University, 
        UK; 
$^{22}$ Institut f\"ur Theoretische Physik, Universit\"at Karlsruhe,
        Germany; 
$^{23}$ Department of Physics and Astronomy, University of Sheffield, 
        UK; 
$^{24}$ LAL, Universit\'e Paris-Sud, IN2P3/CNRS, Orsay, 
        France; 
$^{25}$ Max Planck Institut f\"ur Physik (Werner Heisenberg Insitut),
        M\"unchen, Germany; 
$^{26}$ Soltan Institute for Nuclear Studies, Warsaw, 
        Poland
\end{small}
}

\maketitle 

\begin{abstract}
  This collection of studies on new physics at the LHC constitutes the
  report of the supersymmetry working group at the Workshop `Physics
  at TeV Colliders', Les Houches, France, 2007. They cover the wide
  spectrum of phenomenology in the LHC era, from alternative models
  and signatures to the extraction of relevant observables, the study
  of the MSSM parameter space and finally to the interplay of LHC
  observations with additional data expected on a similar time scale.
  The special feature of this collection is that while not each of the
  studies is explicitely performed together by theoretical and
  experimental LHC physicists, all of them were inspired by and
  discussed in this particular environment.
\end{abstract}

\clearpage

\tableofcontents
\clearpage

\setcounter{figure}{0}
\setcounter{table}{0}
\setcounter{equation}{0}
\setcounter{footnote}{0}
\section*{Aim and structure of this collection \footnote{T.~Plehn}}
\bigskip

With the first LHC data around the corner, the great common goal of
theoretical and experimental high--energy physics appears to be in
close reach: over the coming years, we will have to try to understand
the origins of electroweak symmetry breaking and the role of the TeV
scale from the data rolling in. This effort can only be successful if
theorists and experimentalists work in close collaboration, following
the well--known spirit of the Les Houches workshops. This
collaboration of course starts with the proper understanding of QCD,
the theory which describes any kind of particle production at the LHC,
but which also describes the main backgrounds which Higgs and
new--physics searches have to battle. However, due to the complexity
of LHC data on the one hand, and of the new--physics models at the TeV
scale on the other hand, the interaction between theorists and
experimentalists needs to go much further. Realistically, we expect
that any new--physics search at the LHC will require theorists to
formulate viable and predictive hypotheses which are implemented in
state-of-the-art simulation and extraction tools. Such models can
guide experimental searches towards the of course yet unknown
ultraviolet completion of the Standard Model, even if at least all but
one known models for new physics at the LHC will be soon ruled out.

Over recent years, high--energy theorists have hugely expanded the
number of viable ultraviolet completions of the Standard Model. The
main guiding principle of all of these models is still electroweak
symmetry breaking. There are essentially two paths we can follow to
explain the weak--scale masses of gauge bosons and of
(third--generation) fermions: first, we can assume the minimal Higgs
sector of our Standard Model to hold, which leads to the hierarchy
problem. Without solving this problem, the Standard Model appears to
be incomplete as a fundamental theory valid to all mass scales up to
the Planck scale. Such ultraviolet completions are particularly
attractive if they allow us to incorporate dark--matter candidates or
unification scenarios.  Supersymmetry is in particular in the
experimental community definitely the most carefully studied
completion, but extra dimensions or little--Higgs models are
alternatives worth studying. The alternative path to describe
electroweak symmetry breaking are strongly interacting models, which
avoid predicting a fundamental Higgs boson. Such models have recently
become more viable, if combined for example with extra
dimensions.\bigskip

As indicated by its title, this working group focuses on
supersymmetry as one example for an ultraviolet completion of the
Standard Model with the usual Higgs mechanism. However, simply writing
down one supersymmetric version of the Standard Model does not suffice
in view of the almost infinite number of LHC analyses which would be
possible to study such models. As a matter of fact, signatures which
until recently were thought to be typical for supersymmetry, namely
jets plus missing energy plus maybe like--sign dileptons are by now
mainstream signals for extra dimensions, little--Higgs models, or even
strong interactions. Therefore, this collection of projects should
first be considered as studies of models which lead to typical
supersymmetry signatures, mostly beyond the naive inclusive `missing
energy plus jets' analysis. The obvious question is how with the LHC
running we would go about to understand what the underlying theory of
such signatures could be. Secondly, we include studies on version of a
supersymmetric Standard Model which deviate from the simple MSSM.
Which means that even supersymmetry as an underlying principle does
not have to look exactly like we naive think it should look. It is a
healthy development that theoretical physics has moved beyond its
focus on the minimal supersymmetric ultraviolet extension of the
Standard Model, while at the same time, not all alternatives in and
beyond supersymmetry need to be studied including full detector
simulations.

Independent of our `preferred' models, theorists need to carefully
formulate viable TeV--scale models, including variations of key models
which allow us to test predictions for example of the MSSM. These
alternatives can for example be driven by the similarity of signatures
or by the aim to test certain underlying theory structures. In
supersymmetry such models obviously involve Dirac gauginos (altering
the Majorana nature for example of the gluino or the dark--matter
agent). Naively matching the two degrees of freedom of an on-shell
gluon with those of a gluino only allows us to write down a Majorana
gluino.  In extended supersymmetric models this requirement does not
need to exist. On the other hand, even assuming minimal supersymmetry
an extended Higgs sector like in the NMSSM, should be tested
carefully. Moreover, supersymmetric spectra with a hierarchy between
gauginos and scalars maintain many advantages of supersymmetry, like
unification and a dark--matter agent, while avoiding flavor
constraints at the expense of introducing fine tuning. Due to the
organization of the complete document, in this chapter of the Les
Houches proceedings we limit ourselves to variations of the MSSM,
deferring models like extra dimensions or little--Higgs to another
collection of articles in the same volume.\bigskip

In a second step, theorists and experimentalists need to develop
strategies to extract information on TeV--scale physics from LHC data.
In the standard MSSM scenarios, studying the kinematics of cascade
decays has been shown to be a spectacularly successful, even in the
presence of missing energy from a dark--matter agent escaping the
detector unseen. There are, however, many more or less experimentally
complex ways to use LHC observables to extract information on the
masses of new states from LHC data. The information can come from the
general underlying mass scale is event samples including physics
beyond the Standard Model, or from any combination with cascade
information. Of course, such studies are not limited to supersymmetry,
but they can in principle be used for any new--physics signatures with
decays from strongly interacting new--physics states down to a weakly
interacting dark--matter particle.

Whatever we are looking for at the LHC, technically correct
simulations of new--physics events are crucial, if we ever want to
extract the fundamental parameters from their comparison with data.
There is no good reason to try to extract new physics from
21st-century LHC data using 20th-century Monte--Carlos and methods.
The past years have seen impressive progress in incorporating
new--physics signals in modern Monte--Carlo tools, including for
example the proper simulation of many--particle final states beyond a
naive narrow--width approximation. This particular problem is being
studied in the supersymmetric framework, but it is at least as
relevant for new--physics models which predict more degenerate mass
spectra, like for example generic universal extra dimensions.\bigskip

Obviously, LHC data on TeV--scale physics will not come into a
data--free world. There is a wealth of information we have already
collected on such physics models, and during the LHC era we expect
much more of it. The long list of current and future complementary
data includes electroweak precision data, the muon's anomalous
magnetic moment, precision--flavor physics, dark--matter measurements
linked with big--bang nucleosynthesis, and most importantly at some
stage the high--precision data from a future ILC. In particular, when
it comes to measuring as many model parameters of the TeV--scale
Lagrangian as possible, the proper combination of all these pieces of
information is crucial to our understanding of the ultraviolet
extension of the Standard Model. Only once we can claim a solid
understanding of the TeV scale we should attempt to extrapolate our
physics picture to very large mass scales, to finally determine if our
underlying theory can really be a fundamental theory of Nature. After
all, the LHC is not going to be the last, but the first major
experiment allowing us to carefully study the TeV scale and determine
the fundamental parameters of physics beyond the weak scale over the
coming years!

The last section in this collection follows a great tradition of the
Les Houches workshops: the successful definition and implementation of
interfaces between computer tools used by the theoretical and
experimental high--energy community. 

At this stage, the conveners of the SUSY session would like to express
their gratitude of course to the organizers of this inspiring and
enjoyable workshop.  Moreover, we would like to thank all the young
collaborators in Les Houches and elsewhere, who have made possible the
impressive studies presented in this collection.

\clearpage
\newpage

\setcounter{figure}{0}
\setcounter{table}{0}
\setcounter{equation}{0}
\setcounter{footnote}{0}
\section{A model for dirac and majorana gaugino masses
\protect\footnote{K.~Benakli and C.~Moura}}

\subsection{Introduction}

Massive fermions can appear either as Majorana or Dirac.  Because  the latter allow charged states, they are easier to  detect . And, in fact, all identified fermionic masses are of Dirac type.  The nature of neutrinos masses remains unknown, and unveiling it is the main challenge for double beta decays experiments. It is then legitimate to ask about the form of the masses of new fermions that could be detected by LHC.

In the minimal extension (MSSM) masses of gauginos are of Majorana type. Obtaining Dirac ones requires pairing up with new fermions that should then arise as components of extra chiral fields in the adjoint representation. The easiest way to incorporate these new states is to make the gauge fields as parts of $N=2$ multiplets. Such a scheme is  present in an extra-dimensional picture where the $N=2$ fields appear as bulk states (in the absence of a projection)  while the chiral matter appear as localized states in $N=1$ representations. In such a set-up Dirac masses have been shown to appear naturally in the presence of an anomalous $U(1)$ as a result of new operators that mix the MSSM fields with the anomalous $U(1)$  \cite{Fox:2002bu, Chacko:2004mi, Carpenter:2005tz,  Hisano:2006mv, Nojiri:2007jm, Hsieh:2007wq}.  It was later shown that such operators can be generated by loop effects if the supersymmetry breaking sector is in an $N=2$ representation 
\cite{Antoniadis:2005em, Antoniadis:2006eb, Antoniadis:2006uj}, and that the quartic tree-level Higgs potential is in fact also modified \cite{Antoniadis:2006uj}. Such models where studied as primarely based on the only presence of $D$-term breaking suffer from two issues. First, in the minimal set-up the adjoint scalar  have tachyonic masses\cite{Chacko:2004mi,Antoniadis:2006uj}. Second, typical supersymmetry breaking model would lead to both $D$ and $F$-terms. The latter will fix the first issue, but also turn on new sources for sof terms, in particular Majorana masses for the gauginos.

Here, we will take a different, more phenomenological approach. We will provide with the corresponding Lagrangian containing an $N=2$ extended gauge sector, and we will not address the origin of the soft masses.

\subsection{Primer}

We will start by fixing our conventions for the spinorial notation as well as for the MSSM, and then we will proceed to extend the gauge sector.

\subsubsection*{Spinors notation}

A Dirac fermion $\Psi_{D}$ has four components which can be assembled into   two-component spinors :
\begin{equation}
\Psi_{D} =  \left(\begin{array}{c}
\overline{\psi}^{\dot{\alpha}} \\ 
\chi_{\alpha}
\end{array}\right)
\label{diracgauginos_DiracSpinor}
\end{equation}
where $\overline{\psi}^{\dot{\alpha}} = \epsilon^{\dot{\alpha} \dot{\beta}} \overline{\psi}_{\dot{\beta}}$ and 
$\overline{\psi}_{\dot{\alpha}} = (\psi_{\alpha})^{*}$. Here $\epsilon^{\alpha \beta}$ is the 
completely antisymmetric tensor and $\epsilon^{1 2} = 1$.
We adopt for the $\gamma$ matrices the following representation: 
\begin{equation}
\gamma^{\mu} = 
\left( \begin{array}{c c}
0  & {\overline{\sigma}^{\mu}} \\
{\sigma^{\mu}} & 0
\end{array}  \right) 
,  \quad
\gamma^{5} = i \gamma^{0} \gamma^{1} \gamma^{2} \gamma^{3} =
\left( \begin{array}{c c}
1  & 0 \\
0 & -1
\end{array} \right) 
\label{diracgauginos_GammaRepresentation}
\end{equation}
where $\overrightarrow{\sigma}$ are the Pauli matrices and the notation $\sigma^{\mu} = (1,\overrightarrow{\sigma})$ and $\overline{\sigma}^{\mu} = (1,-\overrightarrow{\sigma})$ is used.
A Dirac mass term here takes the form:
\begin{equation}
\overline{\Psi}_{D} \Psi_{D} = \chi \psi + \overline{\psi} \overline{\chi} ,
\label{diracgauginos_DiracMass}
\end{equation}
with the spinor products $\chi \psi$ and  $\overline{\chi} \overline{\psi}$ 
 defined as $\chi \psi = \chi^{\alpha} \psi_{\alpha}$ and 
$\overline{\chi} \overline{\psi} = \overline{\chi}_{\dot{\alpha}} \overline{\psi}^{\dot{\alpha}}$, 
where, again, we use the notation 
$\chi^{\alpha} = \epsilon^{\alpha \beta} \chi_{\beta}$.
A Majorana fermion can also be written as:
\begin{equation}
\Psi_{M} = 
\left( \begin{array}{c}
\overline{\chi}^{\dot{\alpha}}\\
\chi_{\alpha}
\end{array} \right) ,
\label{diracgauginos_MajSpinor}
\end{equation}
with the Majorama mass term:
\begin{equation}
\overline{\Psi}_{M} \Psi_{M} = \chi \chi + \overline{\chi} \overline{\chi},
\label{diracgauginos_MajMass}
\end{equation}
Below we will always use left handed fermions. The Dirac fermion representing a lepton is:
\begin{equation}
\Psi^{(l^-)}_{D} = 
\left(\begin{array}{c}
\overline{\chi}^{(l^+)} \\
\chi^{(l^-)}
\end{array}\right)
\label{diracgauginos_ElectronSpinor}
\end{equation}
where $\chi^{(l^-)}$ is the left-handed lepton field and $\overline{\chi}^{(l^+)}$ is the 
charge conjugated of the left-handed anti-lepton field: 
$\overline{\chi}^{(l^+)\dot{\alpha}} = \epsilon^{\dot{\alpha} \dot{\beta}} (\chi^{(l^+)}_{\alpha})^{*} $. 
Then, its Dirac mass will be written 
$m_{D} [  \chi^{(l^-)} \chi^{(l^+)} + \overline{\chi}^{(l^+)} \overline{\chi}^{(l^-)} ] $.

\subsubsection*{Supersymmetry}
\label{diracgauginos_secSUSY}

The generic supersymmetric Lagrangian density for a gauge theory discussed here take the form
\footnote{ In this subsection and the next we follow closely the presentation of Refs \cite{Aitchison:2005cf}.}.
\begin{equation}
{\cal L} = {\cal L}_{gauge} + {\cal L}_{chiral} + {\cal L}_{minimal~coupling}.
\label{diracgauginos_SusyLagrangian}
\end{equation}
Here, the gauge kinetic Lagrangian is given by:
\begin{equation}
{\cal L}_{gauge} = -\frac{1}{4} F_{\mu\nu}^{a} F^{\mu\nu \, a} 
+ i \overline{\lambda}^{a} \overline{\sigma}^{\mu} (D_{\mu} \lambda)^{a} 
+ \frac{1}{2} D^{a} D_{a}
\label{diracgauginos_GaugeLagrangian}
\end{equation}
where $F_{\mu\nu}$ is the gauge boson field strength, $\lambda$ and $D^{\alpha}$ are the associated  gaugino and and auxiliary field, respectively. $D_{\mu}$ is the gauge covariant derivative. Here,  the 
index $a$ is a gauge symmetry group index corresponding to the generator $T^{a}$.

The chiral Lagrangian is written as:
\begin{equation}
{\cal L}_{chiral} = D_{\mu}\phi^*_i D^{\mu}\phi_i 
+ i \overline{\chi}_i \overline{\sigma}^{\mu} D_{\mu} \chi_i 
+ F_i^* F_i + \left( \frac{\partial W}{\partial \phi_i} F_i 
-\frac{1}{2} \frac{\partial^2 W}{\partial \phi_i \partial \phi_i } \chi_i \chi_j + h.c. \right).
\label{diracgauginos_ChiralLagrangian}
\end{equation}
Here the chiral fermion $\chi_i$ , the 
boson $\phi_i$, and the auxiliary field $F_i$ belong to the same gauge group representation  and form an $N=1$ multiplet . 
The index $i$ labels the different chiral multiplets. The superpotential $W$ is an holomorphic function of the 
fields $\phi_i$.

Finally, the last piece in the supersymmetric Lagrangian density is:
\begin{equation}
{\cal L}_{minimal~coupling} = -g(\phi^*_i T^{a} \phi_i)D^{a} 
-\sqrt{2} g \left[ ( \phi^*_i T^{a} \chi_i ) \lambda^{a} + 
\overline{\lambda}^{a} (\overline{\chi}_i T^{a} \phi_i) \right] 
\label{diracgauginos_CouplingLagrangian}
\end{equation}
where g is the gauge coupling contant. The last two terms in ( \ref{diracgauginos_CouplingLagrangian}) will 
be important  to us as the scalar field takes a vacuum expectation value (the Higgs multiplets), 
producing   bilinears in the fermions, thus mass terms.

\subsection{MSSM}

\begin{table}[htb]
\begin{center}
\begin{tabular}{c|c|c|c|c|c}
\hline
Names  &                 & Spin 0                  & Spin 1/2 & Spin 1 & $SU(3)$, $SU(2)$, $U(1)_Y$ \\ 
\hline
left-handed  & $Q$   & $(\tilde{u}_L,\tilde{d}_L)$  & $(u_L,d_L)$ & & \textbf{3}, \textbf{2}, 1/3 \\ 
quarks       & $u^c$ & $\tilde{u}^c_L$              & $u^c_L$     & & $\overline{\textbf{3}}$, \textbf{1}, -4/3 \\ 
($\times 3$ families) & $d^c$ & $\tilde{d}^c_L$     & $u^c_L$     & & $\overline{\textbf{3}}$, \textbf{1}, 2/3  \\ 
\hline
leptons & $L$ & ($\tilde{\nu}_{eL}$,$\tilde{e}_L$) & $(\nu_{eL},e_L)$ & & \textbf{1}, \textbf{2}, -1 \\ 
($\times 3$ families) & $e^c$ & $\tilde{e}^c_L$    & $e^c_L$          & & \textbf{1}, \textbf{1}, 2  \\ 
\hline
Higgs & $H_u$ & $(H_u^+ , H_u^0)$ & $(\tilde{H}_u^+ , \tilde{H}_u^0)$ & & \textbf{1}, \textbf{2}, 1  \\ 
      & $H_d$ & $(H_d^0 , H_d^-)$ & $(\tilde{H}_d^0 , \tilde{H}_d^-)$ & & \textbf{1}, \textbf{2}, -1 \\
\hline
gluons & g & & $\tilde{g}$                       & $g$              & \textbf{8}, \textbf{1}, 0 \\ 
$W$    & W & & $\tilde{W}^{\pm} , \tilde{W}^{0}$ & $W^{\pm} , W^0$  & \textbf{1}, \textbf{3}, 0 \\ 
$B$    & B & & $\tilde{B}$                       & $B$              & \textbf{1}, \textbf{1}, 0  \\ 
\hline
\end{tabular}
\caption{Chiral and gauge multiplet fields in the MSSM}
\label{diracgauginos_MSSMFields}
\end{center}
\end{table}

The field content of the MSSM is presented in table \ref{diracgauginos_MSSMFields}.
Note that all the chiral fermions  are left-handed, the charge conjugation
label $c$  allows to use the appropriate
antiparticles. At the renormalisable level,  the MSSM has the superpotential:
\begin{equation}
W = y^{ij}_u u^c_i Q_j \cdot H_u - y^{ij}_d d^c_i Q_j \cdot H_d 
- y^{ij}_e e^c_i L_j \cdot H_d + \mu H_u \cdot H_d .
\label{diracgauginos_MSSMSuperpotential}
\end{equation}
The indices $i,j$ are family indices and runs from $1$ to $3$. The $3 \times 3$ $y$ matrices are the
Yukawa couplings  and the parameter  $\mu$ is a Dirac mass for the Higgsinos. The "$\cdot$" denotes 
the $SU(2)$ invariant couplings, for example: 
$Q \cdot H_u = \tilde{u}_L H_u^0 - \tilde{d}_L H_u^+ $.

\subsubsection*{Soft breaking}

The breaking of supersymmetry is parametrized by a set of terms, labelled soft as they preserve the absence of 
quadratic divergences. The possible soft breaking terms in the MSSM 
are quite limited, there are gaugino masses for each gauge group, squarks 
mass terms, sleptons mass terms, Higgs mass terms and triple scalar couplings. We are primarily interested the gaugino masses given by: 
\begin{equation}
-\frac{1}{2} ( M_3 \tilde{g}^{\alpha} \tilde{g}^{\alpha} + M_2 \tilde{W}^{\alpha} \tilde{W}^{\alpha} 
+  M_1 \tilde{B} \tilde{B} + h.c.)
\label{diracgauginos_SoftGauginoMasses}
\end{equation}
Note that these are Majorama masses. Two of these terms for $\tilde {W}^{1,2}$ combine as a Dirac mass for $\tilde{W}^{\pm}$.

\subsubsection*{Neutralino masses}
\label{diracgauginos_secNeutralino}

 The neutral fermions of interest  are the
higgsinos $\tilde{H}^0_u$ and $\tilde{H}^0_d$ and the gauginos, bino $\tilde{B}$ and wino $\tilde{W}^0$. 
The mass terms for these fields in the MSSM have three origins:
\begin{itemize}
\item The soft breaking terms for the bino and wino,
\begin{equation}
-\frac{1}{2} ( M_2 \tilde{W}^0 \tilde{W}^0 
+  M_1 \tilde{B} \tilde{B} + h.c.) .
\label{diracgauginos_NeutralinoMasses1}
\end{equation}
\item The two last terms in equation (\ref{diracgauginos_CouplingLagrangian}) generate a mixing  between 
$(\tilde{B} , \tilde{W}^0)$ and $(\tilde{H}^0_u , \tilde{H}^0_d)$.  These mass terms are parametrized by the vev
 of the Higgs scalars $<H^0_u> \equiv v_u$ and $<H^0_d>\equiv v_d$ . Using $\tan \beta = v_u / v_d$, one can then 
express them in terms of $\beta$, the weak gauge bosons masses $m_W$ and $m_Z$ 
and the weak mixing angle $\theta_W$ as
\begin{equation}
- m_Z \left[ 
\cos \theta_W  ( \cos \beta \,  \tilde{H}^0_d \tilde{W}^0
- \sin \beta \, \tilde{H}^0_u \tilde{W}^0 ) 
+ \sin \theta_W ( \sin \beta \,  \tilde{H}^0_u \tilde{B} 
-  \cos \beta  \, \tilde{H}^0_d \tilde{B} )
+ h.c.\right]
\label{diracgauginos_NeutralinoMasses2}
\end{equation}
\item The $\mu$ term in the superpotential $W$  contributes 
to the higgsinos masses ,
\begin{equation}
\mu \tilde{H}^0_u , \tilde{H}^0_d + h.c.
\label{diracgauginos_NeutralinoMasses3}
\end{equation}
\end{itemize}

\subsubsection*{Chargino masses}
\label{diracgauginos_secChargino}

Here we consider the charged higgsinos $\tilde{H}^+_u$ and $\tilde{H}^-_d$ and the charged 
gauginos $\tilde{W}^+$ and $\tilde{W}^-$. In the MSSM the origin of the chargino mass terms is completely analogous 
to those presented in subsection \ref{diracgauginos_secNeutralino}, they take here the following form:
\begin{equation}
- M_2 \tilde{W}^+ \tilde{W}^- 
\label{diracgauginos_CharginoMasses1}
\end{equation}
\begin{equation}
- \sqrt{2} m_W \sin \beta \tilde{H}^+_u \tilde{W}^- 
- \sqrt{2} m_W \cos \beta \tilde{H}^-_d \tilde{W}^+ + h.c.
\label{diracgauginos_CharginoMasses2}
\end{equation}
\begin{equation}
- \mu \tilde{H}^+_u , \tilde{H}^-_d + h.c.
\label{diracgauginos_CharginoMasses3}
\end{equation}

\subsection{Extended susy gauge sector}

We now consider the scenario where the gauge sector arise in multiplets of $N = 2$ supersymmetry while matter states 
are in $N = 1$ SUSY representations. Moreover, the Higgs multiplets $H_u$ and $H_d$ are assumed to form an $N = 2$ hypermultiplet.

The field content for the gauge sector is described in table \ref{diracgauginos_ExtendedFields}. Note 
that for each $N = 1$ gauge multiplet present in the MSSM one need to introduce one extra scalar 
and  fermionic fields. The latter are differentiated by a symbol $'$ (see  
table \ref{diracgauginos_ExtendedFields}).

\begin{table}[htb]
\begin{center}
\begin{tabular}{c|c|c|c|c}
\hline
Names  & Spin 0 & Spin 1/2 & Spin 1 & $SU(3)$, $SU(2)$, $U(1)_Y$ \\ 
\hline
gluons &  $\Sigma_g$  & $\tilde{g}$ , $\tilde{g}'$ & $g$ & \textbf{8}, \textbf{1}, 0 \\ 
$W$    & $\Sigma_W^{\pm},\Sigma_W^0$ & $\tilde{W}^{\pm},\tilde{W}^{0},\tilde{W}'^{\pm},\tilde{W}'^{0}$ & $W^{\pm},W^0$ & \textbf{1}, \textbf{3}, 0 \\ 
$B$    & $\Sigma_B$ & $\tilde{B}$, $\tilde{B}'$    & $B$ & \textbf{1}, \textbf{1}, 0  \\ 
\hline
\end{tabular}
\caption{N = 2 gauge supermultiplets fields}
\label{diracgauginos_ExtendedFields}
\end{center}
\end{table}

In addition to the Majorana masses, Dirac ones can now be written. We will extended the MSSM soft terms to  include 
\begin{equation}
-\frac{1}{2} ( M'_3 \tilde{g}'^{\alpha} \tilde{g}'^{\alpha} 
+ M'_2 \tilde{W}'^{\alpha} \tilde{W}'^{\alpha} 
+ M'_1 \tilde{B}' \tilde{B}' )
- ( M^D_3 \tilde{g}^{\alpha} \tilde{g}'^{\alpha}
+ M^D_2 \tilde{W}^{\alpha} \tilde{W}'^{\alpha} 
 + M^D_1 \tilde{B} \tilde{B}' )
+ h.c.
\label{diracgauginos_ExtendedSoftMasses}
\end{equation}
where $M_ i$ are Majorana  and $M^D_ i$ are Dirac masses.

The $N = 2$ supersymmetry in the gauge sector introduces new couplings analogous 
to the two last terms in equation (\ref{diracgauginos_CouplingLagrangian}). These lead to
 new bilinear mixing terms between gauginos and higgsinos when 
the Higgs scalars $H^0_u$ and $H^0_d$ acquire vevs. 

\begin{itemize}
\item Neutralinos:
\begin{equation}
- m_Z \left[ 
\sin \theta_W ( \sin \beta \tilde{H}^0_d \tilde{B}' 
+  \cos \beta \tilde{H}^0_u \tilde{B}' )
- \cos \theta_W  ( \cos \beta \tilde{H}^0_u \tilde{W}'^0
+ \sin \beta \tilde{H}^0_d \tilde{W}'^0  )
+ h.c.\right]
\label{diracgauginos_ExtendedNeutralinoMasses}
\end{equation}
\item Charginos:
\begin{equation}
- \sqrt{2} m_W \cos \beta \tilde{H}^+_u \tilde{W}'^- 
+ \sqrt{2} m_W \sin \beta \tilde{H}^-_d \tilde{W}'^+ + h.c.
\label{diracgauginos_ExtendedCharginoMasses}
\end{equation}
\end{itemize}

\subsection{Fermionic mass matrix}
\label{diracgauginos_secMasses}

We now put all the previous terms together and describe the resulting mass matrices for 
both neutral and charged gauginos  and higgsinos when both Majorana and Dirac term are present.

\subsubsection*{Neutralinos}
\label{diracgauginos_secExtendedNeutralinos}

The neutralino mass terms are presented in equations (\ref{diracgauginos_NeutralinoMasses1} )-(\ref{diracgauginos_NeutralinoMasses3}), 
(\ref{diracgauginos_ExtendedSoftMasses}) and (\ref{diracgauginos_ExtendedNeutralinoMasses}).  The 
neutralino mass matrix, in the 
$(\tilde{B}', \tilde{B}, \tilde{W}'^0, \tilde{W}^0, \tilde{H}^0_d, \tilde{H}^0_u)$ basis is:
\begin{equation}
M_{Neut} = 
\left(\begin{array}{c c c c c c}
 M'_1  & M^D_1 & 0     & 0     &  m_Z s_W s_\beta &   m_Z s_W c_\beta  \\
 M^D_1 & M_1   & 0     & 0     & -m_Z s_W c_\beta &   m_Z s_W s_\beta  \\
 0     & 0     & M'_2  & M^D_2 & -m_Z c_W s_\beta & - m_Z c_W c_\beta  \\
 0     & 0     & M^D_2 & M_2   &  m_Z c_W c_\beta & - m_Z c_W s_\beta  \\
 m_Z s_W s_\beta & -m_Z s_W c_\beta & -m_Z c_W s_\beta &  m_Z c_W c_\beta & 0    & -\mu \\
 m_Z s_W c_\beta &  m_Z s_W s_\beta & -m_Z c_W c_\beta & -m_Z c_W s_\beta & -\mu & 0    \\
\end{array}\right) 
\label{diracgauginos_NeutralinoMassarray}
\end{equation}
where $c_W = \cos \theta_W$, $s_W = \sin \theta_W$, $c_\beta = \cos \beta$ and $s_\beta = \sin \beta$.

Clearly this $6 \times 6$ matrix will provide in general very long expressions for the neutralino 
mass eigenstates. A simple  case is  when the $m_Z$ dependent terms 
in (\ref{diracgauginos_NeutralinoMassarray}) are relatively small compared to the other entries, and can be 
treated as perturbations. Moreover, the  gaugino Majorama masses 
are symmetric in the primed and unprimed fermions: $M'_1 = M_1$ and $M'_2 = M_2$.
In this case the higgsino mass eigenstates are given 
(approximately) by the combinations 
\begin{equation}
\tilde{H}^0_S \simeq \frac{1}{\sqrt{2}} (\tilde{H}^0_u + \tilde{H}^0_d) \quad,\quad
\tilde{H}^0_A \simeq \frac{1}{\sqrt{2}} (\tilde{H}^0_u - \tilde{H}^0_d) 
\label{diracgauginos_HiggsinoEigenstates}
\end{equation}
both having mass squared $\mu^2$. 
The neutral gaugino mass eigenstates are given, 
to leading order in $m_Z/M_i$, by  
\begin{equation}
\tilde{B}_S = \frac{1}{\sqrt{2}} (\tilde{B} + \tilde{B}') \quad,\quad
\tilde{B}_A = \frac{1}{\sqrt{2}} (\tilde{B} - \tilde{B}' ) 
\label{diracgauginos_BinoEigenstates}
\end{equation}
\begin{equation}
\tilde{W}^0_S = \frac{1}{\sqrt{2}} (\tilde{W}^0 + \tilde{W}'^0) \quad,\quad
\tilde{W}^0_A = \frac{1}{\sqrt{2}} (\tilde{W}^0 - \tilde{W}'^0 ) 
\label{diracgauginos_WinoEigenstates}
\end{equation}
with masses 
\begin{eqnarray}
m_{\tilde{B}_S} \simeq M_1 + M^D_1, 
&& m_{\tilde{B}_A} \simeq M_1 - M^D_1 \\
m_{\tilde{W}^0_S} \simeq M_2 + M^D_2 
&& m_{\tilde{W}^0_A} \simeq M_2 - M^D_2. 
\end{eqnarray}
One can express the ratio between Dirac and Majorama masses by the angle $\theta^D$ 
defined by $\tan \theta^D = M^D / M$. Alternatively, the angle $\theta^D$ is measured by 
\begin{equation}
\sin 2 \theta^D_B = \frac{m^2_{\tilde{B}_S} 
- m^2_{\tilde{B}_A}}{m^2_{\tilde{B}_S} + m^2_{\tilde{B}_A}}
\label{diracgauginos_BMassAngle}
\end{equation}
and
\begin{equation}
\sin 2 \theta^D_{W} = \frac{m^2_{\tilde{W}^0_S} 
- m^2_{\tilde{W}^0_A}}{m^2_{\tilde{W}^0_S} + m^2_{\tilde{W}^0_A}} .
\label{diracgauginos_WMassAngle}
\end{equation}

\subsubsection*{Charginos}
\label{diracgauginos_secExtendedCharginos}

The mass terms for the charginos can be expressed in the form 
\begin{equation}
- \frac{1}{2} ( (v^-)^T M_{Ch} v^+ + (v^+)^T M_{Ch}^T v^- + h.c)
\label{diracgauginos_CharginoMassLagrangian}
\end{equation}
where we addopted the basis $v^+ = (\tilde{W}'^+,\tilde{W}^+,\tilde{H}^+_u)$, 
$v^- = (\tilde{W}'^-,\tilde{W}^-,\tilde{H}^-_d)$. 
Collecting all the terms presented in equations (\ref{diracgauginos_CharginoMasses1})-( \ref{diracgauginos_CharginoMasses3}), 
(\ref{diracgauginos_ExtendedSoftMasses}) and  (\ref{diracgauginos_ExtendedCharginoMasses})  leads to the 
chargino mass matrix :
\begin{equation}
M_{Ch} = 
\left(\begin{array}{c c c}
 M'_2   & M^D_2 & \sqrt{2} m_W \cos \beta \\
M^D_2 & M_2   & \sqrt{2} m_W \sin \beta \\
 - \sqrt{2} m_W \sin \beta & \sqrt{2} m_W \cos \beta & \mu \\
\end{array}\right) .
\label{diracgauginos_CharginoMassarray}
\end{equation}
This nonsymmetric mass array is diagonalized by separate unitary transformations in 
the basis $v^+$ and $v^-$, $M_{Ch}^{diag} = U^{\dagger} M_{Ch} V$, where the matrices $U$ and $V$ are unitary.

For the  simple case considered in the previous subsection, where the $m_Z$ 
dependent terms can be treated as perturbations and $M'_2 = M_2$, 
the higgsino mass eigenstates will be given 
approximately by $\tilde{H}^+ _u$ and $\tilde{H}^-_d$, both 
having mass squared $\mu^2$. 
The charged gaugino mass eigenstates will be given, to leading order in $m_Z/M_i$, by the combinations 
\begin{equation}
\tilde{W}^+_S \simeq \frac{1}{\sqrt{2}} (\tilde{W}^+ + \tilde{W}'^+) \quad,\quad
\tilde{W}^+_A \simeq \frac{1}{\sqrt{2}} (\tilde{W}^+ - \tilde{W}'^+) 
\label{diracgauginos_WinoPlusEigenstates}
\end{equation}
\begin{equation}
\tilde{W}^-_S \simeq \frac{1}{\sqrt{2}} (\tilde{W}^- + \tilde{W}'^-) \quad,\quad
\tilde{W}^-_A \simeq \frac{1}{\sqrt{2}} (\tilde{W}^- - \tilde{W}'^-) 
\label{diracgauginos_WinoMinusEigenstates}
\end{equation}
with squared masses 
\begin{eqnarray}
m^2_{\tilde{W}^+_S} \simeq (M_2 + M^D_2)^2 & & m^2_{\tilde{W}^+_A} \simeq (M_2 - M^D_2)^2 \\
 m^2_{\tilde{W}^-_S} \simeq (M_2 + M^D_2)^2 & & m^2_{\tilde{W}^-_A} \simeq(M_2 - M^D_2)^2 
\end{eqnarray}
Note that, in this limit, the winos have approximate  degenerate masses: 
$m^2_{\tilde{W}^0_S} \approx m^2_{\tilde{W}^+_S} \approx m^2_{\tilde{W}^-_S}$ and 
$m^2_{\tilde{W}^0_A} \approx m^2_{\tilde{W}^+_A} \approx m^2_{\tilde{W}^-_A}$.

\subsubsection*{Gluinos}
\label{diracgauginos_secExtendedGluinos}

Since gluinos $\tilde{g}$ and $\tilde{g}'$ are in color octet representation, they 
cannot mix with any other fermion, the only possible gluino masses are the soft ones 
presented in (\ref{diracgauginos_SoftGauginoMasses}) and  (\ref{diracgauginos_ExtendedSoftMasses}). In the basis 
$(\tilde{g}', \tilde{g})$ the gluino mass matrix is simply 
\begin{equation}
M_{Glu} = 
\left(\begin{array}{c c}
 M'_3   & M^D_3 \\ 
 M^D_3 & M_3   \\
\end{array}\right) .
\label{diracgauginos_GuinoMassarray}
\end{equation}

We will illustrate  two limits. The first one is when gaugino Majorama masses 
are symmetric in the primed and unprimed fermions: $M'_3 = M_3$.
The analysis of the gluino mass matrix in this case follows closely the discussion 
after (\ref{diracgauginos_HiggsinoEigenstates}). The gluino mass eigenstates are 
\begin{equation}
\tilde{g}_S = \frac{1}{\sqrt{2}} (\tilde{g} + \tilde{g}') \quad,\quad
\tilde{g}_A = \frac{1}{\sqrt{2}} (\tilde{g} - \tilde{g}' ) 
\label{diracgauginos_GluinoEigenstates}
\end{equation}
with masses \begin{equation}
\label{diracgauginos_ }
m_{\tilde{g}_S} = M_3 + M^D_3, 
\qquad m_{\tilde{g}_A} = M_3 - M^D_3,
\end{equation}
and the ratio between Dirac and Majorama masses is parameterized 
by the angle  $\tan \theta^D_g = M^D_3 / M_3$,  alternatively by 
\begin{equation}
\sin 2 \theta^D_g = \frac{m^2_{\tilde{g}_S} 
- m^2_{\tilde{g}_A}}{m^2_{\tilde{g}_S} + m^2_{\tilde{g}_A}}
\label{diracgauginos_GMassAngle}.
\end{equation}

The second limit is when one of the Majorama masses, say $M'_3$, is very small 
compared to the other entries of the mass matrix   (\ref{diracgauginos_GuinoMassarray}). In this limit, the gluino mass eigenstates 
are given approximately by
\begin{equation}
\tilde{g}_1 \simeq \cos \alpha ~ \tilde{g} - \sin \alpha ~ \tilde{g}' \quad,\quad
\tilde{g}_2 \simeq \sin \alpha ~ \tilde{g} + \cos \alpha ~ \tilde{g}'  
\label{diracgauginos_GluinoEigenstates2}
\end{equation}
where
\begin{equation}
\tan \alpha = - \frac{1}{2} \cot \theta^D_ g \left(  1 + \sqrt{1 + 4 \tan^{2} \theta^D_ g } \right) 
\label{diracgauginos_EigenAngle2}
\end{equation}
and, as before, $\tan \theta^D_g = M^D_3 / M_3$. 
The gluino masses are 
\begin{equation}
m_{\tilde{g}_1} \simeq \frac{M}{2} \left(  1 + \sqrt{1 + 4 \tan^{2} \theta^D_ g } \right) \quad,\quad
m_{\tilde{g}_2} \simeq \frac{M}{2} \left(  1 - \sqrt{1 + 4 \tan^{2} \theta^D_ g } \right) 
\label{diracgauginos_GluinoMasses2}
\end{equation}
and the ratio between Dirac and Majorana masses is parameterized
\begin{equation}
\sin^2 \theta^D_g = \frac{ | m_{\tilde{g}_1} m_{\tilde{g}_2} | }
{m^2_{\tilde{g}_1} + m^2_{\tilde{g}_2} + | m_{\tilde{g}_1} m_{\tilde{g}_2} | } .
\label{diracgauginos_GMassAngle2}
\end{equation}

\subsection{Interactions}
\label{diracgauginos_secInteractions}

We turn now to the interactions between   the extended supersymmetric sector and the MSSM fields. First, we remind
the interactions between the MSSM gauginos, the higgsinos and the scalar Higgs:
\begin{equation}
- \frac{g}{\sqrt{2}} \left( H^*_u \sigma^i \tilde{H}_u \tilde{W}^i 
+ H^*_d \sigma^i \tilde{H}_d \tilde{W}^i  \right) 
-  \frac{g'}{\sqrt{2}} \left( H^*_u  \tilde{H}_u \tilde{B} 
- H^*_d  \tilde{H}_d \tilde{B}  \right) 
\label{diracgauginos_MSSMInteraction}
\end{equation}
where $g$ and $g'$ are the $SU(2)$ and $U(1)_Y$ coupling constants. 

Due to the fact that the two Higgs $H_u$ and $H_d$ form a $N = 2$ hypermultiplet, their 
interactions with the new fermions $W'$ and $B'$ are 
\begin{equation}
- \frac{g}{\sqrt{2}} \left[ H_u \cdot ( \sigma^i \tilde{H}_d ) \tilde{W}'^i 
+ H_d \cdot ( \sigma^i \tilde{H}_u ) \tilde{W}'^i   \right] 
- \frac{g'}{\sqrt{2}} \left( H_d \cdot \tilde{H}_u  \tilde{B}' 
- H_u \cdot \tilde{H}_d \tilde{B}'   \right).
\label{diracgauginos_ExtendedInteraction}
\end{equation}
One can straightforwardly verify that these interactions lead upon electroweak breaking to the gaugino-higgsino mixing 
present in the neutralino and chargino mass matrices .

\subsection{New scalars}
\label{diracgauginos_secScalars}

The $N=2$ vector multiplets include, in addition to the new fermions dicussed above, scalar fields in the adjoint representation.
We  denote these states as $\Sigma_g$, $\Sigma_W$ and $\Sigma_B$.
They couple to the  Higgs chiral fields (now in an $N = 2$  hypermultiplet ) in the superpotential, 
and through their $F$-term modify the tree-level  Higgs scalars  quartic terms in the potential by the new terms:
\begin{equation}
- \frac{g^2}{2} \sum_{i} \left| H_u \cdot \sigma^i H_d \right|^2 
- \frac{g'^2}{2} \left| H_u \cdot H_d \right|^2.
\label{diracgauginos_QuarticHiggs}
\end{equation}

These scalar should  not remains massless, but get soft terms as: 
\begin{equation}
- \frac{1}{2} m^2_{3S}  \Sigma^{\alpha}_g \Sigma^{*\,\alpha}_g 
- \frac{1}{2} m^2_{3A} \Sigma^{\alpha}_g \Sigma^{\alpha}_g 
- \frac{1}{2} m^2_{2S} \Sigma^{\alpha}_W \Sigma^{*\,\alpha}_W 
- \frac{1}{2} m^2_{2A} \Sigma^{\alpha}_W \Sigma^{\alpha}_W 
- \frac{1}{2} m^2_{1S} \Sigma_B \Sigma^*_B.
- \frac{1}{2} m^2_{1A} \Sigma_B \Sigma_B.
\label{diracgauginos_NewSoftPotential}
\end{equation}
with $ m^2_{iS} >m^2_{iA}$.  If the masses in (\ref{diracgauginos_NewSoftPotential}) are 
big compared to the Higgs mass,  these fields can be integrated out and 
in the low energy theory the scalar potential is the one in the MSSM plus the 
contributions coming from the quartic terms (\ref{diracgauginos_QuarticHiggs})  \cite{Antoniadis:2006uj}. Note that the latter contribution disappears, if instead, 
the integration out is supersymmetric (due to a large supersymmetric mass).

\subsection*{Acknowledgements}
We wish to thank the organizers of the "Les Houches: TeV Colliders" for their hospitality.


\clearpage
\newpage

\setcounter{figure}{0}
\setcounter{table}{0}
\setcounter{equation}{0}
\setcounter{footnote}{0}
\section{NMSSM in disguise: discovering singlino dark matter with soft leptons
\protect\footnote{S.~Kraml and A.R.~Raklev}}

\subsection{Introduction}

The Next-to-Minimal Supersymmetric Standard Model (NMSSM) provides an
elegant solution to the $\mu$ problem of the MSSM by the addition of a
gauge singlet superfield $\hat S$ \cite{Fayet:1974pd,Nilles:1982dy,
Frere:1983ag,Derendinger:1983bz}. The superpotential of the Higgs
sector then has the form $\lambda\hat S(\hat H_d\cdot\hat
H_u)+\frac{1}{3}\kappa\hat S^3$. When $\hat S$ acquires a vacuum
expectation value, this creates an effective $\mu$ term,
$\mu\equiv\lambda\langle S\rangle$, which is automatically of the
right size, {\it i.e.}\ of the order of the electroweak scale.

The addition of the singlet field leads to a larger particle spectrum
than in the MSSM: in addition to the MSSM fields, the NMSSM contains
two extra neutral (singlet) Higgs fields --\,one scalar and one
pseudoscalar\,-- as well as an extra neutralino, the singlino. Owing
to these extra states, the phenomenology of the NMSSM can be
significantly different from the MSSM; see chapter~4 of
\cite{Accomando:2006ga} for a recent review and references. In
particular, the usual LEP limits do not apply to singlet and
singlino states. Moreover, the singlino can be the lightest
supersymmetric particle (LSP) and a cold dark matter candidate.

In this contribution, we investigate the LHC signature of a SPS1a-like
scenario, supplemented by a singlino LSP. In such a setup, gluinos and
squarks have the `conventional' SUSY cascade decays into the bino-like
neutralino, ${\tilde\chi^0_2}\sim\tilde B$, which then decays into the
singlino LSP, ${\tilde\chi^0_1}\sim\tilde S$, plus a pair of
opposite sign same-flavour (OSSF) leptons. (The ${\tilde\chi^0_2}$
decay proceeds dominantly through an off-shell slepton.) A dark matter
relic density of $\Omega h^2\sim 0.1$
is obtained if the ${\tilde\chi^0_1}$ and/or ${\tilde\chi^0_2}$
annihilate through pseudoscalar exchange in the s-channel.

One peculiar feature of this scenario is that the mass difference
between ${\tilde\chi^0_1}$ and ${\tilde\chi^0_2}$ is always small; it
reaches at most $\sim 12$ GeV, and is often much smaller. The leptons
originating from the ${\tilde\chi^0_2}\to {\tilde\chi^0_1}l^+l^-$
decay hence tend to be soft. In the standard SUSY analysis, requiring
$p_T(l^\pm)>20$~GeV, there is a risk of missing these leptons and
wrongly concluding to have found the MSSM instead of the NMSSM, with
${\tilde\chi^0_2}$ as the LSP and dark matter candidate (discovery of
the additional Higgs states will also be very difficult at the LHC in
this scenario). The aim of this contribution is to show the
feasibility of detecting the ${\tilde\chi^0_2}\to
{\tilde\chi^0_1}l^+l^-$ decay and measuring the singlino--bino mass
difference by looking for soft di-leptons.

We use the {\tt NMHDECAY}~\cite{Ellwanger:2004xm,Ellwanger:2005dv}
program to compute the NMSSM mass spectrum and Higgs branching ratios,
and to evaluate the LEP bounds; {\tt SPHENO}~\cite{Porod:2003um} is used 
to calculate the sparticle branching ratios, and 
{\tt MICROMEGAS}~\cite{Belanger:2005kh,Belanger:2006is} for the relic density.
The SUSY-breaking parameters of our scenario are listed in
Table~\ref{tab:KramlRaklev_MSSMparameters}. The main difference to the original
SPS1a~\cite{Allanach:2002nj} is that we choose $M_1=0.5M_2=120$~GeV,
leading to a $\tilde\chi^0_2$ mass of $\simeq115$~GeV, in order to
evade LEP bounds when adding the singlino and singlet Higgses. To
obtain a singlino LSP, we choose $\lambda \sim 10^{-2}$ and
$\kappa\sim 0.1\lambda$. This way $\tilde\chi^0_1\sim 99\%\,\tilde S$,
and $m_{\tilde\chi^0_2}$ hardly varies with $\lambda$ and $\kappa$
($\sim 0.1$~GeV). In addition, the trilinear Higgs couplings
$A_\lambda$ and $A_\kappa$ are chosen such that
$m_{\tilde\chi^0_i}+m_{\tilde\chi^0_j}\sim m_{A_2}$ for at least one
combination of $i,j=1,2$, to achieve $0.094\le\Omega h^2\le
0.135$~\cite{Hamann:2006pf}. We thus obtain a set of NMSSM parameter
points with varying $\Delta m \equiv
m_{\tilde\chi^0_2}-m_{\tilde\chi^0_1}$.

\begin{table}[b]
\begin{center}
\begin{tabular}{ccccccccccccc}
\hline
   $M_1$ & $M_2$ & $M_3$ & $\mu_{\rm eff}$ 
                 & $M_{\tilde L_{1,3}}$ & $M_{\tilde E_1}$ & $M_{\tilde E_3}$
                 & $M_{\tilde Q_1}$ & $M_{\tilde U_1}$ & $M_{\tilde D_1}$ 
                 & $M_{\tilde Q_3}$ & $M_{\tilde U_3}$ & $M_{\tilde D_3}$\\    
  120 & 240 & 720 & 360 & 195 & 136 & 133 & 544 & 526 & 524 & 496 & 420 & 521\\
\hline
\end{tabular}
\caption{Input parameters in [GeV] for our SPS1a-like scenario.  
   The NMSSM-specif\/ic parameters are given in Table~\ref{tab:KramlRaklev_NMSSMpoints}.}
\label{tab:KramlRaklev_MSSMparameters}
\end{center}
\end{table}
 

\begin{table}[t]
\begin{center}
\begin{tabular}{cccccccccccc}
\hline
  Point & $\lambda\,[10^{-2}]$ & $\kappa\,[10^{-3}]$ & $A_\lambda$ & $A_\kappa$ 
           & $m_{\tilde\chi^0_1}$  & $m_{A_1}$ & $m_{A_2}$ & $m_{S_1}$   
           & $\Omega h^2$ &  $\Gamma({\tilde\chi^0_2})$\\
\hline
 A & $1.49$ & $2.19$ & $-37.4$ & $-49.0$& $105.4$ & $88$ & $239$ & $89$  & $0.101$  & $7\times10^{-11}$ \\ 
 B & $1.12$ & $1.75$ & $-42.4$ & $-33.6$& $112.1$ & $75$ & $226$ & $100$  & $0.094$  & $9\times10^{-13}$ \\ 
 C & $1.20$ & $1.90$ & $-39.2$ & $-53.1$& $113.8$ & $95$ & $256$ & $97$  & $0.094$  & $1\times10^{-13}$ \\ 
 D & $1.47$ & $2.34$ & $-39.2$ & $-68.9$& $114.5$ & $109$ & $259$ & $92$  & $0.112$  & $4\times10^{-14}$ \\ 
\hline 
\end{tabular}
\caption{NMSSM benchmark points used in this study. Masses and other
dimensionful quantities are in [GeV].
}
\label{tab:KramlRaklev_NMSSMpoints}
\end{center}
\end{table}

The four points used for this study are summarised in
Table~\ref{tab:KramlRaklev_NMSSMpoints}. Points A--D have $\Delta
m=9.7$, $3.0$, $1.5$ and $0.9$ GeV, respectively. The SM-like second
neutral scalar Higgs, $S_2$, has a mass of 115~GeV for all these
points, consistent with the LEP limit.  On the other hand, the
lightest neutral scalar $S_1$ and the lighter pseudoscalar $A_1$ are
mostly singlet states, and can hence be lighter than 114~GeV.
Concerning the neutralino annihilation, for Point A the dominant
channel is $\tilde\chi^0_2\tilde\chi^0_2\to b\bar b$, contributing
$88\%$ to $\langle\sigma v\rangle$. For Point B,
$\tilde\chi^0_1\tilde\chi^0_1$, $\tilde\chi^0_1\tilde\chi^0_2$ and
$\tilde\chi^0_2\tilde\chi^0_2$ annihilation to $b\bar b$ contribute
$10\%$, $15\%$, and $50\%$, respectively.  Point C has again
dominantly $\tilde\chi^0_2\tilde\chi^0_2$, while Point D has about
50\% $\tilde\chi^0_2\tilde\chi^0_2$ and 35\%
$\tilde\chi^0_1\tilde\chi^0_2$ annihilation.

\begin{figure}
\begin{center}
\includegraphics[width=0.5\textwidth]{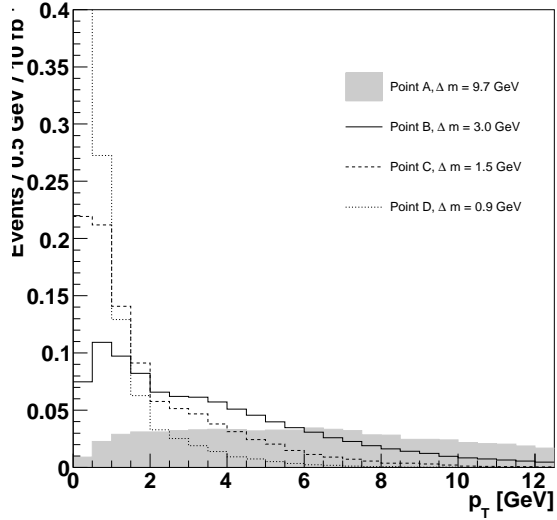}
\caption{$p_T$ distributions for leptons from the decay
$\tilde\chi^0_2\to\tilde\chi^0_1l^+l^-$ in benchmark points A--D. All
distributions are normalised to unity over the whole momentum range.}
\label{fig:KramlRaklev_truept}
\end{center}
\end{figure}

Figure~\ref{fig:KramlRaklev_truept} shows the resulting $p_T$
distributions for leptons from decays to singlinos for all four
benchmark points. Clearly, cuts on lepton transverse momentum of even
10 GeV will remove the wast majority of events for points
B--D. However, one should notice that the distributions have
considerable tails beyond the simple mass difference $\Delta m$, due
to the boost of the $\tilde\chi^0_2$.

\subsection{Monte Carlo analysis}

We perform a Monte Carlo simulation of the benchmarks described above
by generating both SUSY signal and SM background events with {\tt
PYTHIA~6.413}~\cite{Sjostrand:2006za}. The generated events are then
put through a fast simulation of a generic LHC detector, {\tt
AcerDET-1.0}~\cite{Richter-Was:2002ch}. Although {\tt PYTHIA} does not
contain a framework for generating NMSSM events {\it per se}, it has
the capability to handle the NMSSM spectrum and its decays. Since our
scenario predicts the same dominant cross section as in the MSSM,
namely gluino and squark pair-production, with negligible interference
from the non-minimal sector, we use the built-in MSSM machinery for
the hard process, and generate only squark and gluino pair-production.

The detector simulation is done with standard {\tt AcerDET} settings,
with one exception: for detecting decays to the singlino the
detector response to soft leptons is vital. We therefore parametrise
the efficiency of electron and muon identification as a function of
lepton $p_T$. For muons we base ourselves on the efficiencies shown in
Figure~8-5 and 8-9 of the ATLAS~TDR~\cite{unknown:1999fq}; for
electrons we use the same parametrisation scaled down by 0.82. While
this is certainly not a perfect description of the real ATLAS or CMS
efficiencies during data taking, it incorporates some of the most
important effects in an analysis, such as an absolute lower limit for
lepton identification, at around 2--3~GeV for muons, and a difference
in electron and muon efficiencies. However, it does not address other
important issues, e.g.\ mis-identification of charged pions as
electrons. To improve on these simple assumptions one would need a
full simulation of the detectors, or efficiencies from
data, which is clearly beyond the scope of this contribution.

We generate events corresponding to 10~fb$^{-1}$ of both signal and
background (some with weights). Our background consists of large
$p_T$-binned samples of QCD $2\to 2$ (10M), $W$+jet (4M), $Z$+jet
(3M), $WW/WZ/ZZ$ (1M) and $t\bar t$ (5M) events. For the
signal, {\tt PYTHIA} gives a LO cross section of 24~pb for squark and
gluino pair-production, thus 240~000 events are generated per
benchmark point.

\begin{figure}
\begin{center}
\includegraphics[width=\textwidth]{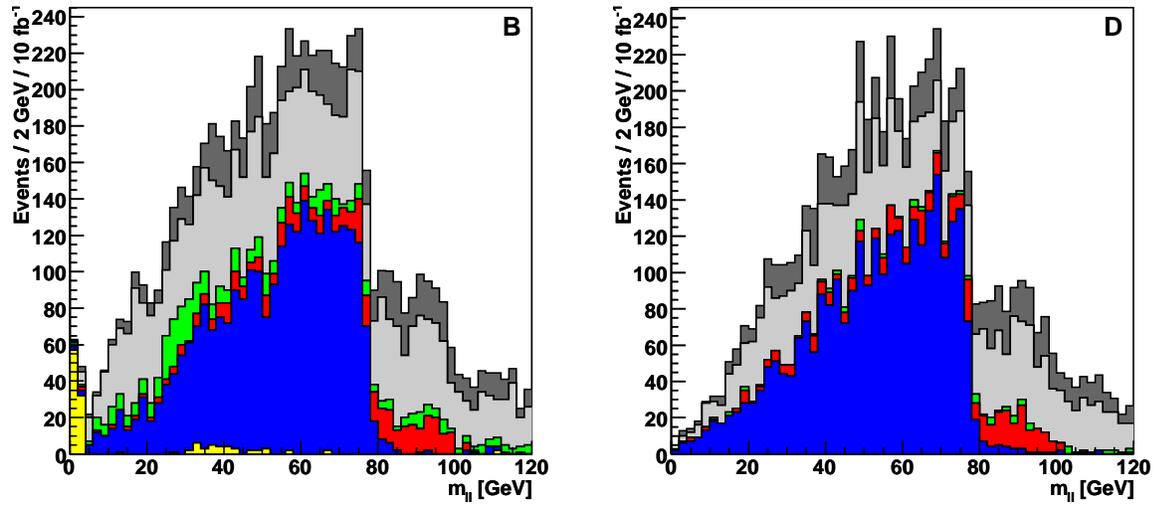}
\caption{Di-lepton invariant mass distributions for point B (left) and
point D (right) with standard lepton $p_T$ cuts. See text for colour
coding.}
\label{fig:KramlRaklev_content_stdlepc}
\end{center}
\end{figure}

We begin our analysis along the lines of the `standard' di-lepton
edge analysis. To isolate the SUSY signal from SM background we apply
the following cuts:
\begin{itemize}
\item
Require at least three jets with $p_T>150,100,50$~GeV.
\item
Require missing transverse energy $\not\!\! E_T > \max(100~{\rm
GeV},0.2M_{\rm eff})$, where the effective mass $M_{\rm eff}$ is the
sum of the $p_T$ of the three hardest jets plus missing energy.
\item
Require two OSSF leptons with $p_T>20,10$~GeV.
\end{itemize}
After these cuts the background is small and consists mainly of
$t\bar{t}$, with some vector boson events surviving. The resulting
di-lepton invariant mass distributions for points B and D can be seen
in the left and right panels, respectively, of
Fig.~\ref{fig:KramlRaklev_content_stdlepc}. The contribution from
decays of $\tilde\chi_3^0$ to $\tilde\chi_2^0$ via right and
left-handed sleptons are shown in red and blue, other SUSY events,
where the leptons mainly come from chargino or stau decays, in light
grey, and the remaining SM background in dark grey. For the B
benchmark point there is a small excess of events coming from decays
of $\tilde\chi_2^0$ to singlinos (yellow) at low invariant masses,
that survives due to other harder leptons in the event. However, all
such events are removed for benchmark point D because of the hard
lepton $p_T$ cut.\footnote{Green denotes lepton combinations with one
lepton from a slepton decay chain and the other from a decay to a
singlino. The few pure singlino events at higher invariant masses are
due to mis-combinations of leptons from different $\tilde\chi_2^0$
decays.} In this case one would miss the singlino and take the
$\tilde\chi_2^0$ to be the LSP dark matter candidate.

\begin{figure}
\begin{center}
\includegraphics[width=\textwidth]{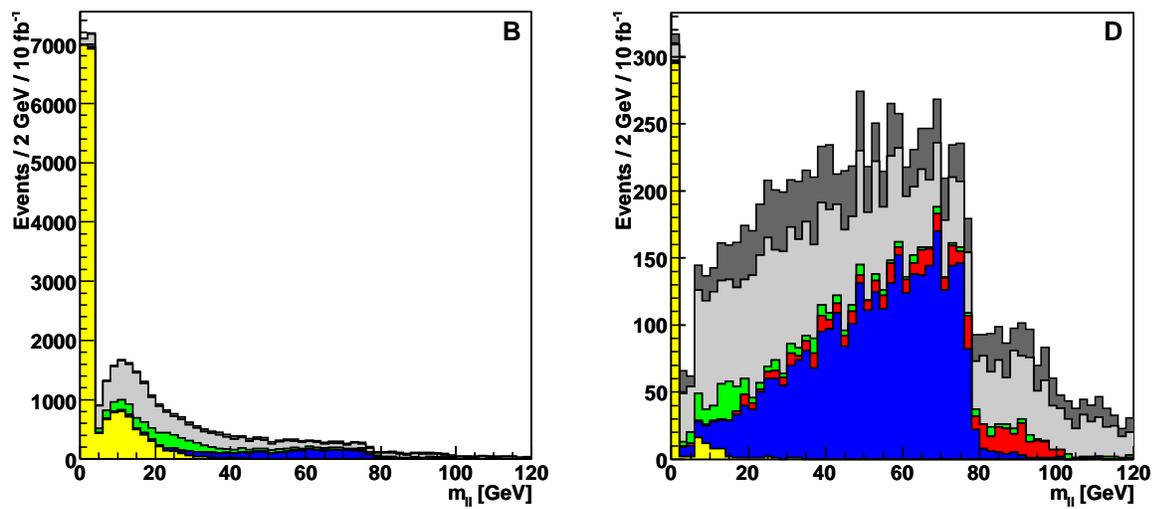}
\caption{Di-lepton invariant mass distributions for point B (left) and
point D (right) without lepton $p_T$ cuts.}
\label{fig:KramlRaklev_content_nolepc}
\end{center}
\end{figure}

It is clear that to increase sensitivity to the disguised NMSSM
scenario, one needs to lower the lepton $p_T$ cuts. However, this
opens the possibility for large increases in background. While most of
this background, from uncorrelated leptons, can in principle be
removed by subtracting the corresponding opposite sign
opposite-flavour (OSOF) distribution, assuming lepton universality,
large backgrounds will increase the statistical error and a soft
lepton sample is more vulnerable to non-universality from e.g.\ pion
decays. The result of completely removing the $p_T$ requirement on the
leptons is shown for benchmark points B and D in the left and right
panels of Fig.~\ref{fig:KramlRaklev_content_nolepc},
respectively. While there is indeed an increase in backgrounds, the
effect on the signal is much more significant. For both benchmarks,
the decay to the singlino is now visible as a large excess at low
invariant masses. We have also tested scenarios with smaller values of
$\Delta m$, and find that we have a significant excess down to $\Delta
m\simeq 0.6$~GeV, with the assumptions on lepton efficiencies
described above.\footnote{In fact, for such small mass differences we
may also see displaced vertices due to the long lifetime of the
$\tilde\chi_2^0$.}

\begin{figure}
\begin{center}
\includegraphics[width=\textwidth]{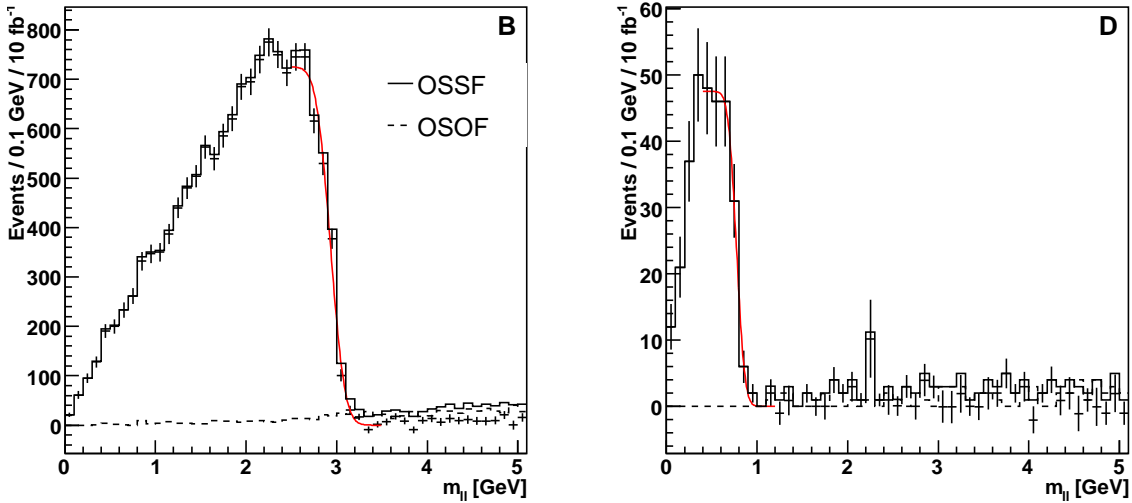}
\caption{Di-lepton invariant mass distributions for point B (left) and
point D (right) after OSOF subtraction.}
\label{fig:KramlRaklev_subtracted}
\end{center}
\end{figure}

In the standard di-lepton analysis the edges of the red and blue
distributions shown in Fig.~\ref{fig:KramlRaklev_content_stdlepc} can
be used to determine the relationship between the neutralino and
slepton masses, in our scenario $m_{\tilde\chi^0_3}^2-m_{\tilde l}^2$
and $m_{\tilde l}^2-m_{\tilde\chi^0_2}^2$. We extract additional
information by also determining the position of the edge at low
invariant masses, fitting a Gaussian-smeared step function to the OSOF
subtracted distribution, shown in
Fig.~\ref{fig:KramlRaklev_subtracted}. In subtracting the OSOF
distribution we have taken into account the asymmetry induced by the
the difference in electron and muon efficiencies. This fit determines
the mass difference $\Delta m$, since
$m_{ll}^{\max}=m_{\tilde\chi^0_2}-m_{\tilde\chi^0_1}$. For point B the
result of the fit is $m_{ll}^{\max}=2.93\pm 0.01$~GeV, to be compared
with the nominal value of $3.05$~GeV, while for point D the result is
$m_{ll}^{\max}=0.77\pm 0.02$~GeV, with a nominal value of
$0.87$~GeV. Both results are significantly on the low side with
respect to the small statistical errors. We speculate that this
systematic error is at least in part due to the step function used in
the fit to the edge, and that a more sophisticated description will
give results closer to the nominal values.

A final comment is in order
concerning early discoveries: in fact, since all SUSY cascades will
contain the decay to a singlino, the lower edge in the di-lepton
distribution may appear much earlier than the `standard' decay
through a slepton, if at all present, provided that the soft leptons
are searched for.

\subsection{Conclusions}

We have demonstrated that lowering the requirements on lepton
transverse momentum in the standard di-lepton edge of SUSY searches
may reveal unexpected features, such as the NMSSM in disguise. While
our numerical results are sensitive to the exact lepton efficiencies,
to be measured at the experiments, and while there may be additional
backgrounds not simulated, such as multi gauge boson and/or multi jet
final states, the OSOF subtraction procedure ensures that the
background is removable and the NMSSM scenario in question is
discoverable down to very small mass differences $\Delta
m=m_{\tilde\chi^0_2}-m_{\tilde\chi^0_1}$.

\subsection*{Acknowledgements}
ARR wishes to thank the members of the Cambridge Supersymmetry Working
Group for many useful discussions, and acknowledges funding from the
UK Science and Technology Facilities Council (STFC).

\clearpage
\newpage

\setcounter{figure}{0}
\setcounter{table}{0}
\setcounter{equation}{0}
\setcounter{footnote}{0}

\section{The MSSM with decoupled scalars at the LHC
\protect\footnote{R.~Lafaye, T.~Plehn, M.~Rauch, E.~Turlay, and D. Zerwas}}



\subsection{Introduction}
Assuming a large soft--breaking scale for all MSSM
scalars~\cite{ArkaniHamed:2004fb,Giudice:2004tc,Wells:2004di,Kilian:2004uj,Bernal:2007uv}
pushes squarks, sfermions and heavy Higgses out of the reach of the
LHC without affecting the gaugino sector. Even though the hierarchy
problem will not be solved without an additional logarithmic fine
tuning in the Higgs sector, such models can be constructed to provide
a good dark--matter candidate and realize grand unification while
minimizing proton decay and FCNCs. We investigate their LHC
phenomenology, with all scalars decoupled from the low--energy
spectrum. We focus on gaugino--related signatures to estimate the
accuracy with which its underlying parameters can be
determined~\cite{Lafaye:2007vs,Bechtle:2005vt}.

\subsection{Phenomenology}
The spectrum at LHC mass scales is reduced to the Standard--Model with
a light Higgs, plus gauginos and Higgsinos. At the high scale $M_S$
the effective theory is matched to the full MSSM and the usual
renormalization group equations apply. The Higgsino mass parameter
$\mu$ and the ratio $\tan\beta$ in the Higgs sector correspond to
their MSSM counter parts. The gaugino masses $M_{1,2,3}$ and the
Higgs-sfermion-sfermion couplings unify, and $M_S$ replaces the
sfermion and the heavy Higgs' mass parameters.  This set resembles the
mSUGRA parameter set except for $\tan\beta$ now playing the role of a
matching parameter (with all heavy Higgses being decoupled) rather
than that of an actual vev ratio~\cite{Drees:2005cp}.

We select our parameter point lead by three constraints: first, we
minimize the amount of fine tuning necessary to bring the light Higgs
mass into the 100 to 200~GeV range and reduce $M_S$ to 10~TeV, which
is still outside the LHC mass range.  Another reason for this low
breaking scale is that we want the gluino to decay inside the detector
(preferably at the interaction point) instead of being
long--lived~\cite{Kraan:2004tz,Kilian:2004uj}.  Heavier sfermions
increase the life time of the gluino such that it creates a displaced
vertex or even hadronizes~\cite{Farrar:1978xj}.

Secondly, we obtain the correct relic dark--matter density $\Omega
h^2=0.111^{+0.006}_{-0.008}$~\cite{Spergel:2006hy} by setting
$\mu=290$ GeV and $M_2(M_{\rm GUT})=132.4$~GeV or $M_2(M_{\rm
  weak})=129$~GeV. This corresponds to the light--Higgs funnel
$m_{\rm LSP} \approx M_2/2 \approx m_h/2$, where the $s$-channel Higgs
exchange enhances the LSP annihilation rate. And finally, $m_h$ needs
to be well above the LEP limit, which we achieve by choosing
$\tan\beta=30$. We arrive at a parameter point with $m_h = 129$~GeV,
$m_{\tilde{g}}=438$~GeV, chargino masses of 117 and 313~GeV, and
neutralino masses of 60, 117, 296, and 310~GeV, using a modified
version of SuSpect which decouples the heavy scalars from the MSSM
RGEs~\cite{Bernal:2007uv,Djouadi:2002ze}. The neutralinos/charginos
$\tilde\chi^{0}_2$ and $\tilde\chi^{\pm}_1$ as well as
$\tilde\chi^{0}_4$ and $\tilde\chi^{\pm}_2$ are degenerate in mass.
All neutralinos/charginos and most notably the gluino are much lighter
than in the SPS1a parameter point, which greatly increases all LHC
production cross sections. It is important to note that this feature
is specific to our choice of parameters and not generic in
heavy--scalar models.

\begin{table}[b]
\begin{small}
\begin{center}
\begin{tabular}{c|r||c|r}
\hline
$\tilde g\tilde g$                 & 63 pb & $\tilde\chi^{\pm}\tilde g$     & 0.311 pb \\
$\tilde\chi^{\pm}\tilde\chi^{0}$   & 12 pb & $\tilde\chi^{0}\tilde g$       & 0.223 pb \\
$\tilde\chi^{\pm}\tilde\chi^{\mp}$ &  6 pb & $\tilde\chi^{0}\tilde\chi^{0}$ & 0.098 pb  \\
\hline
\multicolumn{2}{l}{Total} & \multicolumn{2}{r}{82 pb}\\
\end{tabular}
\caption{\label{splitsfitter_xsec} NLO cross sections for SUSY pair
  production at the LHC. Branching ratios are not included.}
\end{center}
\end{small}
\end{table}

Table~\ref{splitsfitter_xsec} shows the main (NLO) cross sections at
the LHC~\cite{Beenakker:1996ch,Beenakker:1999xh}. The SUSY production
is dominated by gluino pairs whose rate is eight times that of the
SPS1a point: the lower gluino mass enlarges the available phase space,
while in addition the destructive interference between $s$ and
$t$--channel diagrams is absent. The second largest process is the
$\tilde\chi^{\pm}_1\tilde\chi^{0}_2$ production, which gives rise to a
145~fb of hard-jet free, $e$ and $\mu$ trilepton signal, more than a
hundred times that of the SPS1a parameter choice.

\subsection{Observables}
The first obvious observable is the light Higgs mass $m_h$. Although
slightly higher than in most MSSM points, $m_h$ can still be measured
in the Higgs decay to two photons~\cite{Bettinelli}. The systematic
error on this measurement is mainly due to the uncertainty of the
electromagnetic energy scale.

A measurement of the gluino pair production cross section appears
feasible and can be very helpful to determine
$M_3$~\cite{Lester:2005je}. Most gluinos (85\%) will decay through a
virtual squark into a chargino or a neutralino along with two jets.
The chargino will in turn decay mostly into the LSP plus two leptons
or jets. Such events would then feature at least 4 hard jets, a large
amount of missing energy and possibly leptons.  The main backgrounds
are $t\overline{t}$ pairs (590~pb) and $W+$jets (4640~pb) as well as
$Z+$jets (220~pb). Despite these large cross sections, we have checked
using a fast LHC-like simulation that most of the background can be
eliminated by requiring a minimal number of hard jets or by applying
standard cuts on the missing energy or the effective mass $M_{\rm
  eff}=\slash\hskip-3mm E_T+\sum p_{T j}$.  The main source of
systematic errors on the cross section is the 5\%~error on the
luminosity. The theory error on the cross section we estimate to 20\%.

The next relevant observable is the trilepton signal. After gluino
pairs, the second--largest rate comes from the direct production of
$\tilde\chi^{\pm}_1\tilde\chi^{0}_2$, with 22\% of
$\tilde\chi^{\pm}_1$s decaying through a virtual $W$ into an electron
or muon, a neutrino and the LSP. Similarly, 7\% of $\tilde\chi^{0}_2$s
decay through a virtual $Z$ into an opposite--sign--same--flavor
lepton pair (OSSF) and the LSP. The resulting signal features three
leptons (two of them with OSSF), missing energy from the LSPs and the
neutrino, and no jet from the hard process. The backgrounds are mainly
$WZ$ (386~fb) and $ZZ$ (73~fb), the latter with one missed lepton.
Taking into account all branching ratios~\cite{Muhlleitner:2003vg},
the trilepton signal has a rate of 145~fb. Without any cuts, the
identification efficiencies of 65\% ($e$) and 80\% ($\mu$) leave us
with 110 to 211~fb for the background and 40 to 74~fb for the signal,
depending on the number of electrons and muons in the final state. A
dedicated study with the appropriate tools would evidently provide a
better understanding of signal and background. As in the previous
case, the main source of systematic errors is the luminosity.  We also
take the theory error on the value of the trilepton cross section to
be roughly 20\%.

For the trilepton signal we can define a kinematic observable: 10\%
of $\tilde\chi^{0}_2$s decay into an OSSF lepton pair and the LSP. The
distribution of the invariant mass of the leptons features a kinematic
upper edge at $m_{\tilde\chi^{0}_2}-m_{\tilde\chi^{0}_1}$. Such an
observable gives precious information on the neutralino sector and
hence on $M_1$. Its systematic error is dominated by the lepton energy
scale. The statistical error we estimate to be of the order of 1\%,
from a ROOT fit of the $M_{\ell\ell}$ distribution. Finally, we use
the ratio of gluino decays including a $b$ quark to those not
including a $b$. We roughly assume a systematic error of 5\% due to
the $b$ tagging and 20\% on the theory prediction.

\begin{table}[t]
\begin{small}
\begin{center}
\begin{tabular}{l|r|r|l|r|r}
\hline
\multicolumn{2}{c|}{observables} & 
\multicolumn{2}{c|}{systematic error} & statistical error & theory error\\
\hline
$m_h$                                      &128.8 GeV&0.1\%&energy scale&0.1\%& 4\%\\
$m_{\tilde\chi^{0}_2}-m_{\tilde\chi^{0}_1}$&   57 GeV&0.1\%&energy scale&0.3\%& 1\%\\
$\sigma(3\ell)$                            & 145.2 fb&  5\%&luminosity  &  3\%&20\%\\
$g\rightarrow b/{\rm not}(b)$                &     0.11&  5\%&$b$ tagging &0.3\%&20\%\\
$\sigma(\tilde g\tilde g)$                 &  68.2 pb&  5\%&luminosity  &0.1\%&20\%\\
\hline
\end{tabular}
\caption{\label{splitsfitter_obs}Summary of all observables and their errors. 
  We assume an integrated luminosity of $100~{\rm fb}^{-1}$}.
\end{center}
\end{small}
\end{table}

Table~\ref{splitsfitter_obs} summarizes the central values and errors
for all observables we use. The third and fourth columns give the
experimental systematic errors and their sources, the fifth column
gives the statistical errors corresponding to a few years of the LHC's
nominal luminosity ($100~{\rm fb}^{-1}$). The last column gives a
conservative estimate of the theory uncertainties.

\subsection{Parameter determination}
To study the effects of the different error sources, we first look at
a low--statistics scenario and ignore all theory uncertainties. Then,
we choose the limit of high statistics to estimate the ultimate
precision barrier imposed by experimental systematical errors.
Finally, we look at the effect of theory errors by including them into
the previous set. We expect the theory errors to dominate, based on
the currently available higher--order calculations. We use the
parameter extraction tool SFitter~\cite{Lafaye:2007vs}, which in
parallel to Fittino~\cite{Bechtle:2005vt} was developed for this kind
of problem.

With no information on the squark and sfermion sector, except for
non-observation, we are forced to fix $M_S$ and $A_t$ in our fits.
Moreover, we set $M_2 = M_1$ at the unification scale $M_{\rm GUT}$,
lacking enough information from the neutralino/chargino sector.  Using
Minuit, we then fit the remaining parameters to the LHC observables.
The $\chi^2$ minimum we identify using Migrad, while Minos determines
the approximately Gaussian errors. Note that in a more careful SFitter
study we would use flat theory
errors~\cite{Hocker:2001xe,Lafaye:2007vs}, but given the huge
difference in computing time we employ a Gaussian approximation in
this preliminary study. Our distant starting point is
$(M_1,M_3,\tan\beta,\mu) = (100,200,10,320)$.

\begin{table}[b]
\begin{small}
\begin{center}
\begin{tabular}{l|r|r|r|r|r|r|r|r}
\hline
parameter & nominal & fitted & \multicolumn{2}{|c}{low stat.} & \multicolumn{2}{|c}{$\infty$
  stat.}&\multicolumn{2}{|c}{$\infty$ stat.$+$theory}\\
\hline
$M_2$       & 132.4 GeV & 132.8 GeV & 6   &   5\% &0.24&0.2\%&21.2&16\%  \\
$M_3$       & 132.4 GeV & 132.7 GeV & 0.8 & 0.6\% &0.16&0.1\%& 5.1& 4\%  \\
$\mu$       &   290 GeV & 288 GeV   & 3.8 & 1.3\% &1.1 &0.4\%&  48&17\%  \\
$\tan\beta$ & 30        & 28.3      & 60  & undet.&1.24&  4\%& 177&undet.\\
\hline
$M_1$       & 132.4 GeV &132.8 GeV  & \multicolumn{6}{|c}{$=M_2$} \\
$A_t$ & \multicolumn{2}{|c}{0}      & \multicolumn{6}{|c}{fixed}  \\
$M_S$ & \multicolumn{2}{|c}{10 TeV} & \multicolumn{6}{|c}{fixed}  \\
\hline
\end{tabular}
\caption{\label{splitsfitter_fit}Result of the fits. Errors on the 
  determination of the parameter are given for the three error sets
  described in the text.}
\end{center}
\end{small}
\end{table}

Table~\ref{splitsfitter_fit} shows the result of the different fits.
It is interesting to note that $\tan\beta$ is undetermined except in
the case of infinite statistical and theory's accuracy. This is due to
the fact that only one of the five Higgs masses is measured. We
present a study on the determination of $\tan\beta$ from $(g-2)_\mu$
elsewhere in this volume. The quality of the trilepton and gluino
signals gives very good precision on the determination of $M_1$ and
$M_3$, even with low statistics. Including theory errors indeed
decreases the accuracy but still allows for a determination of the
mass parameters: $M_3$ only depends on the large gluino rate and its
decays, explaining its relative stability for smaller statistics. The
weakly interacting $M_1$ and $M_2$ suffer a larger impact from the
theory errors, because they depend on the trilepton rate and also on
the $b$ to non-$b$ gluino--decay ratio, both of which bear a large
theory error.

\subsection{Outlook}
The MSSM with heavy scalars is built to satisfy current experimental
and theoretical constraints on physics beyond the Standard Model while
keeping some of the features of the TeV--scale MSSM.  At the LHC,
light gauginos and Higgsinos will lead to sizeable production rates,
allowing us to study these new states.

The main observable channels are gluino pairs and the tri-leptons,
whose hard-jet free channel makes it a fairly clean channel with
respect to Standard--Model and SUSY backgrounds. Additional
observables such as the light Higgs mass, the
$(m_{\tilde\chi^{0}_2}-m_{\tilde\chi^{0}_1})$ kinematic edge and the
$b$-to-non-$b$ gluino--decay ratio give us access to most parameters
at the level of a few percent with 100 fb$^{-1}$ luminosity, based on
experimental uncertainties.  Theory errors increase the error bands on
the model parameters to $\mathcal{O}(15\%)$. 

Obviously, the scalar sector including $\tan\beta$ is only poorly
constrained, if at all.  New complementary observables could improve
this limitation.  Similarly, a look at other parameter points would be
needed to remove the specific properties of the point studied and to
provide a more complete view of the LHC discovery potential of a MSSM
with decoupled scalars.

\section*{Acknowledgements}
We would like to thank the organizers of the great Les Houches 2007
workshop {\it Physics at TeV-Scale Colliders}.



\clearpage
\newpage

\setcounter{figure}{0}
\setcounter{table}{0}
\setcounter{equation}{0}
\setcounter{footnote}{0}
\section{
 Finding the SUSY mass scale
\protect\footnote{A.J.~Barr, C.~Gwenlan and C.G.~Lester}}


\subsection{Introduction}

In the past it has been suggested that a good starting point for the
determination of the mass scale of new susy or other exotic particles is the
``effective mass'' distribution
\cite{Hinchliffe:1997iu,usemeffPhysicsTDR}.  There are a number of
slightly different definitions of ${{M_{\rm Eff}}}$ and the phrase ``mass
scale'' \cite{Tovey:2000wk} but a typical definition of
${{M_{\rm Eff}}}$ would be
\begin{eqnarray}
\label{eq:meff}
{{M_{\rm Eff}}} & = & {\bf p}_T^\mathrm{miss} + \sum_{i} {p_{T(i)}},
\end{eqnarray}
in which ${\bf p}_T^\mathrm{miss}$ 
is the magnitude of the event's missing transverse
momentum and where ${p_{T(i)}}$ is the magnitude of the transverse momentum of
the $i$-th hardest jet or lepton in the event.

\par

All definitions of ${{M_{\rm Eff}}}$ are motivated by the fact that new
TeV-scale massive particles are likely to be produced near threshold,
and so by attempting to sum up the visible energy in each event, one
can hope to obtain an estimate of the energy required to form the two
such particles.  Broadly speaking, the peak in the ${{M_{\rm Eff}}}$
distribution is regarded as the mass-scale estimator.

Although the effective mass is a useful variable, and simple to
compute, it has undesirable properties: The desired correlation
between ${{M_{\rm Eff}}}$ and the mass scale relies on the assumption
that the particles are produced near threshold.  While it is true that
the cross sections will usually {\em peak} at threshold, they can have
significant tails extending to $\sqrt{\hat{s}}$ values considerably
beyond the threshold value. It is very hard to make {\em precise}
statements about the mass scale from ${{M_{\rm Eff}}}$ alone.

\par

In this letter we introduce a variable, $m_{TGen}$, which is designed
to make more precise measurements of the mass scale
by using event kinematics, rather than simple energy sums or ad-hoc rules.
The aim is to produce event-by-event lower bound on the mass of 
pair-produced heavy particles. The variable 
has been constructed so that it is our ignorance about the ancestry of the 
final state particles, and the loss of information from the 
invisible massive heavy particles.

A solution already exists for the simplest case of interest -- 
in which the final state consists of only two visible particles 
(plus the invisibles). For this case there is no combinatorial problem,
as one can assume that each visible particle belongs to one of the 
initial heavy particles (e.g. one jet comes from each squark parent).
The variable defined for that case is known as $m_{T2}$ and 
is described in \cite{pubstransversemass,Barr:2003rg}.

\par

The generalisation to the case of arbitrary numbers of final state visible particles is the subject of this letter and is called $m_{TGen}$.
$m_{TGen}$ is defined to be the smallest value of $m_{T2}$
obtained over all possible
partitions of momenta in $F$ into two subsets $\alpha$ and $\beta$ --
each subset representing the decay products of a particular ``side''
of the event.  Note that $m_{T2}$ is itself defined in terms of ${\bf
p}_T^\alpha$ and $m_\alpha$ (respectively the transverse momentum and
mass of one side of the event), $ {\bf p}_T^\beta$ and $m_\beta$
(respectively the transverse momentum and mass of the other side of
the event), and $\chi$ (the mass of each of the unobserved particles
which are supposed to have been produced on each side of the event) as
follows:
\begin{eqnarray}
&m_{T2}&({\bf p}_T^\alpha, {\bf p}_T^\beta,  {\bf p}^\mathrm{miss}_T ,
  m_\alpha, m_\beta, \chi)
  \equiv \nonumber \\
 &  \equiv &{ \min_{{\bf q}_T^{(1)} +
{\bf q}_T^{(2)} = {\bf p}^\mathrm{miss}_T }} {\Bigl[ \max{ \Bigl\{
m_T^2({\bf p}_T^\alpha, {\bf q}_T^{(1)}; m_\alpha, \chi) ,\
m_T^2({\bf p}_T^\beta,  {\bf q}_T^{(2)}; m_\beta, \chi) \Bigr\} }
\Bigr]}.\label{MT2:MT2DEF}
\end{eqnarray}
where
\begin{equation} 
m_T^2 ( {\bf p}^{\alpha}_T, {\bf p}_T^\chi; m_\alpha, \chi) \equiv { m_{\alpha}^2 + \chi^2 +
2 ( E_T^{\alpha} E_T^\chi - {\bf p}_T^{\alpha}\cdot{\bf p}_T^\chi ) }
\label{MT2:MTDEF}
\end{equation}
in which \begin{equation}E_T^{\alpha} = { \sqrt { ({\bf p}_T^{\alpha}) {^2} + m_\alpha^2 }}
\qquad \hbox{ and } \qquad {E_T^\chi} = { \sqrt{ ({\bf
p}_T^\chi){^2} + \chi^2 } }
\label{MT2:ETDEF}\end{equation}
and likewise for $\alpha \longleftrightarrow \beta$.  With the above
definition (in the case $\chi = m_{\tilde{\chi}_1^0}$), $m_{T2}$ generates and
event-by-event lower bound on the mass of the particle whose decay
products made up either of the two sides of the event, under the
assumption that the event was an event represents pair production
followed by decay to the visible particles and an unseen massive
particle on each side.  When evaluated at other values of $\chi$ 
the above properties are retained approximately (see
\cite{pubstransversemass,Barr:2003rg,Cho:2007qv,Gripaios:2007is,Barr:2007hy,Cho:2007dh}).  There exist events which
allow this lower bound to saturate, and so (in the absence of
background) the upper endpoint of the $m_{T2}$ distribution may be used
to determine the mass of the particle being pair produced.

\subsection{Example distributions}
In this letter we show the results for simulations of an example
particle spectrum
for proton-proton collisions at LHC centre-of-mass energy of
$\sqrt{s}=$14~TeV.  The {\tt
HERWIG} \cite{Corcella:2002jc,Moretti:2002eu,Marchesini:1991ch} Monte
Carlo generator was used to produce inclusive unweighted
supersymmetric particle pair production events.  Final state particles
(other than the invisible neutrinos and neutralinos) were then
clustered into jets by the longitudinally invariant $k_T$ clustering
algorithm for hadron-hadron collisions\cite{Catani:1993hr} used in the
inclusive mode with $R=1.0$ \cite{Ellis:1993tq}.  Those resultant jets
which had both pseudo-rapidity ($\eta=-\ln\tan\theta/2$) satisfying
$|\eta|<2$ and transverse momentum greater than 10~GeV/c were used to
calculate $m_{TGen}$ and $M_{\rm Eff}$.

\begin{figure}[tbh]
\begin{center}
\includegraphics[angle=90,width=0.7\textwidth]{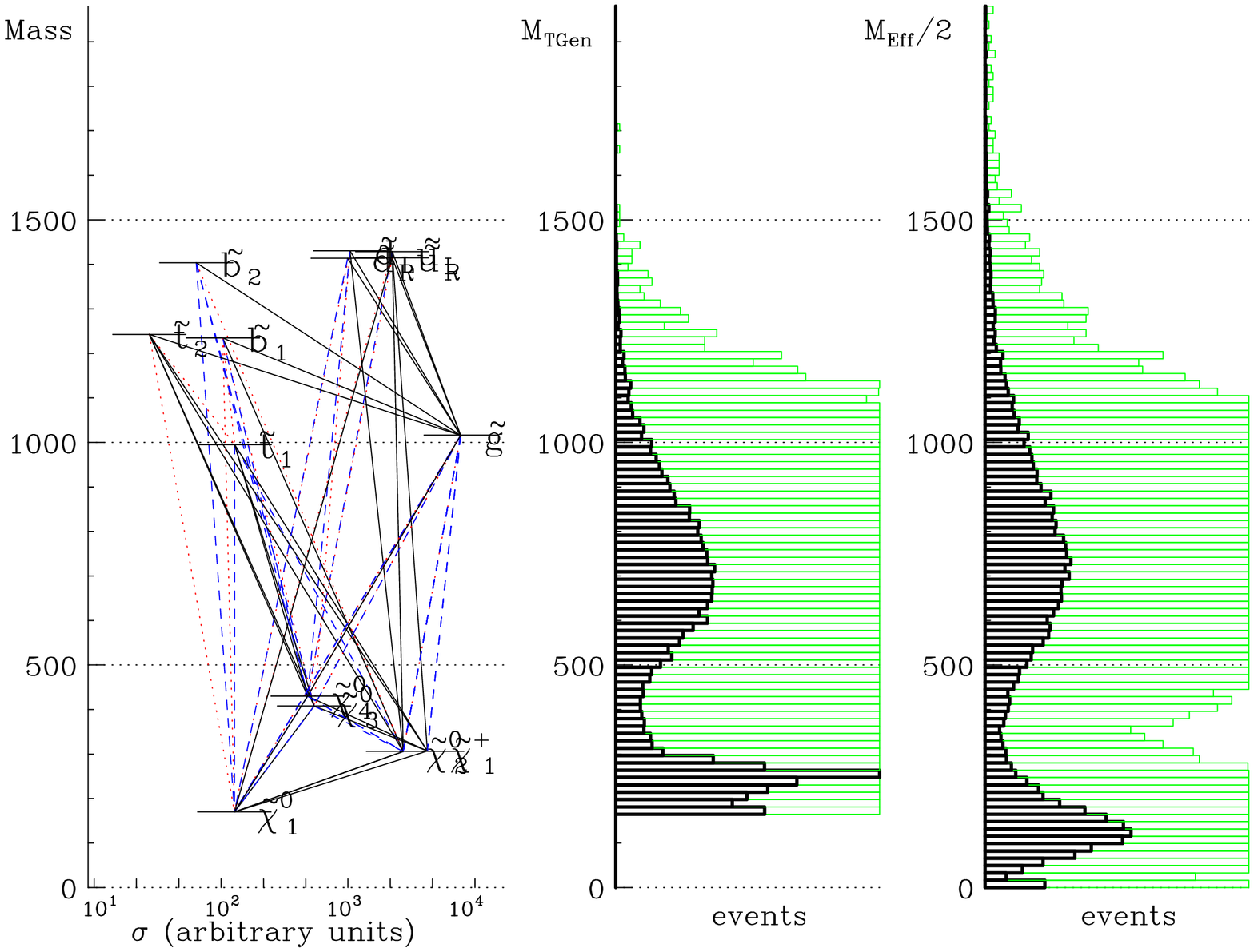}
 \caption{On the left hand side is a graphical representation of the susy mass
spectrum of the point described in the text. The vertical positions 
of the particles indicate their masses.
The horizontal positions of the centres of the bars indicate the 
relative LHC production cross-section (arbitrary units).
The lines joining particles indicate decays with branching fractions
in the following ranges:
greater than $10^{-1}$ solid; 
$10^{-2} \rightarrow 10^{-1}$ dashed;
$10^{-3} \rightarrow 10^{-2}$ dotted.
The middle plot shows the distribution of our variable, $M_{TGen}$, 
with $M_{TGen}$ increasing vertically to ease comparison with the spectrum.
The right hand plot shows the distribution of another variable, 
$M_{\rm Eff}/2$, where $M_{\rm Eff}$ is defined in eq. \ref{eq:meff}.
In both the $M_{TGen}$ and the $M_{\rm Eff}$ plots, the lighter shading
shows the histograms with the number of events
multiplied by a factor of twenty, so that the detail in the upper tail may be seen.
}
\label{fig:search}
\end{center}
\end{figure}

\begin{figure}[tbh]
\begin{center}
    \begin{minipage}[b]{.49\linewidth}
      \begin{center}
        \includegraphics[width=0.98\linewidth]{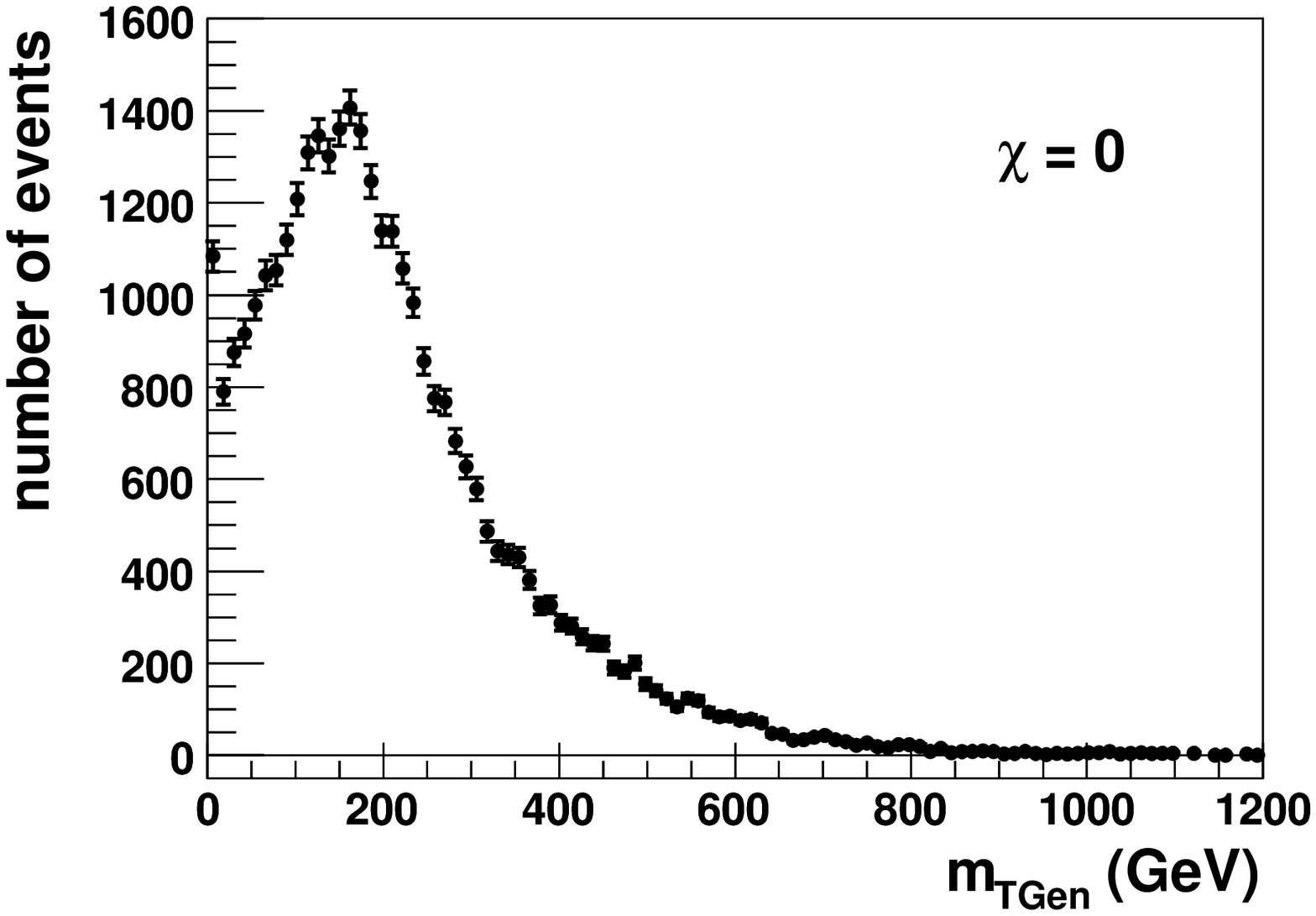}\\
        {\bf (a)}
      \end{center}
     \end{minipage}
    \begin{minipage}[b]{.49\linewidth}
      \begin{center}
        \includegraphics[width=0.98\linewidth]{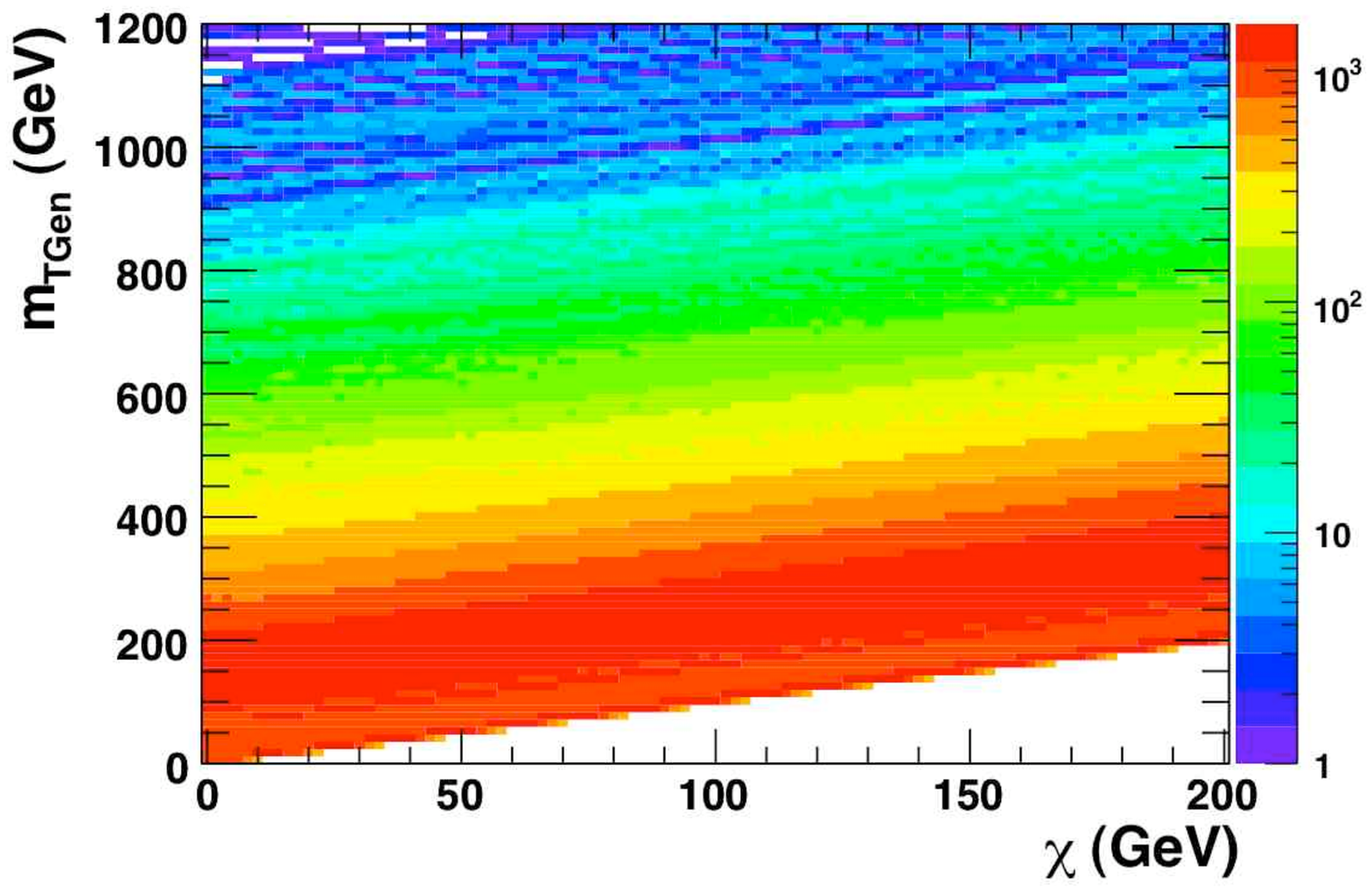}\\
        {\bf (b)}
      \end{center}
     \end{minipage}
\end{center}
\caption{$M_{TGen}$ distributions for a Les Houches blind data sample as described in the text.
{\bf (a)} Invisible particle mass, $\chi=0$. {\bf (b)} The colour scale shows the number of events with a particular value of $M_{TGen}$ (y-axis) as a function of the input invisible particle mass, $\chi$ (x-axis).
\label{fig:partially-sighted}
}
\end{figure}

In figure \ref{fig:search}
we show the distributions of $M_{TGen}$ and $M_{\rm Eff}$ for a sample point 
with a spectrum, defined by the mSUGRA parameters: 
\{$m_0=1200~$GeV, $m_\frac{1}{2}=420~$GeV, $\tan\beta=10$, $m_t=174~$GeV, $\mu<0$\}, 
The spectrum and branching ratios were calculated using {\tt Isajet}\cite{Baer:1999sp} version~7.58.
In these plots we assume it is possible to accurately assign all visible momenta to the correct category $F$ or
$G$, i.e.\ ``interesting final state momenta'' versus ``initial state
radiation''. Plots in which this is not assumed look similar and can be found in 
\cite{Lester:2007fq}.
The {\tt HERWIG} initial state radiation and underlying event have been 
switched off, and the parameter $\chi$ which is required to calculate $M_{TGen}$
has been set to the mass of the lightest supersymmetric particle.

It can be seen that, as intended, the upper edge of the distributions 
gives a very good indication of the mass of the heaviest pair-produced sparticle.
Other supersymmetric points show similar behaviour \cite{Lester:2007fq}. 
This means that the position of the upper edge of $m_{TGen}$
can be used to find out about the mass scale of any 
semi-invisibly decaying, heavy, pair-produced particles.

Furthermore, a change in slope can be observed at lower masses due to 
significant pair production of lower-mass particles 
(e.g. chargino and/or neutralino pairs).
Therefore it is also possible in principle to extract from
information about several different mass scales.

In figure \ref{fig:partially-sighted} we plot the $m_{TGen}$ distribution
for events from the Les Houches 2007 blind data 
sample\footnote{Described in these proceedings}.
We select events which have at least one jet with 
$|\eta|<3.2$ and transverse momentum 
greater than 400~GeV, and also have missing transverse momentum 
greater than 70~GeV. We veto any event containing an electron, muon or photon
with transverse momentum greater than 20~GeV.\footnote{Due to the lack of trigger information in the ``blind sample''.}
There is no evidence for an end-point in the distribution, 
like the one seen in figure \ref{fig:search}
or in the other examples in \cite{Lester:2007fq}.
Neither is there evidence for any ``kinks'' in the $m_{TGen}$ distribution 
when plotted as a function of the invisible particle mass, $\chi$ 
(figure \ref{fig:partially-sighted}b). 
Such end-points and kinks would be expected if the events contained
many-body or cascade decays of strongly interacting objects
to visible and invisible heavy particles
\cite{Cho:2007qv,Gripaios:2007is,Barr:2007hy,Cho:2007dh}, 
and so disfavours a model of this type.

\subsection{Conclusions}
In conclusion, $m_{TGen}$ could be a useful variable for determining mass scales at the LHC. If you are interested in using it, contact the authors for the code and let us know what you find.

\subsection*{Acknowledgements}
The authors thank the scientific organisers of the Les Houches workshop 
for their hospitality and for liasing with the secretariat.
CB and AB would like to thank the UK Science and Technology Facilities Council
for their
for the financial support of their fellowships.
CGL would like to thank the Friends of Nature for extra pillows.



\clearpage
\newpage

\setcounter{figure}{0}
\setcounter{table}{0}
\setcounter{equation}{0}
\setcounter{footnote}{0}


\section{A hybrid method for SUSY masses 
from fully identified cascade decays
\protect\footnote{M.M.~Nojiri, G.~Polesello and D.R.~Tovey}}
\subsection{Introduction}
This letter describes a new hybrid technique for improving the
precision with which SUSY particle masses can be measured at the
LHC. Existing techniques usually make use of the positions of
end-points in experiment-wise distributions of invariant mass
combinations of visible SUSY decay products
\cite{Hinchliffe:1996iu,Allanach:2000kt,Miller:2005zp}, or use
$E_T^{miss}$ constraints from `symmetric' events in which the same
SUSY decay chain has participated in both `legs' of the event
\cite{Barr:2003rg}. In both cases some information regarding the
events is discarded -- in the former case event-wise $E_T^{miss}$
information is not used in the experiment-wise end-point analysis, in
the latter experiment-wise invariant mass end-point constraints are
not used in the event-wise analysis. In this letter we describe a
simple `hybrid' technique which enables optimum use of both
experiment-wise and event-wise information to fully reconstruct SUSY
events and hence improve the mass measurement precision.

\subsection{Description of technique}
The new technique involves conducting a kinematic fit to each selected
SUSY event, with the sparticle masses as free parameters. Crucially,
the $\chi^2$ function of the fit involves both event-wise $E_T^{miss}$
constraints and experiment-wise invariant mass end-point
constraints. It should be appreciated that without the $E_T^{miss}$
constraints each event-wise fit is formally equivalent to solving the
experiment-wise invariant mass end-point constraints for the
individual sparticle masses, and consequently each fit will give the
same value for each mass. The RMS values of the distributions of these
masses will then be consistent with the mass precisions obtained from
the conventional method involving solution of the end-point
constraints. Addition of the event-wise $E_T^{miss}$ constraints
reduces the number of degrees-of-freedom of the kinematic fits and
hence can improve the mass measurement precision. In this case the
widths of the distributions of mass values obtained from different
events for one Monte Carlo experiment are larger than those obtained
when $E_T^{miss}$ constraints are excluded, however the means of the
distributions across many such experiments measure the masses more
accurately.

\subsection{Example: $\tilde{q}_L$ decays in SPS1a}
At mSUGRA point SPS1a there is a significant
branching ratio for the decay chain 
\begin{equation}
\tilde{q}_L \rightarrow \tilde{\chi}^0_2 q \rightarrow \tilde{l}_R lq \rightarrow \tilde{\chi}^0_1 llq.
\label{hybrid_eqn1}
\end{equation} 
This chain provides 5 kinematic end-point mass constraints from
invariant mass combinations of jets and leptons
\cite{Weiglein:2004hn}:

\begin{itemize}

\item $m(ll)^{max}$ = 77.08 $\pm$ 0.08(scale) $\pm$ 0.05(stat) GeV
\item $m(llq)^{max}$ = 431.1 $\pm$ 4.3(scale) $\pm$ 2.4(stat) GeV
\item $m(llq)^{min}$ = 203.0 $\pm$ 2.0(scale) $\pm$ 2.8(stat) GeV
\item $m(lq)_{hi}^{max}$ = 380.3 $\pm$ 3.8(scale) $\pm$ 1.8(stat) GeV
\item $m(lq)_{lo}^{max}$ = 302.1 $\pm$ 3.0(scale) $\pm$ 1.5(stat) GeV

\end{itemize}

For this study unbiased samples equivalent to 100~fb$^{-1}$ (one Monte Carlo
`experiment') of SPS1a signal events and $t\bar{t}$ background events
were generated with {\tt HERWIG 6.4}
\cite{Corcella:2000bw,Moretti:2002eu} and passed to a generic LHC
detector simulation \cite{RichterWas:2002ch}. A lepton reconstruction
efficiency of 90\% was assumed.

Events were selected in which the above decay chain appears in both legs of
the event with the following requirements:

\begin{itemize}

\item $N_{jet}$ $\geq$ 2, with $p_T(j2)$ $>$ 100 GeV,
\item $M_{eff2} = E_T^{miss} + p_T(j1) + p_T(j2)$ $>$ 100 GeV,
\item $E_T^{miss}$ $>$ max(100 GeV,$0.2M_{eff2}$),
\item $N_{lep}$ $=$ 4, where $lep = e/\mu$(isolated) and $p_T(l4)$ $>$ 6 GeV,
\item 2 Opposite Sign Same Flavour (OSSF) lepton pairs. If the pairs
are of different flavour both pairs must have $m(ll)<m(ll)^{max}$. If
both pairs are of the same flavour then one and only one of the two
possible pairings must give two $m(ll)$ values which are both less
than $m(ll)^{max}$. These pairings allocate the leptons to each leg of
the event.
\item One and only one possible pairing of the two leading jets with
the two OSSF lepton pairs must give two $m(llq)$ values less than
$m(llq)^{max}$. These pairings allocate the jets to each leg of the
event.
\item For each inferred leg of the event the maximum(minimum) of the
two $m(lq)$ values must be less than $m(lq)_{hi(lo)}^{max}$. This
ordering allocates the leptons to the $near$ and $far$
\cite{Allanach:2000kt} positions in the decay chain.

\end{itemize}

The requirement of 4-leptons in two OSSF pairs and two high-$p_T$ jets
consistent with kinematic end-points, together with large
$E_T^{miss}$, is effective at removing the majority of SM and SUSY
backgrounds (see below).

Each selected event was fitted with MINUIT \cite{James:1975dr}. Free
parameters were taken to be the four masses appearing in the decay
chain: $m(\tilde{q}_L)$, $m(\tilde{\chi}^0_2)$, $m(\tilde{l}_R)$ and $m(\tilde{\chi}^0_1)$. The mass-shell
conditions and measured momenta of the visible decay products for each
leg were solved to determine the LSP four-momenta, giving two
solutions for each leg. The $\chi^2$ minimisation function was defined
by:
\begin{eqnarray}
\label{hybrid_eqn2}
\chi^2 &=&   \left(\frac{m(ll)^{max}_{evt}-m(ll)^{max}_{expt}}{\sigma_{m(ll)^{max}}}\right)^2 \nonumber \\
       &+& \left(\frac{m(llq)^{max}_{evt}-m(llq)^{max}_{expt}}{\sigma_{m(llq)^{max}}}\right)^2
       + \left(\frac{m(llq)^{min}_{evt}-m(llq)^{min}_{expt}}{\sigma_{m(llq)^{min}}}\right)^2 \nonumber\\
       &+& \left(\frac{m(lq)_{hi;evt}^{max}-m(lq)_{hi;expt}^{max}}{\sigma_{m(lq)_{hi}^{max}}}\right)^2
       + \left(\frac{m(lq)_{lo;evt}^{max}-m(lq)_{lo;expt}^{max}}{\sigma_{m(lq)_{lo}^{max}}}\right)^2 \nonumber\\
       &+& \left(\frac{p_x(\tilde{\chi}_{1}^0(1))+p_x(\tilde{\chi}_{1}^0(2))-E_x^{miss}}{\sigma_{E_x^{miss}}}\right)^2
       + \left(\frac{p_y(\tilde{\chi}_1^0(1))+p_y(\tilde{\chi}_{1}^0(2))-E_y^{miss}}{\sigma_{E_y^{miss}}}\right)^2,
\end{eqnarray}
where $evt$ denotes an expected end-point value derived from the
masses in the event-wise fit with the formulae of
Ref.~\cite{Allanach:2000kt}, and $expt$ denotes a `measured'
experiment-wise end-point value. The uncertainties $\sigma$ in these
`measured' endpoints were those quoted above. The uncertainties on the
measurements of the $x$ and $y$ components of $E_T^{miss}$,
$\sigma_{E_x^{miss}}$ and $\sigma_{E_y^{miss}}$, were given by
$0.5\sqrt{E_T^{sum}}$ where $E_T^{sum}$ is the scalar sum of jet $p_T$
of the event. This function incorporating both event-wise $E_T^{miss}$
constraints and experiment-wise end-point constraints was evaluated
for each of the four pairs of $\tilde{\chi}^0_1$ momentum solutions obtained
from solving the leg mass-shell conditions. Fitted masses were
obtained when $\chi^2$ was minimised for the event. Fitted masses were
used in the subsequent analysis only if MINUIT judged the fit to have
converged and $\chi^2_{min} < 35.0$.

Following application of the selection cuts described above and the
requirements of fit convergence and low fit $\chi^2_{min}$ 38 SUSY
`signal' events with the above decay chain appearing in both legs were
observed. 4 SUSY background events were observed, consisting of the
above decay chain in both legs but with one or two leptonically
decaying staus produced in the decays of the $\tilde{\chi}^0_2$'s. No $t\bar{t}$
background events were observed in 100~fb$^{-1}$ equivalent data. More
SM background events may be expected in a real experiment, given that
effects such as charge and lepton mis-identification are not included
in the fast detector simulation. Nevertheless the contribution is
still expected to be negligible.

\begin{figure}[ht]
\begin{center}
\epsfig{file=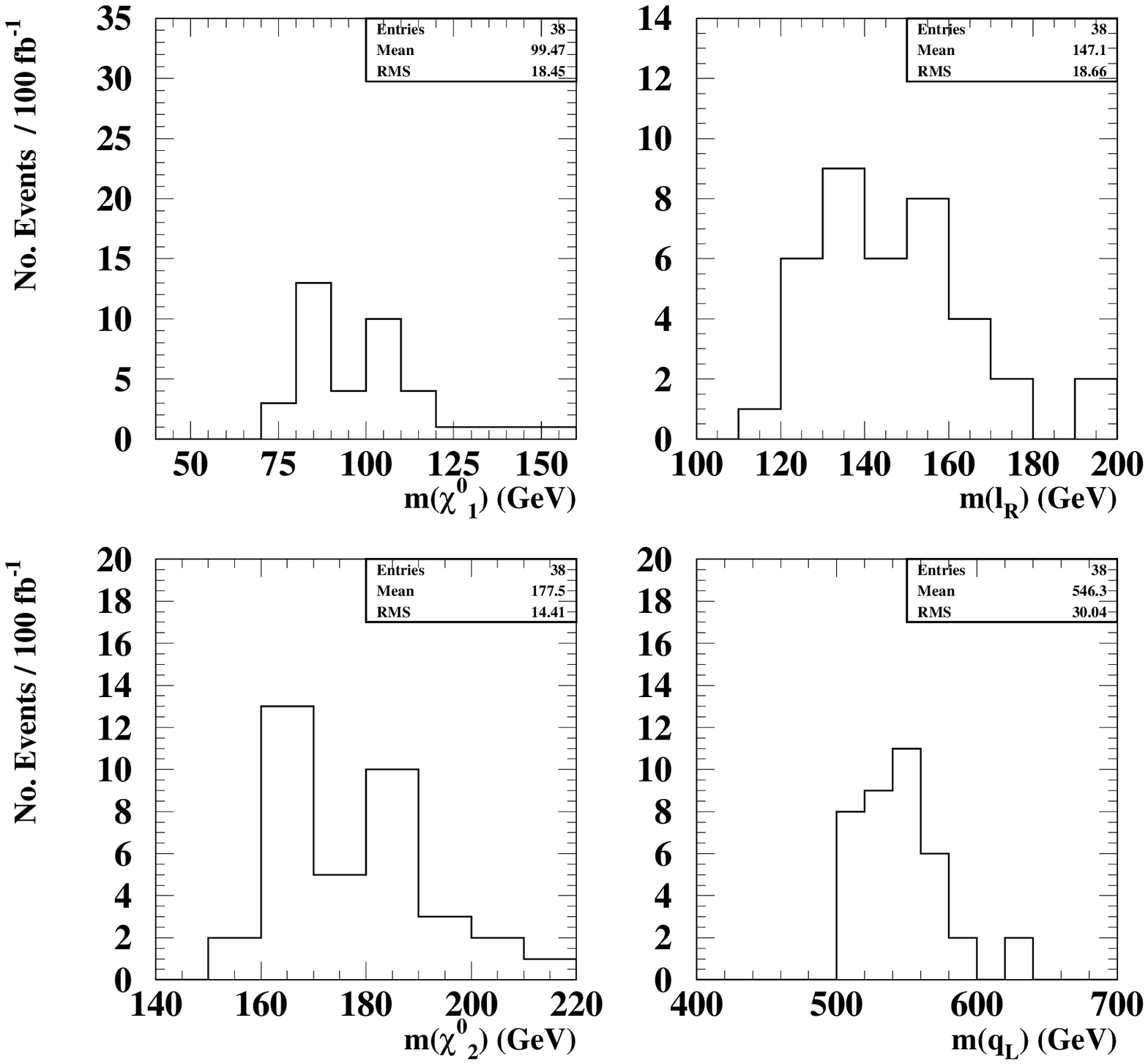,height=3.0in}
\epsfig{file=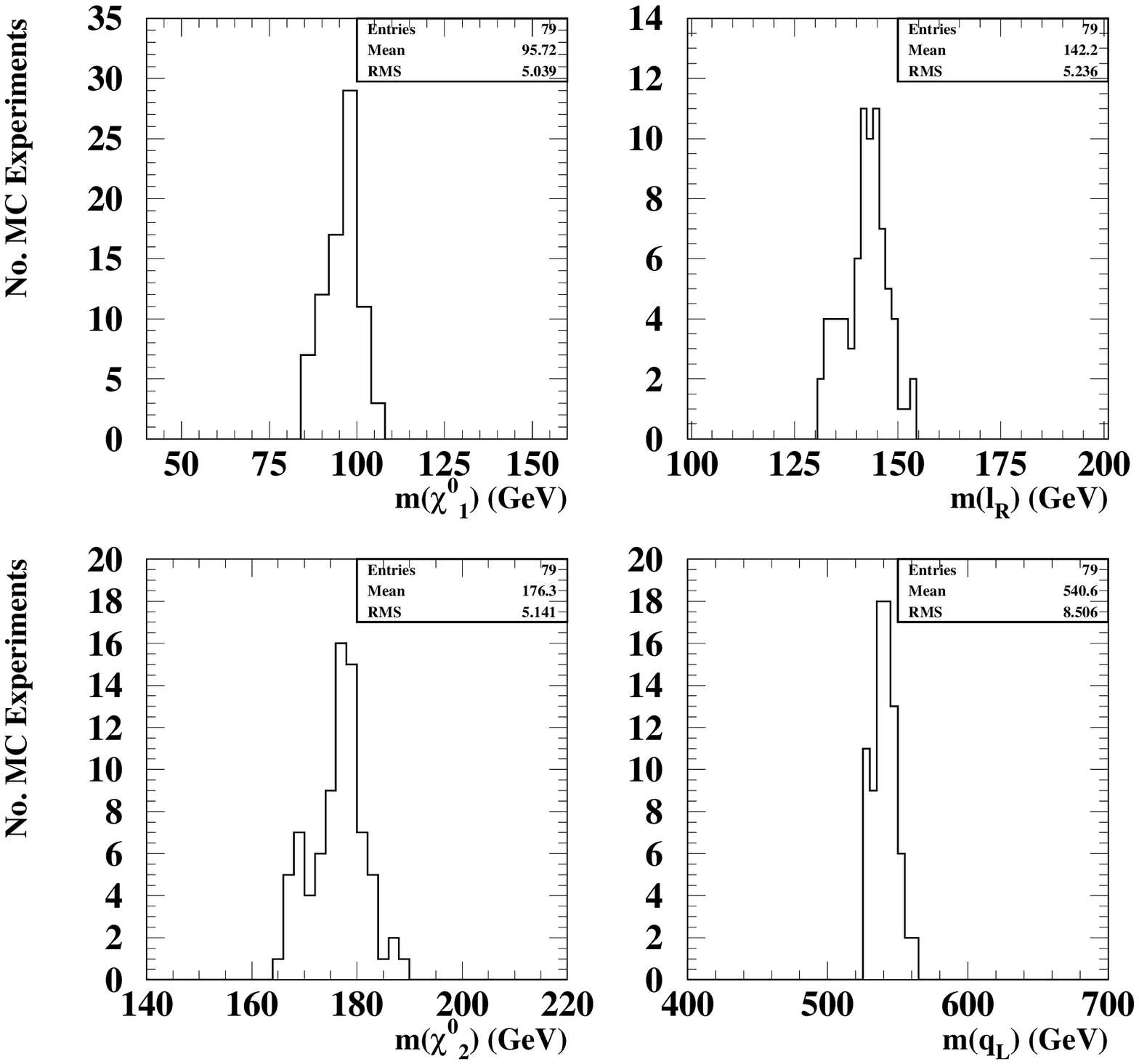,height=3.0in}
\caption{\label{hybrid_fig1} Distributions of sparticle masses. The
four distributions on the left are of masses obtained from event-wise
fits, for one MC experiment. Each entry is obtained by minimising the
$\chi^2$ function shown in Eqn.~\ref{hybrid_eqn2}. The four
distributions on the right are likelihood distributions of sparticle
masses obtained from 100 MC experiments. Each entry is the mean of an
experiment-wise mass histogram such as those on the left.}
\end{center}
\end{figure}

Each event-wise fit generated one set of values for the sparticle
masses, namely those values which minimise the broad $\chi^2$ function
in Eqn.~\ref{hybrid_eqn2}. The distributions of these values for one
Monte Carlo experiment are shown in Fig.~\ref{hybrid_fig1}(left). In
order to demonstrate the performance of the technique and judge the
uncertainties in the measurements the above procedure was repeated for
100 Monte Carlo experiments. For each experiment, kinematic end-point
positions were sampled from gaussians with means and sigmas given by
the means and uncertainties listed above. The five sampled end-point
positions for each experiment were solved simultaneously with a MINUIT
fit to give initial mass values for input to the MINUIT event-wise
kinematic fits. For each experiment relative jet(lepton) energy scale
values were sampled from gaussians of width 1\%(0.1\%) reflecting
likely ultimate energy scale uncertainties at the LHC. Each experiment
generated a set of sparticle mass histograms similar to those shown in
Fig.~\ref{hybrid_fig1}(left). The means of these histograms for the
100 MC experiments were then used to construct likelihood histograms
for the masses, shown in Fig.~\ref{hybrid_fig1}(right). The standard
deviations of these histograms were taken to provide the uncertainties
on the sparticle mass measurements.

Unbiased MC data equivalent to only one 100~fb$^{-1}$ experiment were
available for this study. For this reason the same events were used
for each MC experiment, with just the end-point values and jet/lepton
energy scales varying. The additional uncertainties in the final mass
values expected from varying event samples were estimated from the
mean statistical uncertainties in the mean experiment mass values as
extracted from the event-wise distributions such as those shown in
Fig.~\ref{hybrid_fig1}(left). We evaluated the
experiment-by-experiment spread due to varying event samples as
$\sigma/\sqrt{n}$, where $\sigma$ is the RMS of the event-wise
distributions as shown in Fig.~\ref{hybrid_fig1}(left), and $n$ is the
number of entries in each plot. These additional contributions were
added in quadrature to the uncertainties obtained from the study. This
approximation was checked with a second sample of SPS1a events
equivalent to 100 different MC experiments, biased to force gluinos to
decay to $\tilde{q}_L$, $\tilde{b}$ or $\tilde{t}$, $\tilde{q}_L$ to decay to
$\tilde{\chi}^0_2$ and $\tilde{\chi}^0_2$ to decay to $\tilde{e}$ or $\tilde{\mu}$.

\begin{table}[ht]
\begin{center}
{\small%
\begin{tabular}{|c|c|c|c|c|c|c|c|}
\hline
State &Input &\multicolumn{2}{|c|}{End-Point Fit} &\multicolumn{2}{|c|}{Hybrid Method, $E_T^{miss}$}&\multicolumn{2}{|c|}{Hybrid Method, no $E_T^{miss}$}\\
\cline{3-8}
 & &Mean &Error &Mean &Error &Mean &Error \\
\hline
$\tilde{\chi}^0_1$ &96.05     &96.5 &8.0 &95.8(92.2)   &5.3(5.5) &97.7(96.9)   &7.6(8.0) \\ 
$\tilde{l}_R$ &142.97      &143.3 &7.9 &142.2(138.7) &5.4(5.6) &144.5(143.8) &7.8(8.1) \\
$\tilde{\chi}^0_2$ &176.81   &177.2 &7.7 &176.4(172.8) &5.3(5.4) &178.4(177.6) &7.6(7.9) \\
$\tilde{q}_L$ &537.2--543.0&540.4 &12.6 &540.7(534.8) &8.5(8.7) &542.9(541.4) &12.2(12.7) \\
\hline
\end{tabular}}
\caption{Summary of mass measurement precisions for SPS1a
states. Column 2 lists masses used in the {\tt HERWIG} generator,
Columns 3 and 4 the fitted masses and uncertainties obtained from the
conventional fit to kinematic end-points, Columns 5 and 6 the
equivalent values obtained with the new technique and Columns 7 and 8
the equivalent values obtained with the new technique excluding
$E_T^{miss}$ constraints. Figures in parentheses are those obtained
with the biased sample of non-repeated events. All masses are in
GeV. The quoted mass range for $\tilde{q}_L$ excludes $\tilde{b}$ squarks,
which are produced less readily than the light squarks. \label{hybrid_tab1}}
\end{center}
\end{table}

The results of this study are summarised in Table~\ref{hybrid_tab1}. For
comparison purposes the analysis was initially carried out with the
$E_T^{miss}$ constraints removed from the $\chi^2$ function. The
measurement precisions are consistent with those obtained from the
conventional end-point fitting method, as expected following the
reasoning outlined above. The analysis was then
repeated including the $E_T^{miss}$ constraints, giving an overall
improvement in sparticle mass precisions $\sim$ 30\% for all four
masses considered. A similar improvement was found when using the
biased sample of non-repeated events for different experiments.

\subsection*{Acknowledgements}
The authors wish to thank the organisers of the Les Houches 2007 {\it
Physics at TeV-Scale Colliders} workshop for what was once again an
excellent and stimulating working environment. DRT wishes to
acknowledge STFC for support.

\clearpage
\newpage

\setcounter{figure}{0}
\setcounter{table}{0}
\setcounter{equation}{0}
\setcounter{footnote}{0}
\section{A blind SUSY search at the LHC
\protect\footnote{G.S.~Muanza}}




\subsection{Introduction}
Most of the SUSY prospects \cite{:1999fr}\cite{Ball:2007zza} at the LHC are based on simulations
and analyses where the analyzer knows all the properties of the signal as well as those of the SUSY
and Higgs backgrounds to this signal.
\par\noindent
In constrast with these situations, we describe in this letter a new SUSY data challenge at the LHC
which is called the "Blind SUSY Search Project" \cite{BSSP_web}. As name of this project suggests,
here, the 
analyzer ignores the properties of the searched SUSY signal. 
\par\noindent
We have produced a 100 $pb^{-1}$ pseudo-data ($PsD_{1}$) sample which consists in a randomized mixture of the Standard
Model (SM) backgrounds and the inclusive Higgs and SUSY production of an unrevealed SUSY model. We also
provide separate and independent samples of the SM backgrounds.
\par\noindent
The aim of the challenge is to determine the type of underlying SUSY breaking mechanism as well as the
corresponding parameters of the hidden SUSY model.
\par\noindent
All of the samples are under a simple ROOT \cite{Brun:1997pa} format so as to propose this challenge not only to the
experimental HEP community but also to the theorists.
\par\noindent
The motivations for this challenge lie in the following questions. Let's hypothesize the presence of a SUSY signal
at the LHC, can one:
\par\noindent
$\bullet$ 
determine the excess with respect to the SM expectations and quantify it?

\par\noindent
$\bullet$ 
handle the possible presence of several SUSY and Higgs signals and how does that
affect the measurement of experimental observables (masses, mass differences, cross sections, ratios of 
branching fractions, spins,...)?

\par\noindent
$\bullet$ 
determine the type of underlying SUSY breaking mechanism at play (gravity or gauge or anomaly mediated)

\par\noindent
$\bullet$ 
distinguish different types of phenomenological hypotheses 
(R-parity conservation, phases, high scale unifications,...)?

\par\noindent
$\bullet$ 
evaluate the values of the parameters of the underlying SUSY model?

\par\noindent
Part of such questions has been posed and partially answered in previous SUSY challenges. But most of
them provided either an exclusive signal (ie: without all the decay channels open),
or were missing the Higgs and SUSY background, or part
of the SM backgrounds,... In the current challenge we tried to provide
all of these pieces.

\par\noindent
In section 2, we'll describe in some details how the samples were produced. In sections 3 and 4
respectively we'll explain how to access and how to analyze the data.


\subsection{Samples production}

\subsubsection*{Production tools}
All the processes were generated using Pythia version 6.325 (v6.325) \cite{Sjostrand:2003wg}. The SUSY mass spectrum and decay table
in the v1.0 "SUSY Les Houches Accord" (SLHA)\cite{Skands:2003cj} format was read in by Pythia. The CTEQ6L1 \cite{Kretzer:2003it}
proton parton density functions (PDF) were utilized for all the processes through
an interface to the LHAPDF \cite{Bourilkov:2006cj} v5.2.3 package.
\par\noindent
For the SM backgrounds and the Higgs processes the $\tau$ were decayed by Tauola \cite{Jadach:1990mz} v2.6.
However, because of missing pieces in the Tauola interface, the $\tau$ from SUSY processes were
decayed by Pythia unabling to account for the spin correlations in these cases.
\par\noindent
The ATLAS detector response was simulated using a personal fast simulation based on ATLFAST \cite{ATLFAST} 
v00-02-22. All the reconstructed quantities are simulated using smearing functions and the
corresponding variables
are also used to mimic the trigger conditions\footnote{The participants are asked to present
results only with events passing at least one of these conditions.} as described in the ATLAS High Level Trigger TDR
\cite{HLT}. 
Note that the events failing the trigger conditions were removed from the $PsD_{1}$ set, but not from
the background samples. Obviously the Monte Carlo (MC) truth informations were removed from the 
$PsD_{1}$ set as well.
\par\noindent
The output of the fast simulation is a PAW \cite{Brun:1990vs} ntuple which is subsequently converted into a root-tuple using ROOT v5.14.
\par\noindent
For the samples normalization an integrated luminosity of 100 $pb^{-1}$ was assumed and the Pythia leading
order (LO) cross sections were used for the sake of simplicity. Nevertheless some background (sub)-processes
with very high cross sections had to be arbitrarily taken out of the $PsD_{1}$ sample in order to keep
the number of events to be produced within reasonable limits.
Table \ref{blind_susy_search_tab1} contains the full list of these pruned sub-processes:

\begin{table}[h]
\begin{center}
\begin{tabular}{|c|c|c|}
\hline
Process & Pythia & Kinematical Cut \\
        & Process Index &          \\
\hline\hline
$qq\to qq$ & $MSEL=1$ &  $\hat{p}_{T} < 160$ GeV  \\
($q=u/d/s/g$) &  &    \\
\hline
$c\bar c$       & $MSEL=4$      & $\hat{p}_{T} < 40$ GeV \\
\hline
$b\bar b$       & $MSEL=5$      & $\hat{p}_{T} < 40$ GeV \\
\hline
$\gamma+jets$   & $MSUB(14,29,115)=1$   & $\hat{p}_{T} < 20$ GeV \\
\hline\hline
 Low mass resonances & - & - \\
 ($\Psi,\ \Upsilon,\ \chi,...$) & - & - \\
\hline 
 Elastic Scattering & - & - \\ 
\hline
 Diffraction & - & - \\
\hline
\end{tabular}
\caption{\label{blind_susy_search_tab1} The list of high cross section processes that were removed
from the $PsD_{1}$ sample.}
\end{center}
\end{table}

\par\noindent
The $PsD_{1}$ sample is stored in 7 root-tuples and contains in total 4.5M events for a total size
of 12 Gb.
\par\noindent
Each background root-tuples contains exactly 100k events. In total there are 1593 such root-tuples for
a total statistics of 159.3M events that amount to 424 Gb.
\par\noindent
All the other details about the events generation can be found on the project website: \cite{BSSP_web}.


\subsection{Access to data}

The full dataset was too large to be kept on disk. Therefore it is stored on tape at the Lyon Computing Center,
except for the $PsD_{1}$ sample which can also be downloaded from the project website. 
\par\noindent
For those who have an account at this Computing Facility, the samples are available on HPSS\cite{HPSS} in
the $cchpssd0\:/hpss/in2p3.fr/home/m/muanza/GDR\_SUSY/SUSY\_Blind/$ directory.
And the following sub-directories: $Pseudo\_DATA/final/0/100\_inv\_pb/$, $BKGD/<bkgd\_process>$ and
$ANALYSIS/SKIMMING/mET150/$ respectively contain the $PsD_{1}$, the SM background and the
skimmed ($\rlap{\kern0.25em/}E_{T}>$ 150 GeV) samples.

\par\noindent
For those who don't have an account at the Lyon Computing Center, there's a possible data access using 
SRB\cite{SRB}. The participant has to be registered as an SRB user. This can be done by sending me an email at
$muanza@in2p3.fr$. Then the participant needs to install an SRB client on his (her) computer to be able to list
and copy the root-tuples. The main SRB directory is 
\par\noindent
$/home/smuanza.ccin2p3/GDR\_SUSY/SUSY\_Blind/$ and all the
sub-directories structure is that of HPSS.
\par\noindent
All the useful details about SRB are explained on the projcet website.


\subsection{Data analysis}

\subsubsection*{Template analysis program}
A template analysis tarball is provided on the project website
\par\noindent 
($http://www-clued0.fnal.gov/\%7Emuanza/Blind\_SUSY/Analysis/Run\_Analysis.tar.gz$). It can either be used
as is or be hacked by the participants. In any case, the most useful information for those
who'd like to write their own analysis code are the cross sections of all the SM backgrounds.
They can be found in the "$proc\_xsect$" array at the top of the Analysis.C file.

\subsubsection*{Some illustrations of the $PsD_{1}$ sample}
Here are some plots advertising the project. They were produced after rejecting the events
not passing any trigger requirements\footnote{A code for simulating the trigger conditions can be found in the ana:HLT()
function of the ana.C file in the analysis tarball}.

\begin{figure}[ht]
\begin{center}
\includegraphics[width=0.35\textwidth]{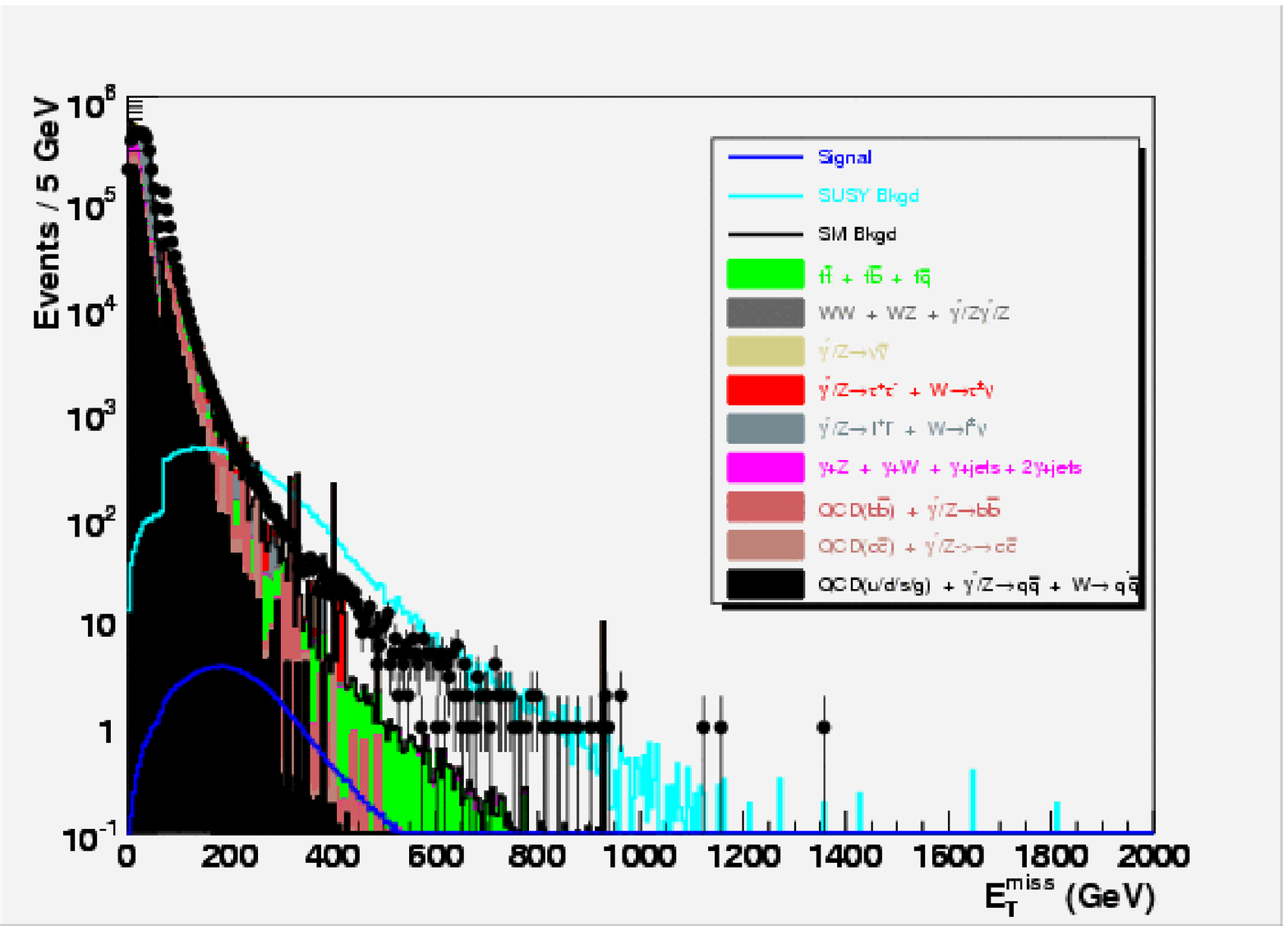}
\includegraphics[width=0.35\textwidth]{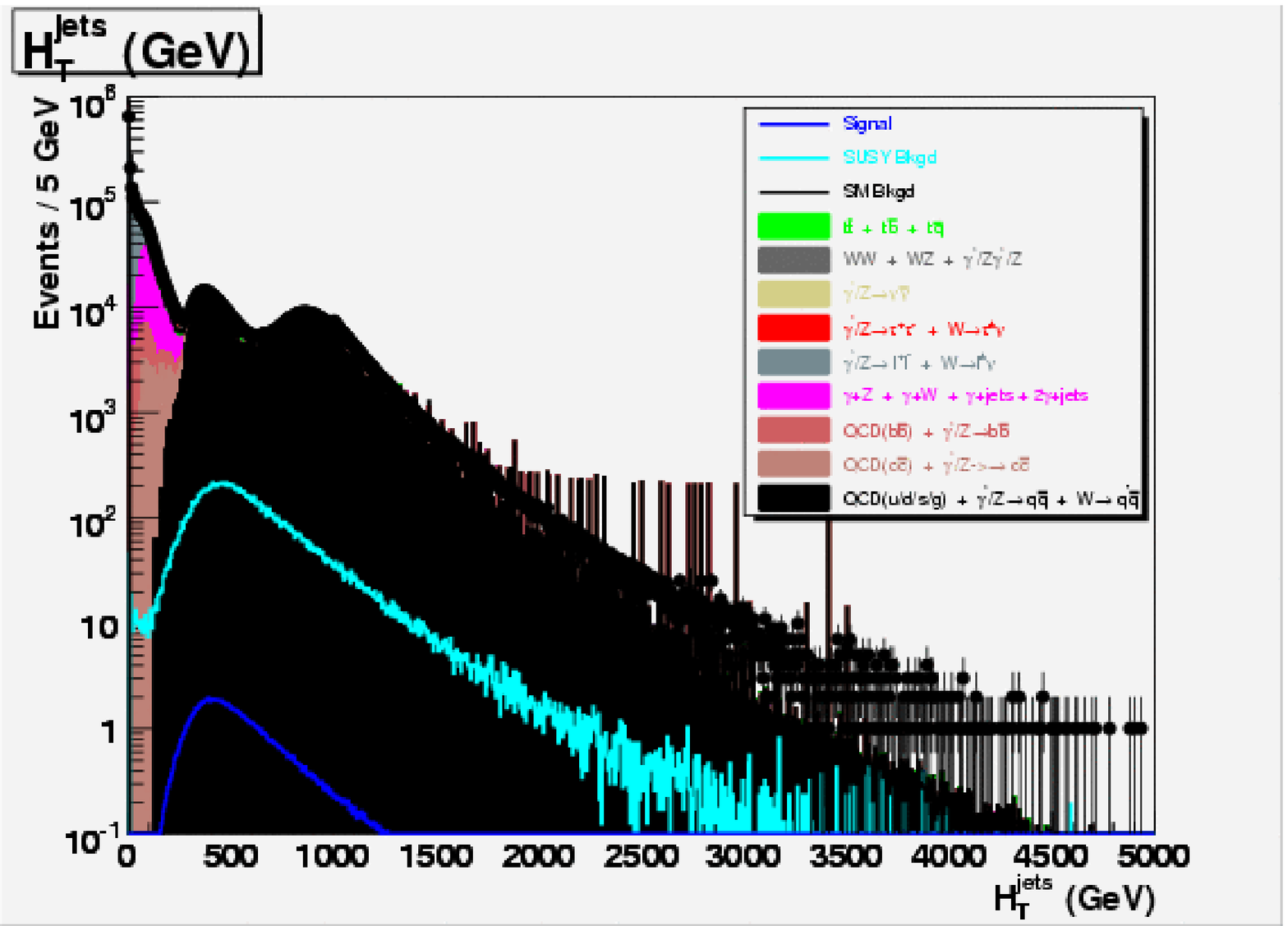}
\caption{The distributions of the missing (left) and scalar (right) $E_{T}$
after rejecting the events failing the trigger requirements.
A SUSY signal (dark blue) studied in \cite{Allanach:2006fy} and its SUSY background (light blue) are superimposed
only for illustrative purposes.}
\label{blind_susy_search_fig1}
\end{center}
\end{figure}


\par\noindent
These plots exhibit an excess of pseudo-data events in the high missing and scalar $E_{T}$ tails...

\subsubsection*{Signal templates production}
A tarball of the production package is available on the project website
\par\noindent
(\small{$http://www-clued0.fnal.gov/\%7Emuanza/Blind\_SUSY/Production/running\_susyblind.tar.gz$}). The analyzers
have to produce their SLHA input cards for their preferred SUSY models and to produce
signal templates using the production tarball. This way they can test ideas about possible
signals that could explain the difference between the $PsD_{1}$ sample and the SM background.


\subsection{Conclusions and prospects}

We have proposed a special SUSY data challenge at the LHC that includes the full Higgs and SUSY
inclusive production for an unrevealed SUSY model on top of the SM backgrounds. 
\par\noindent
The aim of this challenge is to subject to a blind analysis possible strategies to disentangle a given Higgs
or SUSY signal in the presence of simultaneous contributions from different other Higgs and SUSY processes. 
And to see how well these strategies enable to determine the properties of the SUSY model under study.

We look forward for participants to this challenge to present
their analysis of this first pseudo-data sample. We are eager to see
what experimental observables they'll have measured, what will be their uncertainties estimates.
We are expecting their initial best guess for the values of the hidden SUSY model parameters. And
we also suggest they provide the SUSY fitter
groups \cite{Rauch:2007xp}\cite{Bechtle:2005qj} with their observables and uncertainties so as to find out what
global fits could teach us about in this blind analysis context.


\subsection*{Acknowledgements}
We would like to thank the IPN Lyon and the IN2P3 for their financial support.


\clearpage
\newpage

\setcounter{figure}{0}
\setcounter{table}{0}
\setcounter{equation}{0}
\setcounter{footnote}{0}
\section{Off-shell effects for decay processes in the MSSM
\protect\footnote{N.~Kauer and C.F.~Uhlemann}}


\subsection{Introduction\label{ofsdec_sec:intro}}

Theoretical arguments and experimental observations indicate that new particles
or interactions play an important role at the TeV scale, which will become
directly accessible at the Large Hadron Collider (LHC)
and its planned complement, the International Linear Collider.
In the near future we can therefore anticipate ground-breaking discoveries that 
reveal physics beyond the Standard Model (BSM) and allow to gain insight into
the structure of the fundamental theory.  Theoretically appealing extensions
of the Standard Model often feature numerous additional interacting 
heavy particles.  The phenomenology of supersymmetric (SUSY) theories, for example,
is characterized by sparticle production and cascade decays,
which lead to many-particle final states and scattering amplitudes with complex 
resonance structure.
In order to extract the additional Lagrangian parameters of an extended theory 
from collider data, theoretical predictions are required that match the 
experimental accuracies.
In theoretical calculations production and decay stages can be factorized by means of
the narrow-width approximation (NWA),
which effectively results in on-shell intermediate states.
Its main advantage is that sub- and nonresonant as well as nonfactorizable 
amplitude contributions can be neglected in a theoretically consistent way,
resulting in significant calculational simplifications at tree and loop level.
For these reasons, the NWA is employed in nearly all studies of BSM physics.
We note that it is implicitly applied whenever branching ratios are extracted 
from scattering cross sections.  A reliable NWA uncertainty 
determination is therefore crucial.
Given the width $\Gamma$ and mass $M$ of an unstable 
particle, the uncertainty of the NWA is commonly estimated as 
${\cal O}(\Gamma/M)$.
With $\Gamma/M$ frequently $\lesssim\ 2\%$, its uncertainty is expected to be
small in comparison to, for instance, QCD corrections.

\begin{figure}[t]
\begin{center}
\begin{minipage}[c]{.49\linewidth}
\flushleft \includegraphics[height=6.7cm]{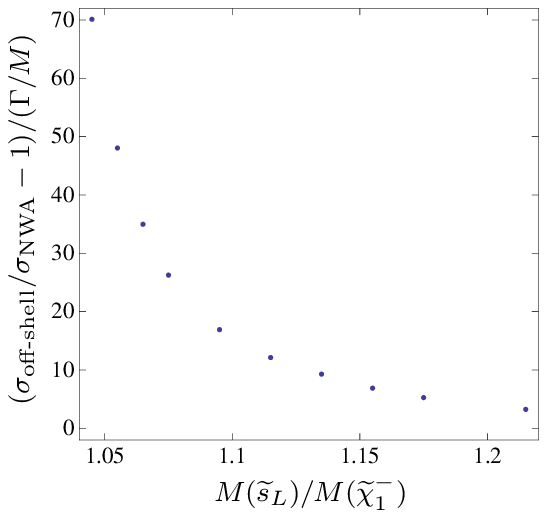}
\end{minipage} \hfill
\begin{minipage}[c]{.49\linewidth}
\flushright \includegraphics[height=6.7cm]{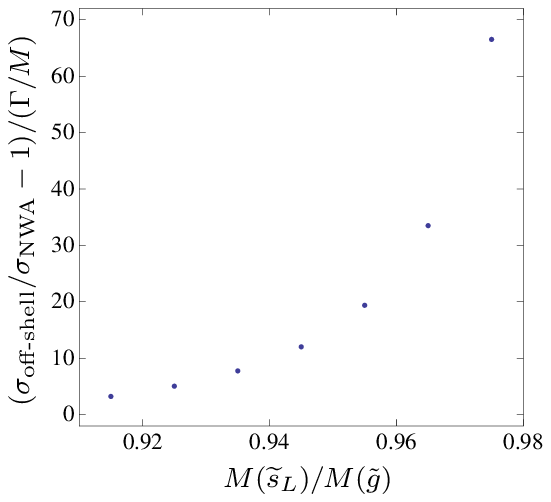}
\end{minipage}\\[0.2cm]
\caption{The accuracy of the NWA cross section normalized to the conventionally 
expected 
uncertainty is shown for $\tilde{g}\,\widetilde{u}_L$ production at the LHC followed 
by the cascade decay $\tilde{g} \to \widetilde{s}_L\bar{s}$ and 
$\widetilde{s}_L \to \widetilde{\chi}^-_1 c$ in the MSSM at SPS1a' for 
a variable strange squark mass that approaches the chargino mass (left)
and the gluino mass (right).
\label{ofsdec_fig:lhcdecay}}
\end{center}
\end{figure}

Recently, two circumstances have been observed in which the NWA
is not reliable: the first involves decays 
where a daughter mass approaches the parent mass \cite{Berdine:2007uv},
the second involves the convolution of parton distribution functions with a resonant
hard scattering process \cite{Kauer:2007zc}.
We are thus motivated to investigate when and why the NWA is not
appropriate in the context of cascade decays in the Minimal Supersymmetric
Standard Model (MSSM).
We first consider a typical example, namely 
$\tilde{g}\,\widetilde{u}_L$ production at the LHC, i.e. in proton-proton collisions
at 14 TeV, with the subsequent cascade decay $\tilde{g} \to \widetilde{s}_L\bar{s}$ 
and $\widetilde{s}_L \to \widetilde{\chi}^-_1 c$ at the SPS1a' benchmark point 
\cite{AguilarSaavedra:2005pw,Allanach:2002nj} in the MSSM parameter space.
Phenomenologically, to consider 
a squark decay into the LSP candidate $\widetilde{\chi}^0_1$ would be more natural,
but the resulting complete Feynman amplitude features a complicated resonance 
structure whose study we leave to future work.  
Even for the gluino decay chain considered here, interference arises from 
$\tilde{g} \to (\widetilde{c}^\ast_L \to \widetilde{\chi}^-_1 \bar{s}) c$.
However, it does not exceed the expected NWA uncertainty and can therefore be 
neglected.
We focus on off-shell effects for the resonant $\widetilde{s}_L$ state
(with $M=570$ GeV and $\Gamma=5.4$ GeV at SPS1a')
and hence treat the chargino as stable and the gluino in NWA with spin correlations.
As shown in Fig.~\ref{ofsdec_fig:lhcdecay}, the NWA error substantially exceeds the
expectation of $\Gamma(\widetilde{s}_L)/M(\widetilde{s}_L) < 1\%$
when the strange squark mass approaches either the chargino or gluino mass 
of $184$ and $607$ GeV, respectively.\footnote{The tools of Ref.~\cite{Reuter:2005us} were used in our calculations.}
Note that the region where the NWA is inappropriate is not restricted to
mass configurations where the Breit-Wigner shape is cut off kinematically,
i.e.~where $M(\widetilde{s}_L) - M(\widetilde{\chi}^-_1) \lesssim 
\Gamma(\widetilde{s}_L)$ or $M(\tilde{g}) - M(\widetilde{s}_L) \lesssim 
\Gamma(\widetilde{s}_L)$.


\subsection{Resonant {\boldmath $1\to 3$} decays in the MSSM\label{ofsdec_sec:1to3decay}}

\begin{figure}
\begin{center}
\begin{minipage}[c]{.49\linewidth}
\flushleft \includegraphics[height=6.7cm]{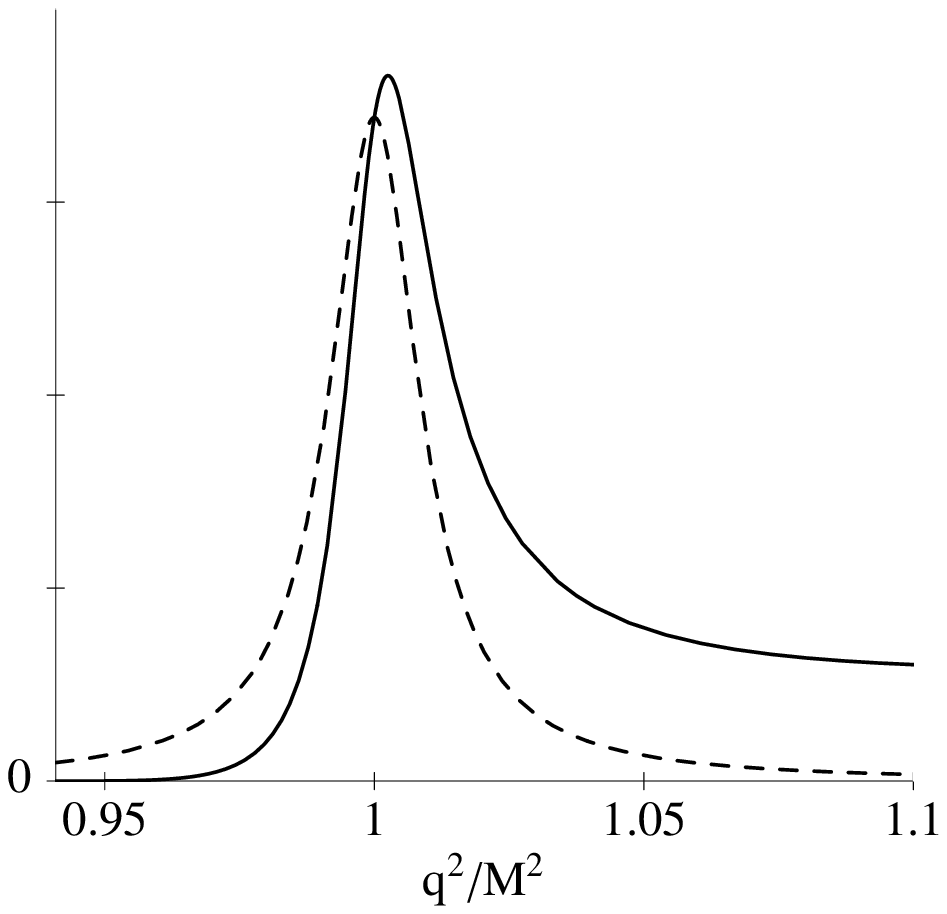}
\end{minipage} \hfill
\begin{minipage}[c]{.49\linewidth}
\flushright \includegraphics[height=6.7cm]{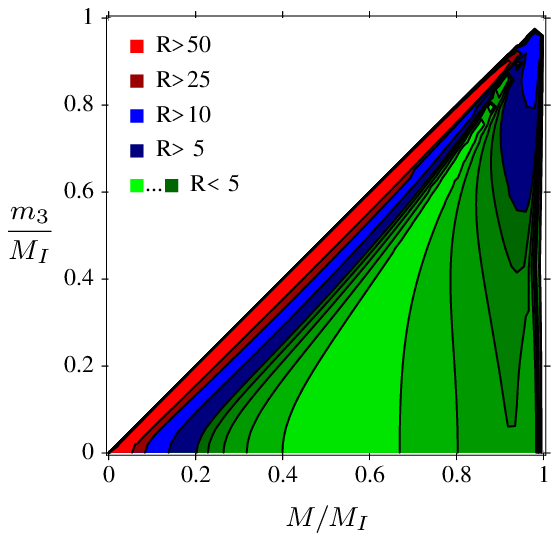}
\end{minipage}\\[0.2cm]
\caption{Resonant $1\to 3$ decay SSS-SSV (see main text) with $\Gamma/M=0.01$: 
The graph displays the 
$q^2$-dependence of the Breit-Wigner (dashed) that is integrated out 
in the NWA and of the complete integrand of Eq.~\protect\ref{ofsdec_eq:ampfac} (solid) 
for $m_3=M-3\Gamma$.  The contour plot shows $R$, the resulting off-shell-NWA 
deviation in units of $\Gamma/M$, as function of $m_3$ and $M$ with $m_1=m_2=0$.
\label{ofsdec_fig:sss-ssv}}
\end{center}
\end{figure}

The example in Sec.~\ref{ofsdec_sec:intro} suggests resonant $1\to 3$ decays
as smallest unit that features the amplified off-shell effects.
Giving type (4-momentum, mass) for each particle, we define a resonant $1\to 3$ decay 
by
\begin{equation}
\label{ofsdec_eq:kin}
 T_I (P_I, M_I) \to T_1 (p_1, m_1),\;  T (q, M)\ \ \text{and}\ \ T (q, M) \to  T_2 (p_2, m_2),\;  T_3 (p_3, m_3)\;.
\end{equation}
The width of the intermediate particle with momentum $q$ is $\Gamma$.  
Type can be scalar (S), fermion (F) or vector boson (V).  In the MSSM, 48 generic
processes exist and are identified with type codes $T_IT_1T$-$TT_2T_3$.
For each process we have systematically scanned the MSSM parameter space
for the maximum deviation $|R|$ of off-shell ($\Gamma_\text{off-shell}$) and 
NWA ($\Gamma_\text{NWA}$) decay rate predictions, where
$R = (\Gamma_\text{off-shell}/\Gamma_\text{NWA}-1)/(\Gamma/M)$.
Note that in $R$, coupling constants typically cancel with the exception of
the relative strength of the chiral components of SFF and VFF vertices, 
which has been varied in addition to the masses and width.
From this survey \cite{Uhlemann:diplomarbeit} 
we conclude that large deviations $|R|$ do not occur for 
configurations with a resonance mass that is very far from kinematical bounds.  The NWA exploits that in the limit $\Gamma\to 0$ the squared propagator 
$D(q^2)\equiv [(q^2-M^2)^2+(M\,\Gamma)^2]^{-1}$ 
is asymptotically equal to $2\pi K_\text{NWA}\delta(q^2-M^2)$ with 
$K_\text{NWA}=1/(2M\,\Gamma)=\int_{-\infty}^\infty D(q^2)\,dq^2/(2\pi)$.
The Breit-Wigner shape is thus effectively integrated out.
The origin of unexpectedly large deviations for configurations 
where kinematical bounds are outside the resonance region 
is that the $q^2$-dependence of the 
residual integrand significantly distorts the peak and tail of $D(q^2)$.
We find that the effect is most pronounced for the decay process SSS-SSV.
We thus use it to demonstrate the distortion.
With $m_1=m_2=0$, the $q^2$-integrand is given by
\begin{equation}
\label{ofsdec_eq:ampfac}
\left(1-\frac{q^2}{M_I^2}\right) \left(1-\frac{m_3^2}{q^2}\right)
\left(\frac{(q^2-m_3^2)^2}{m_3^2}\right)\frac{1}{(q^2-M^2)^2+M^2 \Gamma^2}\;.
\end{equation}
The 1st- and 2nd-stage decay PS elements contribute the first and second factor,
respectively.  The 2nd-stage decay matrix element gives the third factor.
When $m_3^2 \lesssim M^2$ the second and third factor effect a strong 
deformation of $D(q^2)$, which, together with the resulting large deviations,
is displayed in Fig.~\ref{ofsdec_fig:sss-ssv}.
The deviation grows with
increasing power of the deforming factors.
When $M$ approaches the lower kinematic bound, $|R|$ 
is sensitive to the type of the 2nd-stage decay, 
which determines the power of the factor that 
deforms the Breit-Wigner peak.  While this factor enhances the Breit-Wigner 
tail, the factor of the 1st-stage decay suppresses it.
And vice versa for the upper bound.
We find stronger effects for SSV, VSV, FFV, VVV and SVV than for 
FSF, SFF, VFF, VSS and SSS vertices.
Using our generic results we probe resonant $1\to 3$ decays at SPS benchmark points 
\cite{Allanach:2002nj}.  Decays with larger deviations are shown in Table \ref{ofsdec_tab:spsdevs}.

Affected decays generally have a small branching ratio due to similar-mass 
configurations.  For example, BR = 1.3\% for the decay mode 
$\tilde \chi_1^+\rightarrow \tilde \chi_1^0 u \bar{d}$ at SPS1a.  It proceeds via the intermediate states $W^+$ (resonant), 
$\tilde u_L$ and $\tilde d_L^\ast$ (nonresonant).  The resonant $W^+$ contribution
with $R\cdot\Gamma/M=13\%$ (see Table \ref{ofsdec_tab:spsdevs}) induced by the mass 
ratio $(m_1 + M)/M_I = 0.975$ 
dominates, and the off-shell prediction including
the nonresonant contributions deviates by about 11\% from the NWA prediction.  
Since the 1st-decay stage is not affected by QCD corrections, 
this error is particularly significant.
For a detailed discussion of effects at SPS points including cascade decay segments
we refer to Ref.~\cite{Uhlemann:diplomarbeit}.

\subsection{Conclusions}
When the NWA is applied to decay chains 
the approximation error will exceed order $\Gamma/M$ for mass configurations 
in an extended vicinity of segment kinematical bounds due to a significant distortion 
of the Breit-Wigner peak and tail, which is effected by the $q^2$-dependence
of the phase space elements and residual matrix elements.
In phenomenological studies of affected models, fully off-shell
tree-level Monte Carlos \cite{Reuter:2005us} should thus be used even 
though it requires more computing resources.
For decay processes involving strongly interacting particles
QCD corrections are known to be large 
\cite{Beenakker:1996dw,Kraml:1996kz,Djouadi:1996wt} 
and need to be taken into account.
For this purpose, a suggestive NWA improvement is proposed 
in Ref.~\cite{Kauer:2007nt}.
We have chosen the MSSM to illustrate how large off-shell 
effects can occur in extended models, but emphasize that the effects do not 
depend on SUSY.

\begin{table}[t]
\begin{center}
\begin{tabular}{|c|c|c|c|}
\hline
decay process & SPS & $R$ & $\Gamma/M$ $[\%]$ \\
\hline
\rule[0mm]{0mm}{3.5mm}
  $\tilde g\rightarrow d \tilde d_L^\ast\rightarrow d \bar d\tilde \chi_1^0$ & $1a$  &$9.54$& $0.935$  \\
  $\tilde g\rightarrow d \tilde d_L^\ast\rightarrow d \bar d\tilde \chi_1^0$ & $5$ & $ 11.4$& $0.956$  \\
  $\tilde g\rightarrow u \tilde u_L^\ast\rightarrow u \bar u \tilde \chi_1^0$ & $1a$ & $5.98$& $0.976$  \\
  $\tilde g\rightarrow u \tilde u_L^\ast\rightarrow u \bar u \tilde \chi_1^0$ & $5$  & $9.46$ & $0.975$  \\
$\tilde \chi_1^+\rightarrow \tilde \chi_1^0 W^+\rightarrow \tilde \chi_1^0 u \bar{d}$ & $1a$ &$5.21$    &$ 2.49 $ \\
$\tilde \chi_1^+\rightarrow \tilde \chi_1^0 W^+\rightarrow \tilde \chi_1^0 e^+ \nu_e$ & $1a$ &$5.21 $   & $2.49 $\\
  $\tilde g\rightarrow\bar{b}\tilde b_2\rightarrow \bar{b} b \tilde \chi_1^0$ & $4$  & $6.43$& $1.11$  \\
  $\tilde g\rightarrow \bar u \tilde u_L\rightarrow \bar u d\tilde \chi_1^+$ & $9$ & $ 114$& $1.19$  \\
  $\tilde g\rightarrow d \tilde d_L^\ast \rightarrow d \bar u \tilde \chi_1^+$ & $9$  & $ 209$& $1.19$  \\
\hline
\end{tabular}
\end{center}
\caption{$R$, the off-shell-NWA deviation in units of $\Gamma/M$, for resonant 
$1\to 3$ decays at SPS benchmark points.\label{ofsdec_tab:spsdevs}}
\end{table}
\subsection*{Acknowledgements}

N.~Kauer would like to thank the Ecole de Physique des Houches and the
Galileo Galilei Institute for Theoretical
Physics for the hospitality and the INFN for partial support during the
completion of this work.  
This work was supported by the BMBF, Germany (contract 05HT1WWA2).

\clearpage
\newpage

\setcounter{figure}{0}
\setcounter{table}{0}
\setcounter{equation}{0}
\setcounter{footnote}{0}
\section{Supersymmetric corrections to \boldmath{$M_W$} and \boldmath{$\sin^2 \theta^l_w$} in mSUGRA
\protect\footnote{B.C.~Allanach, F. Boudjema and A.M.~Weber}}

\subsection{Introduction}
Specific patterns of supersymmetry (SUSY) breaking provide relationships
between various sparticle masses. In mSUGRA for instance, the
GUT-scale scalar masses are set to $m_0$, the gaugino masses $M_{1/2}$ and the
universal scalar SUSY breaking trilinear coupling $A_0$. These degeneracies
are broken by renormalisation group effects between the GUT scale and $M_Z$.
It is by now well known that various regions of mSUGRA parameter space are
ruled out by direct sparticle search constraints, which place lower
bounds upon the sparticle masses.

Sparticles may appear in loop corrections to electroweak
observables, therefore affecting the values of the latter as
predicted within the Standard Model once a set of independent
physical input parameter is chosen, such as the electromagnetic
couplings, $G_F$ from $\mu$ decay, $M_Z$ and other SM particle
masses. Two such precision observables are the W boson mass, $M_W$,
as measured at LEPII and the Tevatron and the effective
leptonic mixing angle, $\sin^2 \theta^l_{w}$, derived from the
lepton asymmetries measured at LEPI and SLD. 
The former is related to the electric charge $e=\sqrt{4\pi\alpha}$, the weak mixing angle $s_w^2=1-M_W^2/M_Z^2$,
$G_F$, and $M_Z$ via
\begin{equation}
\frac{G_F}{\sqrt{2}}=\frac{e^2}{8 s_{w}^2 M_W^2}(1+\Delta r),
\end{equation}
where the parameter $\Delta r$ is a model dependent quantity which accounts for all higher order corrections to the muon decay (this includes self energies, vertex and box corrections in a given model, see ref.~\cite{Heinemeyer:2006px} for a recent discussion in the context of the MSSM).
The $M_W$ value which solves the above relation constitutes the model specific prediction of $M_W$
for a given fixed set of Standard Model and new physics parameters.
Similarly, one can express the effective leptonic mixing angle as
\begin{equation}
\sin^2 \theta^l_{w}=s_w^2(1+\Delta\kappa),
\end{equation}
where the model dependent higher order corrections to the leptonic $Z$~boson decay, $Z\to l\bar l$, enter via~$\Delta \kappa$ (details concerning $\sin^2 \theta^l_{w}$ in the general MSSM can be found in ref.~\cite{Heinemeyer:2007bw}). The prefactor $s_w^2=1-M_W^2/M_Z^2$, and thus also $\sin^2 \theta^l_{w}$, is furthermore sensitive to radiative corrections via~$M_W$.
%
It is convenient to split the MSSM higher order corrections into Standard Model and SUSY type contributions~\cite{Heinemeyer:2006px,Heinemeyer:2007bw}
\begin{equation}
\Delta r = \Delta r^{\rm SM}|_{M^{\rm SM}_H=M_h} + \Delta r^{\rm SUSY}, \ \ \ \ \ \Delta \kappa = \Delta \kappa^{\rm SM}|_{M^{\rm SM}_H=M_h} + \Delta \kappa^{\rm SUSY},
\end{equation}
with the Standard Model Higgs boson mass $M^{\rm SM}_H$ set to the lightest MSSM Higgs mass $M_h$.
Direct search constraints put lower bounds upon sparticle masses,
limiting the size of the SUSY contributions,
which are propagator suppressed by large sparticle masses.

Empirical constraints on $M_W$ and $\sin^2\theta_w^l$ are very
tight: they are taken here to
be~\cite{Allanach:2007qk,Alcaraz:2007ri}
\begin{equation}
M_W=80.398\pm0.027 \mbox{~GeV},\qquad
\sin^2 \theta_w^l = 0.23153 \pm 0.000175. \label{empitic}
\end{equation}
One may ask how large the SUSY contributions  to the electroweak
observables are, given the current strong constraints upon
sparticle masses from LEP2 and the Tevatron. If the SUSY
contributions are much smaller than experimental errors upon the
relevant observables, then there is no need to include them in any
fit. Here, we use results from a previous fit of mSUGRA to dark
matter and other indirect data (including the direct search
constraints) in order to see how big the SUSY contribution to
$M_W$ and $\sin^2 \theta_w^l$ may be.

\subsection{The fits}

In refs.~\cite{Allanach:2007qk}, multi-dimensional fits to mSUGRA were
presented using the {\tt SOFTSUSY2.0.10}~\cite{Allanach:2001kg} spectrum
calculator and the {\tt   micrOMEGAs1.3.6}~\cite{Belanger:2004yn} dark matter code. $m_0$, $A_0$, $\tan \beta$,
$M_{1/2}$,
$m_t$, $m_b$, $\alpha_s
(M_Z)$ and $\alpha$ were all scanned simultaneously using the Metropolis
algorithm in a Markov Chain Monte Carlo technique. It was assumed that the
WMAP-constrained relic density of dark matter $\Omega_{DM} h^2$ consisted
entirely of the lightest neutralino, which is stable by the assumption of
R-parity.
The following data were
included in the fit: the anomalous magnetic moment of the muon, $(g-2)_\mu$,
$BR(b \rightarrow s \gamma)$, Tevatron $BR(B_s \rightarrow \mu^+ \mu^-)$
constraints, $\Omega_{DM} h^2$, $M_W$, $\sin^2 \theta_w^l$, LEP2 Higgs
constraints as well as other constraints upon sparticle masses from direct
searches. Data on $m_t$, $m_b$, $\alpha_s(M_Z)$ and $\alpha$ were also
included in the likelihood. We refer the reader to ref.~\cite{Allanach:2007qk}
for the details.
We note here that $M_W$ and $\sin^2 \theta_w^l$ were
determined using am embryonic version of the {\tt SUSYPOPE}~\cite{Arne} code,
which is a state-of-the-art MSSM calculation~\cite{Heinemeyer:2006px,Heinemeyer:2007bw} of the
electroweak observables\footnote{The {\tt SOFTSUSY} determination of $M_W$ and
  $\sin^2 \theta_w^l$ is not to a sufficient accuracy, given the tiny
  empirical errors upon   their measured values.}. {\tt SUSYPOPE} is also
capable of calculating the Standard Model value prediction for $M_W$ given
other Standard Model input parameters and the Higgs mass.
A list of $\sim$500~000 weighted mSUGRA parameter space
points,
with weighted frequency
proportional to their combined likelihoods, was the result of the fit. This set
of
points is referred to as {\tt KISMET} ({\bf K}iller {\bf I}nference in
{\bf S}usy {\bf M}eteorology).
The points are presented for public use on URL
\begin{verbatim}
http://users.hepforge.org/~allanach/benchmarks/kismet.html
\end{verbatim}

In ref.~\cite{Allanach:2007qk}, the fits were presented in several ways: the
ways relevant to our discussion here will be the
frequentist fashion (utilising the {\em profile likelihood}) and a Bayesian fit
with flat priors in the inputs listed above.
While the profile likelihood takes into account only the best fit parameters,
the Bayesian fit includes volume effects. Volume effects take into account the
volume of the probability distribution: for instance, a region which has a
very large volume in marginalised (or averaged) parameter directions but a
less good fit can still have an appreciable effect on the marginalised
posterior probability distribution.
However, as is well known, the Bayesian interpretation is dependent upon the
subjective prior choice unless the data are plentiful and precise.
As was pointed out in ref.~\cite{Allanach:2007qk}, the indirect data used in
the fits are currently not plentiful and precise enough. We therefore display
both methods and use the difference between the two as an indication of the
size of inherent uncertainty in our interpretation of the fits.

\subsection{The $W$ mass and the weak mixing angle}

We take the {\tt KISMET} points and re-weight them, taking off the likelihood
contributions from $M_W$ and $\sin^2 \theta_w^l$. By marginalising against
these two variables, we may then examine what size of SUSY contribution to
each is expected from fits to the other indirect data.
\begin{figure}
\begin{center}
\unitlength=1in
\begin{picture}(7,3)(0,0)
\put(-0.5,0){\includegraphics[width=4in]{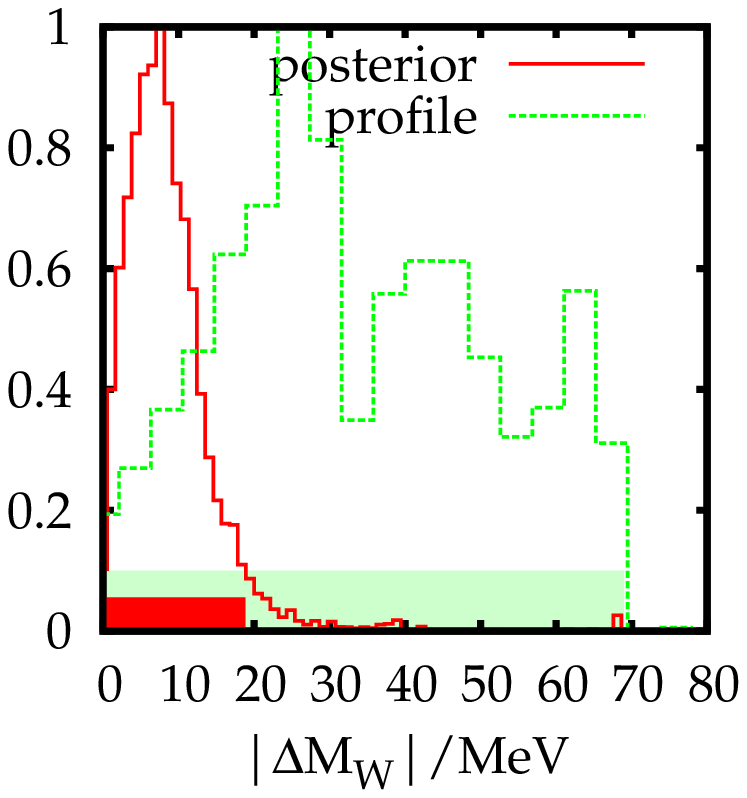}}
\put(2.5,0){\includegraphics[width=4in]{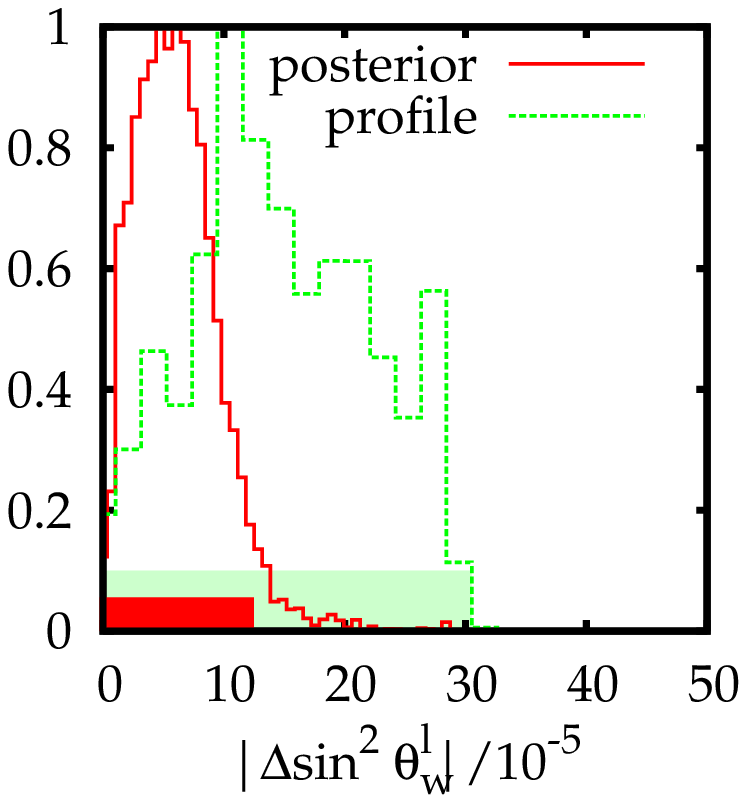}}
\end{picture}
 \caption{Distributions to the SUSY contribution to the
   electroweak observables $M_W$ and $\sin^2 \theta_w^l$ from mSUGRA fits.
   The posterior probability distribution function (pdf) is shown in red
   (dark) whereas the profile likelihood is
   shown in green (light).
   Each histogram has an arbitrary normalisation. Horizontal bands display the
   one-sided 95$\%$ Bayesian credibility (condidence level) regions for the
   posterior pdf (profile likelihood).
}
\label{search}
\end{center}
\end{figure}
In order to calculate the one-sided 95$\%$ limits upon the {\em magnitude} of
the SUSY corrections to each observable, we re-bin in terms of  $|\Delta
M_W|$ and $|\Delta\sin^2 \theta_w^l|$. $\Delta M_W$ is the difference between the
mSUGRA prediction and the Standard Model one with $\alpha_s, \alpha, m_t, m_b$
and $M_h$
identical to those associated with the mSUGRA point in question.
The posterior pdf with flat priors gives us
a 95$\%$ Bayesian credibility interval of
\begin{equation}
| \Delta M_W | < 19 \mbox{~MeV}, \qquad
| \Delta\sin^2 \theta_w^l | < 12 \times 10^{-5}.
\end{equation}
The 95$\%$ upper confidence limit in the frequentist interpretation is
obtained in each case from the profile likelihood:
\begin{equation}
| \Delta M_W | < 70 \mbox{~MeV}, \qquad
| \Delta\sin^2 \theta_w^l | < 30 \times 10^{-5}.
\end{equation}
Thus, the frequentist constraints are somewhat (two and a half times) more
relaxed than the Bayesian constraints with flat priors.
These numbers are to be compared with the empirical uncertainties quoted in
Eq.~\ref{empitic} of 17.5 MeV and $27 \times 10^{-5}$, respectively.
Even the tighter 95$\%$ Bayesian constraints are around the same values and
so we conclude that the SUSY contribution to the likelihood cannot be
neglected.

\subsection*{Acknowledgements}
This work has been partially supported by the STFC.

\clearpage
\newpage

\setcounter{figure}{0}
\setcounter{table}{0}
\setcounter{equation}{0}
\setcounter{footnote}{0}


\section{LHC and the muon's anomalous magnetic moment
\protect\footnote{M.~Alexander, S.~Kreiss, R.~Lafaye, T.~Plehn, M.~Rauch, and D.~Zerwas}}



\subsection{Introduction}

The strongest hint for a TeV-scale modification of the Standard Model
originates from the anomalous magnetic moment of the muon. This
parameter has been evaluated both in experiment and in theory to
unprecedented precision. We use the experimental value~\cite{Bennett:2006fi}
\begin{equation}
a_\mu^\mathrm{(Exp)} \equiv (g-2)_\mu/2 = 116 592 080(63) \times 10^{-11}
\quad .
\label{gmtwosfitter:gm2exp}
\end{equation}
In contrast, the Standard Model prediction is
smaller~\cite{Miller:2007kk,Hagiwara:2006jt}
\begin{equation}
a_\mu^\mathrm{(SM)} = 116 591 785(61) \times 10^{-11}
\quad ,
\label{gmtwosfitter:gm2SM}
\end{equation}
where we use the number using $e^+e^-$ data, as the discrepancy with
tau data is still large~\cite{Davier:2007ua}.
This difference corresponds to a $3.4\sigma$ deviation between theory
and experiment, suggesting a new--physics effect. Supersymmetry
provides a particularly attractive explanation of this discrepancy.
Contributions from the supersymmetric partners of the muon, the
muon-neutrino and the gauge and Higgs bosons modify the
Standard--Model prediction. The masses of the new particles
responsible for this signal should be of the order of several
hundred~GeV, a mass range well accessible to the
LHC~\cite{Czarnecki:2001pv,Stockinger:2006zn}. Therefore, LHC will be
able to test the hypothesis that this discrepancy is caused by
TeV-scale supersymmetry.

However, the benefit of such a test will go in both ways: after the
LHC will have measured observables like the masses of supersymmetric
particles or kinematic edges involving such
particles~\cite{Bachacou:1999zb,Allanach:2000kt}, the question will
arise what the fundamental parameters of the Lagrangian
are~\cite{Lafaye:2007vs,Bechtle:2005vt}. Since ($g-2$) at leading
order is proportional to $\tan\beta$ it includes useful
information~\cite{Hertzog:2007hz}, which can significantly improve the
extraction of $\tan\beta$.

As an example we use the experimentally well-studied parameter point
SPS1a~\cite{Allanach:2002nj}. Its theory prediction for $a_\mu$ is
$a_\mu^\mathrm{(SPS1a)} = a_\mu^\mathrm{(SM)} + 282 \cdot 10^{-11}$.
This leads to a deviation from the experimentally observed value of
$\Delta a_\mu = -13 \cdot 10^{-11}$, which is well below the
experimental error bounds. Therefore, we can safely use the
experimental value without further modifications.

Our analysis uses the parameter extraction tool
SFitter~\cite{Lafaye:2007vs} where we have added the necessary modules
to calculate the anomalous magnetic moment. To obtain a good balance
between precision and required time to perform the scans we use the
one-loop expression for ($g-2$) with additional leading two-loop
QED-logarithms. Note that in this study we are not mainly interested
in the best--fitting MSSM parameter point, but in the errors of the
MSSM parameters, so this simplification is appropriate.

\subsection{Weak--scale MSSM analysis}

The determination of the weak--scale MSSM Lagrangian at the LHC is
clearly preferable to tests of SUSY--breaking assumptions, as long as
we have enough information available at the LHC. The extracted model
parameter can then be run to a higher scale, to test for example
unification patterns~\cite{Lafaye:2007vs}. However, some model
parameters can be fixed, because neither LHC nor ($g-2$) will include
any information on them. 
Properly including the
top-quark mass as a free parameter we assume a 19-dimensional
weak--scale MSSM parameter space listed in
Table~\ref{gmtwosfitter:mssm_theo}. As we have shown in
Ref.~\cite{Lafaye:2007vs}, even this reduced parameter space cannot be
determined completely at the LHC. In the parameter point SPS1a, we for
example find an eightfold degeneracy in the gaugino--higgsino
sub-sector. Because only three of the neutralinos and none of the
charginos can be observed at the LHC, the connection between their
masses and the model parameters $M_1$, $M_2$ and $\mu$ is not unique.
In addition the sign of $\mu$ is not determined by LHC data
alone~\cite{Allanach:2006cc}.

This is where the anomalous magnetic moment of the muon adds
important information. First, the deviation from the Standard Model
prediction is proportional to the sign of $\mu$, the parameter which
couples the two Higgs superfields in the superpotential. Including
($g-2$) data will clearly favor one sign of $\mu$, namely the
$\mu>0$, thereby reducing the degeneracy by a factor of two.

When reconstructing the fundamental parameters of the Lagrangian the
central values have to be accompanied with the correct error bars.
There are three different types of experimental errors on the
observables: a statistical error and the two (correlated) systematic
errors for the jet and lepton energy scales. All experimental errors
are Gaussian shaped. In addition, we include flat theory errors of
$1\%$ for all colored particle masses and $0.5\%$ for all others.
For ($g-2$) we use the values given in
Eqs.(\ref{gmtwosfitter:gm2exp},~\ref{gmtwosfitter:gm2SM}). The
convolution of these errors is described in Ref.~\cite{Lafaye:2007vs}.
To determine the errors on the model parameter we randomly smear the
nominal values for SPS1a.  The corresponding random numbers obey a
distribution according to the associated errors. Then we minimize
$\chi^2$ for each pseudo--measurement and repeat this procedure 10000
times. The emerging distribution of the parameters is simply the
result of the correct error propagation. Using a Gaussian fit we then
extract the central value and the $1\sigma$ standard deviation of each
parameter.

\begin{table}[t]
\begin{center} \begin{small}
\begin{tabular}{|l|r@{$\pm$}rr@{$\pm$}r|r@{$\pm$}rr@{$\pm$}r|r|}
\hline
                     & \multicolumn{4}{c|}{only experimental errors} & \multicolumn{4}{c|}{including flat theory errors}&  SPS1a\\
\hline
                     & \multicolumn{2}{c}{LHC}     &\multicolumn{2}{c|}{LHC $\otimes (g-2)$}&  \multicolumn{2}{c}{LHC}    &\multicolumn{2}{c|}{LHC $\otimes (g-2)$}&  \\
\hline                                                                           
\boldmath{$\tan\beta$} &\bf    9.8 &\bf 2.3          &\bf    9.7 &\bf 2.0          &\bf   10.0 &\bf 4.5          &\bf   10.3 &\bf 2.0          &\bf  10.0 \\
$M_1$                &     101.5 & 4.6             &     101.1 & 3.6             &     102.1 & 7.8             &     102.7 & 5.9             &    103.1 \\
$M_2$                &     191.7 & 4.8             &     191.4 & 3.5             &     193.3 & 7.8             &     193.2 & 5.8             &    192.9 \\
$M_3$                &     575.7 & 7.7             &     575.4 & 7.3             &     577.2 & 14.5            &     578.2 & 12.1            &    577.9 \\
$M_{\tilde{\tau}_L}$ &     196.2 & $\mathcal{O}(10^2)$     &     263.4 & $\mathcal{O}(10^2)$     &     227.8 & $\mathcal{O}(10^3)$     &     253.7 & $\mathcal{O}(10^2)$     &    193.6 \\
$M_{\tilde{\tau}_R}$ &     136.2 & $\mathcal{O}(10^2)$     &     156.8 & $\mathcal{O}(10^2)$     &     164.1 & $\mathcal{O}(10^3)$     &     134.1 & $\mathcal{O}(10^2)$     &    133.4 \\
$M_{\tilde{\mu}_L}$  &     192.6 & 5.3             &     192.3 & 4.5             &     193.2 & 8.8             &     194.0 & 6.8             &    194.4 \\
$M_{\tilde{\mu}_R}$  &     134.0 & 4.8             &     133.6 & 3.9             &     135.0 & 8.3             &     135.6 & 6.3             &    135.8 \\
$M_{\tilde{e}_L}$    &     192.7 & 5.3             &     192.2 & 4.5             &     193.3 & 8.8             &     194.0 & 6.7             &    194.4 \\
$M_{\tilde{e}_R}$    &     134.0 & 4.8             &     133.6 & 3.9             &     135.0 & 8.3             &     135.6 & 6.3             &    135.8 \\
$M_{\tilde{q}3_L}$   &     478.2 & 9.4             &     476.1 & 7.5             &     481.4 & 22.0            &     485.6 & 22.4            &    480.8 \\
$M_{\tilde{t}_R}$    &     429.5 & $\mathcal{O}(10^2)$     &     704.0 & $\mathcal{O}(10^2)$     &     415.8 & $\mathcal{O}(10^2)$     &     439.0 &$\mathcal{O}(10^2)$      &    408.3 \\
$M_{\tilde{b}_R}$    &     501.2 & 10.0            &     502.4 & 7.8             &     501.7 & 17.9            &     499.2 & 19.3            &    502.9 \\
$M_{\tilde{q}_L}$    &     523.6 & 8.4             &     523.0 & 7.5             &     524.6 & 14.5            &     525.5 & 10.6            &    526.6 \\
$M_{\tilde{q}_R}$    &     506.2 & 11.7            &     505.8 & 11.4            &     507.3 & 17.5            &     507.6 & 15.8            &    508.1 \\
$A_\tau$             &\multicolumn{2}{c}{fixed 0}  &\multicolumn{2}{c|}{fixed 0}  &\multicolumn{2}{c}{fixed 0}  &\multicolumn{2}{c|}{fixed 0}  &   -249.4 \\
$A_t$                &    -500.6 & 58.4            &    -519.8 & 64.3            &    -509.1 & 86.7            &    -530.6 & 116.6           &   -490.9 \\
$A_b$                &\multicolumn{2}{c}{fixed 0}  &\multicolumn{2}{c|}{fixed 0}  &\multicolumn{2}{c}{fixed 0}  &\multicolumn{2}{c|}{fixed 0}  &   -763.4 \\
$m_A$                &     446.1 & $\mathcal{O}(10^3)$     &     473.9 & $\mathcal{O}(10^2)$     &     406.3 & $\mathcal{O}(10^3)$     &     411.1 & $\mathcal{O}(10^2)$     &    394.9 \\
$\mu$                &     350.9 & 7.3             &     350.2 & 6.5             &     350.5 & 14.5            &     352.5 & 10.8            &    353.7 \\
$m_t$                &     171.4 & 1.0             &     171.4 & 1.0             &     171.4 & 1.0             &     171.4 & 0.90            &    171.4 \\
\hline                 
\end{tabular}          
\end{small} \end{center}
\caption[]{Result for the general MSSM parameter determination at 
  the LHC in SPS1a. The left part neglects all theory errors, the
  right one assumes flat theory errors. In all cases a set
  of 20 kinematic endpoints and the top-quark and lightest Higgs--mass 
  measurements have been used. In the third and fifth column 
  we include the current measurement of ($g-2$).
  All masses are given in GeV.}
\label{gmtwosfitter:mssm_theo}
\end{table}

Table~\ref{gmtwosfitter:mssm_theo} shows the result of our SPS1a
analysis. For comparison and to make the effect of the additional
($g-2$) data easily visible, we include the result without ($g-2$)
data from Tables VIII and IX of Ref.~\cite{Lafaye:2007vs}. We give
results with experimental errors only (columns 2 and 3) and including
theory errors (columns 4 and 5). The effect of the additional
information on the accuracy of the parameter determination is clearly
visible. It is particularly significant for $\tan\beta$, which is not
well determined by the measurements of kinematic endpoints at the LHC.
The best source of information on $\tan\beta$ is the light MSSM Higgs
mass~\cite{Frank:2006yh}, but this observable strongly relies on the
assumed minimal structure of the Higgs sector, on the knowledge of
many other MSSM parameters, and on the estimate of the theory errors
due to higher orders.  Because of a lack of complementary measurements
(for example $A_t$) a change in $\tan\beta$ can always be compensated
by an appropriate change in other MSSM parameters, leaving the value
of all LHC observables unchanged. Additional sources of a $\tan\beta$
measurement are the production rate for heavy Higgs
bosons~\cite{Kinnunen:2004ji} and rare decays like $B_s \to \mu^+
\mu^-$, which we study elsewhere in this volume, but both of them only
work for large enough values of $\tan\beta$.

The ($g-2$) prediction has a leading linear dependence on $\tan\beta$.
Therefore, the improvement of the $\tan\beta$ errors by more than a
factor of two can be easily understood. This improved accuracy of
$\tan\beta$ influences those parameters which must be re-rotated when
$\tan\beta$ is changed to reproduce the same physical observables.
Correlations and loop corrections propagate the improvement over
almost the complete parameter space. 

\subsection{SLHAio}

The link between the main SFitter modules and the ($g-2$) module is
provided by SLHAio.  SLHAio is a library which allows for a smooth
communication between different programs according to the SUSY Les
Houches Accord (SLHA)~\cite{Skands:2003cj} and its extension
SLHA2~\cite{Allanach:2007zz}.  With such an interface, for example a
($g-2$) code can easily be used in a large--scale program like
SFitter.

The separation of actual calculations and SLHA interface creates a
simple structure. SLHAio itself has an easy instruction set with a
concept similar to the XML Path Language (XPath)~\cite{xpath}. Each
data field in SLHA is associated with a leaf in a tree. The leafs are
accessed and manipulated with a path. The path itself is similar to
the path used for navigation in file systems.

The tree has no predefined structure. Its size is determined by the
amount of information stored.  It can of course grow beyond the fields
defined in SLHA and SLHA2. This feature we use for the smuon mass
matrix, where SLHA restricts itself to third--generation particles.
There are two data types for each field: string and double.
Conversions are done automatically. Setting a leaf with a string and
afterwards reading a double is possible, as long as the string can be
converted.  The library assures the highest precision possible.
Strings which are never changed through doubles remain strings,
because this representation has the highest precision. Conversions to
strings are done in the format defined in SLHA. One part of SLHAio is
a templated container class for matrices. This class is fully
integrated into SLHAio, so matrices can be read, stored and printed
with a single command.

As discussed above, the ($g-2$) code can be used stand--alone and
within SFitter. When used with SFitter, data is shared directly via
SLHAio. This is a huge increase in performance, because no files need
to be written and doubles do not need to be converted to strings and
back again. It also means that once the tree is set up, parameter
changes are as fast as the access to a double pointer. At this stage,
no SLHAio functions are involved. While SLHAio is used in the SFitter
extraction tool, it will in the future become publicly available.

\subsection{Conclusions}

Supersymmetry provides a particularly convincing explanation for the
currently observed $3.4 \sigma$ discrepancy between the experimental
value for the anomalous magnetic moment of the muon and its 
Standard--Model prediction. If this signal proves to be correct, there exist
new particles in a mass range accessible to the LHC. When we attempt
to reconstruct weak--scale MSSM parameters from LHC observables, the
numerical value of ($g-2$) provides an attractive additional handle on
the MSSM parameters. Using the parameter point SPS1a as an example, we
have shown that ($g-2$) essentially determines the sign of $\mu$ in
the weak--scale MSSM, cutting the number of discrete solutions obtained
in the general MSSM in half. In addition, many of the parameter errors
are reduced, most notably the error on the extraction of the
notoriously difficult parameter $\tan\beta$.

\section*{Acknowledgements}
We would like to thank the organizers of the great Les Houches 2007
workshop {\it Physics at TeV-Scale Colliders}.



\clearpage
\newpage

\setcounter{figure}{0}
\setcounter{table}{0}
\setcounter{equation}{0}
\setcounter{footnote}{0}


\section{Towards combining cascade decays and flavor physics
\protect\footnote{S.~J\"ager, T.~Plehn and M.~Spannowsky}}



\subsection{Determining supersymmetric parameters}

For many years it has been known that the kinematics of cascade decays
is particularly well suited to extract the masses of the particles
involved~\cite{Bachacou:1999zb,Allanach:2000kt}. More recently, we
have seen how these masses can be used to determine the (running)
TeV--scale Lagrangian parameters, with the ultimate goal of evolving
these parameters to higher energy scales and extracting information on
the mechanism of supersymmetry breaking. SFitter~\cite{Lafaye:2007vs}
and Fittino~\cite{Bechtle:2005vt} are two computer tools specifically
designed to determine TeV--scale supersymmetric parameters with the
proper experimental and theory errors.

Most studies of cascade decays are based on the decays of (five)
light--flavor squarks and gluinos. Hence, for example in SPS1a we
would have convincing control over these squark masses and over the
neutral gaugino masses. A crucial and yet notoriously hard parameter
to extract is $\tan \beta$, both in the neutralino/chargino sector and
in the Higgs sector. In most analyses, we rely on the light MSSM Higgs
mass for information on $\tan \beta$~\cite{Frank:2006yh}.  This
extraction depends on a large number of supersymmetric parameters, on
the strict MSSM assumption, and on a reliable estimate of the theory
errors. Elsewhere in this volume we show how a measurement of
$(g-2)_\mu$ can be used to determine $\tan\beta$ from the lepton
sector. In a former Les Houches project it was shown that $\tan \beta$
can be extracted from the production rate of heavy MSSM Higgs bosons,
which for large values of $\tan \beta$ is typically proportional to
$m_b^2 \tan^2 \beta$.  Combining all errors entering the
cross--section measurements this study predicts a total error of
$12\%$ to $16 \%$ on the $bbA/bbH$ Yukawa coupling, which is
proportional to $\tan\beta$~\cite{Kinnunen:2004ji}.

Another strategy for an indirect $\tan\beta$ measurement are
flavor--physics observables. For example the rare decay rate for $B_s
\to \ell \ell$ is proportional to $\tan^6 \beta/m_A^4$.  This steep
behavior makes it a prime suspect to extract $\tan
\beta$~\cite{Buchalla:2008jp}. A major problem of such an extraction
is the correct estimate of the theory error on the observable. The
second problem is the dependence of the effective $bs\{h,H,A\}$
couplings on the stop and chargino masses appearing in the loop.  In
this article we briefly report on a pre-study done for SFitter, to
give a first estimate if these two problems will leave channels like
$B_s\to\ell \ell$ promising candidates to be included in the SFitter set
of observables.

Unfortunately, the usual parameter point SPS1a with $\tan\beta=10$ is
not well suited to study the determination of $\tan\beta$. Even if
there should be a sensitivity from a measurement of $B_s \to \ell \ell$,
it is unclear if we will observe any of the heavy Higgs bosons at the
LHC.  We therefore modify this parameter point in the direction of
SPS1b, simply choosing a range of larger $\tan\beta$ values. For
simplicity we assume that the set of cascade observables is not
altered by this change, including the sbottom mass determination.
While this assumption might be quantitatively naive, it will serve our
purpose of estimating the odds of combining different sources of
information on $\tan\beta$.

\subsection{Combining flavor and cascades}

FCNC processes involving down-type quarks in the Standard Model are
both highly suppressed and sensitive to the mass and couplings of the
top. This is because, beyond their loop suppression, the unitarity and
hierarchical structure of the CKM matrix and the hierarchy $m_{t,W}
\gg m_{u,c}$, entail strong GIM cancellations between the
light--flavor loop contributions. Turning around this argument, they
are sensitive probes of new--physics effects. In particular when the
supersymmetric flavor structure is (close to) minimally
flavor--violating \cite{Hall:1990ac,Buras:2000qz,D'Ambrosio:2002ex},
they provide a handle on stop and chargino masses.  Of special
interest are FCNC mediated by neutral Higgs exchange, which exhibit a
double enhancement: first, they involve the large bottom Yukawa
coupling $y_b \propto y_b^{\rm SM} \tan\beta$. Secondly, the
loop--induced contribution of $v_u \equiv \langle H_u \rangle \gg v_d
$ to the down-type fermion mass matrix destroys the alignment of the
mass matrix with the Yukawa couplings and renders the latter
flavor-nondiagonal~\cite{Hall:1993gn,Hempfling:1993kv,Carena:1994bv,Blazek:1995nv,Hamzaoui:1998nu,Carena:1999py,Babu:1999hn},
leading to an additional factor $\tan\beta$. For minimal flavor
violation, the corresponding FCNC Higgs couplings have the
form~\cite{Babu:1999hn,Isidori:2001fv,Buras:2002vd}
\begin{equation}  \label{eq:hvert}
  {\cal L}_{\rm eff} \supset \bar b_R s_L \phi_i
          \; x_i\, V_{tb}^* V_{ts} \;
          \frac{y_t^2 y_b}{\cos\beta} \;
   \frac{\epsilon_Y }
        {(1 + (\epsilon_0 + y_t^2 \epsilon_Y) \tan\beta)(1 + \epsilon_0 \tan\beta)},
\end{equation}
where $x_H=-\sin(\alpha-\beta)$ and $x_A=i$. The FCNC couplings
of $h$ are suppressed by a factor $x_h=\cos(\alpha-\beta)$, offsetting
the $\tan\beta$ enhancement. The parameters $\epsilon_0$ and
$\epsilon_Y$ parameterize loop-induced ``wrong-Higgs'' contributions
to the down-quark mass matrix. The sensitivity to MSSM parameters
becomes most transparent in the limit $v \ll M_{\rm SUSY}$, when
\begin{equation}
   \epsilon_Y = - \frac{1}{16\,\pi^2} \frac{A_t}{y_t \mu} 
     \Big[\frac{x \ln x}{(1-x)(x-y)} + \frac{y \ln y}{(1-y)(y-x)} \Big],
\end{equation}
with $x=m_{\tilde t_L}^2/\mu^2$ and $y=m_{\tilde t_R}^2/\mu^2$.  The
observables most sensitive to these couplings are $B_{s,d} \to \ell^+
\ell^-$, where the tree-level $H,A$ exchange contributes at the
amplitude level
as~\cite{Babu:1999hn,Isidori:2001fv,Buras:2002vd,Choudhury:1998ze,Bobeth:2001sq}
\begin{equation}
{\cal A}(B_q \to \ell^+ \ell^-)^{H,A} \propto \frac{y_b\, y_\ell}{\cos\beta\,
    m_{A}^2} \stackrel{\tan\beta \gg 1}{\propto} \frac{m_b m_\ell \tan^3\beta}{m_A^2} .
\end{equation}
Among the modes accessible at LHCb, ATLAS, and CMS, $B_s \to \mu^+
\mu^-$ has the largest branching fraction~\cite{Buchalla:2008jp},
which can be dominated by the neutral-Higgs contribution.

\begin{table}[b]
\begin{center}
\begin{small}
\begin{tabular}{|c|cc|cc|}
\hline
$\tan\beta$ & \multicolumn{2}{c|}{30} &
                \multicolumn{2}{c|}{40} \\
\hline
  & value & error & value & error \\
\hline\hline
$m_h$    & 112.6   & 4.0 & 112.6 &  4.0 \\
$m_t$ & 174.5 & 2.0 & 174.5 &  2.0\\
$m_{H^{\pm}}$ & 354.2 & 10.0 & 307.2 & 10.0 \\
$m_{\chi^0_1}$ & 98.4 & 4.8 & 98.7 & 4.8  \\
$m_{\chi^0_2}$ & 183.1 & 4.7 & 183.5 & 4.7  \\
$m_{\chi^0_3}$ & 353.0  & 5.1 & 350.7 & 5.1  \\
$m_{\chi^{\pm}_1}$  & 182.8  & 50.0 & 183.1 & 50.0   \\
$m_{\tilde{g}}$ & 607.7  & 8.0 & 607.6 &  8.0 \\
${\rm BR}(B_s\to \mu \mu)$ & 7.3 $\cdot 10^{-9}$   & $\sqrt{N} \otimes 15\%$ & 3.2$\cdot 10^{-8}$ & $\sqrt{N} \otimes 15\%$ \\
$m_{tb}$ & 404.2  & 5.0 & 404.2 & 5.0  \\
\hline
\end{tabular}
\end{small}
\end{center}
\caption{Set of toy measurements. The simple combined (absolute) errors 
  are SPS1a--inspired.
\label{tgbsfitter_data}
}
\end{table}

The usual cascade measurements we assume with this simple study
include $\tilde{g} \rightarrow \tilde{t}_1 t \rightarrow t \bar{t}
\chi^0_1$, $\tilde{g} \rightarrow \tilde{t}_1 t \rightarrow t \bar{b}
\chi^+_1$, $ \tilde{q}_L \rightarrow \chi^0_3 q \rightarrow \chi^0_2 Z
q \rightarrow \tilde{\tau}_R \tau Z q \rightarrow \chi^1_0 \tau \tau q
Z$, $\tilde{q_L} \rightarrow \chi^+_2 q \rightarrow \chi^+_1 Z q
\rightarrow q W \chi^0_1 Z q \rightarrow q q' q'' \chi^0_1 Z$, where
the $\chi^+_2$ cascade is strictly speaking not necessary for our
analysis. In our simple parameter points $M_0 = 150$, $M_{1/2} = 250$,
$A_0 = -100$, $\mu > 0$, and $\tan \beta =
(30,40)$~\cite{Djouadi:2006bz}, these cascades should be visible. In
our Minuit fit we include 10 observables listed in
Table~\ref{tgbsfitter_data}, including ${\rm BR}(B\to \mu \mu)$ and
the edge measurement $m_{tb}$~\cite{Hisano:2003qu}.  In addition, we
need to include some very basic information on the chargino--stop
sector. The measurement of the three neutralino masses gives us
information on the chargino mass parameters $M_2$ and $\mu$.  The
left--handed stop mass is linked to the left--handed sbottom mass via
SU(2). However, the right--handed stop mass as well as the
off--diagonal entry into the mass matrix are not determined by cascade
decays. The latter is dominated by the trilinear coupling $A_t$,
which enters for example the calculation of the light Higgs
mass~\cite{Hahn:2007fq}.  Extracting $A_t$, however, requires a
measurement of the dominant heavy Higgs mass parameter, which again
limits us to reasonably large values of $\tan\beta$.  We assume $m_A$
to be known either directly or via the charged Higgs mass, similarly
to Ref.~\cite{Kinnunen:2004ji}.

The modified SPS1a parameter point for two example values of
$\tan\beta$ is specified in Table~\ref{tgbsfitter_fit}, together with
the best-fit values and the errors from our Minuit fit.  For $B_s\to
\mu \mu$, LHCb alone expects about 100 events at the Standard--Model
rate after 5 years of running~\cite{Buchalla:2008jp}. For our study we
assume an integrated luminosity of $10~{\rm fb}^{-1} $ for the $B_s$
sample. The Higgs--mediated contributions always increases the
corresponding events number. As a consequence, the theory error, which
at present ranges around $30\%$, will soon dominate the total
uncertainty, unless it can be reduced.  It is mainly due to the decay
constant $f_{B_s}$, which can be calculated using numerical
lattice-QCD methods (see Ref.~\cite{DellaMorte:2007ny} for a recent
review). For our study we simply assume a reduction of the error on
$f_{B_s}$ to $7\%$, about half its present value. Such a reduction is
commonly believed to be realistic over the next five years.

For our study, we perform two sets of fits, one ignoring the theory
error and one combining it in quadrature with the statistical error.
A more refined treatment of the theory error is in progress and will
use the proper Rfit ansatz~\cite{Hocker:2001xe}, as implemented in
Sfitter~\cite{Lafaye:2007vs}. We indeed see that without taking into
account the theory error, $\tan\beta$ will be determined to $10\%$
from the combined toy data sample. Including a realistic theory error
increases this number to $15 \cdots 20\%$.  The errors on the
remaining parameters, shown in Tab.~\ref{tgbsfitter_fit} remains
largely unchanged. Slight shifts in either direction are at his stage
well within the uncertainly on the determination of the error bars,
and the central fit value for the top--mass parameter seems to be
consistently lower than the input value (by roughly half a standard
deviation). Comparing our error estimates on the mass parameters for
example with the SFitter analysis~\cite{Lafaye:2007vs}, we expect the
situation to improve for all model parameters once we include a more
extensive set of measurements and properly correlated errors.

\begin{table}[t]
\begin{center}
\begin{small}
\begin{tabular}{|c||ccc|cc||ccc|cc|}
\hline
& \multicolumn{3}{c|}{no theory error} & 
  \multicolumn{2}{c||}{$\Delta {\rm BR}/{\rm BR}=15\%$} & 
  \multicolumn{3}{c|}{no theory error} &  
  \multicolumn{2}{c|}{$\Delta {\rm BR}/{\rm BR}=15\%$} \\
\hline
                         & true  &  best   & error  & best   & error   & true     & best     & error  & best     & error \\
\hline\hline
$\tan \beta$  & 30      & 29.5    & 3.4    & 29.5    & 6.5       & 40        & 39.2     & 4.4   & 39.2   & 5.8    \\
$M_A$          & 344.3 & 344.4 & 33.8  & 344.3  & 31.2    & 295.5  & 304.4   & 35.4   & 295.6 & 33.9  \\
$M_1$          & 101.7 & 100.9 & 16.3  & 100.9  & 16.4    & 101.9  & 101.0   & 16.3   & 101.0 & 16.3 \\
$M_2$          & 192.0 & 200.3 & 18.9  & 200.3  & 18.8    & 192.3   & 200.3   & 20.0   & 200.7 & 18.9 \\
$\mu$           & 345.8 & 325.6 & 20.6  & 325.6  & 20.6    & 343.5   & 322.9   & 20.7   & 323.3 & 20.6  \\
$M_3$          & 586.4 & 575.8 & 28.8  & 575.8  & 28.7    & 586.9   & 576.0   & 28.7     & 575.8 & 29.0 \\
$M_{\tilde{Q}_L}$ &  494.4  & 494.4 &  78.1   &  494.3  &  78.0   & 487.1    &  487.6 & 79.4 & 487.5 & 78.9 \\
$M_{\tilde{t}_R}$ & 430.0   & 400.4  & 79.5   & 399.8    & 79.5    & 431.5    & 399.2  & 86.7 & 399.1 & 82.6 \\
\hline
%
\end{tabular}
\end{small}
\end{center}
\caption{The modified SPS1a point and the errors from the parameter fit 
  for the two values of $\tan\beta=30,40$.  Dimensionful quantities
  are in units of GeV. For the measurement of ${\rm BR}(B_s \to \ell
  \ell)$ we assume either no theory error or an expected improvement
  to $15\%$, as compared to the current status.
\label{tgbsfitter_fit}
}
\end{table}

\subsection{Caveats}

Note that the detailed results of this study should not be used at
face value.  First of all, it is not clear if the stop--mass
measurement can be achieved in the SPS1a parameters point or with an
increased value of $\tan\beta$. Secondly, for the charged Higgs mass
we only use a toy measurement. And last but not least, we do not (yet)
take into account error correlations at this stage. None of these
omissions we expect to move the result of a complete analysis into a
definite direction, but there is certainly room for the final error
bars to move.

This study shows, however, that the parameter $\tan\beta$ can indeed
be extracted from a combined cascade and flavor data sample.  Already
at this stage we can conclude that the combination of cascade--decay
and flavor observables will crucially depend on the quality of the
theory predictions in the flavor sector.  In particular, an improved
understanding of non-perturbative QCD effects in $B_s \to \ell \ell$
decays is needed to meaningfully exploit this highly promising link.
From the high--$p_T$ point of view we also generally see that to
measure $\tan\beta$ we need to improve the analysis of the
stop--chargino sector in the classic decay--kinematics analyses.

The problem with the measurement of $\tan\beta$ from the light MSSM
Higgs mass or from $(g-2)_\mu$ or from rare $B$ decays is that each of
these indirect measurement rely on assumptions about the flavor and
Higgs sectors.  Moreover, these different measurements point to
different parameters, not only from a renormalization point of view,
but also because of large QCD effects distinguishing between them. The
more direct extraction from cross sections times branching ratios of
heavy Higgs bosons is at the same time plagued by large theory
uncertainties, due to QCD corrections and uncertainties in the
bottom--parton picture~\cite{Plehn:2002vy,Boos:2003yi}. If we should
indeed find evidence for the MSSM in the LHC era we obviously expect a
serious jigsaw approach to the $\tan\beta$ determination.

\subsection*{Acknowledgments}

We are grateful to Mihoko Nojiri and Giacomo Polessello for their
ongoing encouragement.  And of course we thank the organizers of the
great Les Houches 2007 workshop {\it Physics at TeV-Scale Colliders}!



\clearpage
\newpage

\setcounter{figure}{0}
\setcounter{table}{0}
\setcounter{equation}{0}
\setcounter{footnote}{0}

\section{BBN lithium problem consequences at LHC
\protect\footnote{P.~Zalewski}}

\maketitle

\subsection{Lithium problem}

Recent measurements of the fluctuations
of the microwave background radiation 
allowed for determination of the contribution of baryons to critical density
$\Omega_bh^2\approx0.0224$ \cite{Spergel:2006hy} 
which is the only free parameter
of the standard BBN. SBBN previsions are in agreement with experimental 
estimates of abundances for deuterium and and helium~4 but 
have problem in explaining lithium 7 proliferation, which is about 3 times 
to high.

There is no agreement whether stated above discrepancy
is a~real problem because apparent primordial $^7$Li abundance 
is derived only from the observation of so called Spite 
plateau~\cite{Spite:1982} 
in low metallicity POP\,II stars.
It is believed, that gas present in atmospheres of these very old starts 
have not changed composition since the BBN era. $^7$Li is, however, fragile so 
in principle its depletion could be explained by some stellar evolution model 
but no fully satisfactory model have been 
proposed~\cite{Asplund:2005yt,Jedamzik:2007cp}.   
Recently preliminary observation of similar plateau was done also for 
$^6$Li~\cite{Asplund:2005yt} 
which was produced during 
SBBN but below detectable level.
If confirmed, the plateau could suggest pre-galactic $^6$Li origin 
corresponding to the primordial abundance at least an order of magnitude
higher than SBBN predictions.  Moreover, since $^6$Li 
is by far more fragile than $^7$Li, any model 
of its destruction will aggravate $^6$Li problem.

Although above sketched lithium problems could have standard explanation
it is very interesting to note beyond standard model solutions
that alters BBN. All of them postulate long lived massive particles with
lifetimes around 1ks or more. Their decays allow for late $^6$Li production
or late $^7$Li destruction or both without altering abundances 
of other isotopes. If these particles could be negatively 
charged than also bound states must be taken into account.
In the present letter there is no place for review all proposed 
solutions. We would concentrate on two, which are in agreement
with cosmological constraints, and could have very 
interesting consequences for the LHC phenomenology. In both, an existence of 
long-lived stau is postulated. In the 
first~\cite{Jedamzik:2005dh} 
stau lifetime is of the order of
1ks. The solution for both lithium problems is found within CMSSM for
stau mass of the order of 1\,TeV which is, unfortunately, out of reach 
at LHC.
However, solution of only $^6$Li problem is possible for stau mass around
few hundreds GeV. The second solution of both lithium problems is
possible for very long stau lifetime of 1\,Ms and stau mass of about 
300\,GeV~\cite{Jedamzik:2007cp}.   
  
\subsection{Possible discovery at LHC}

Both 
ATLAS and CMS
experiments have developed strategies to look for 
charged massive
particles (CHAMP) if they decay lengths exceed detector sizes. 

Methods tested on full detector simulation
are based on TOF measurements in the muon systems 
(CMS drift tubes, ATLAS drift tubes and RPCs) or
specific ionization measurement in the tracker (CMS). 
Since in both experiments at least two independent measurements are
performed it is possible to evaluate misidentification probabilities directly
from data. If simulated performances will be confirmed almost background
free selections with efficiency of the order of 10\% could be designed.
After collecting  10/fb of data this allows for the discovery if the cross
section exceeds 10\,pb which corresponds to stau mass around 
300\,GeV~\cite{Zalewski:2007up}.

This provoke natural question about lifetime
measurement of such CHAMP. Any estimate of it could be of crucial importance. 
The problem is, however, that interesting range of lifetimes is above
100\,s. It is obvious that decays in flight are not only by far inefficient
but also insensitive. 

However, the first proposition in this direction
was made already a~decade ago~\cite{Kazana:1999}. The point was about using
CMS muon system as a~late electromagnetic calorimeter (so called $\mu$CAL)
in the following way.
If significant fraction of energy release is electromagnetic and if the
decay happens inside iron yoke at a~distance to the next muon station
equivalent to few radiation lengths, then developing electromagnetic cascade
causes large accumulation of hits in the station. By design the cascade
ends in the next yoke section. So the signal is large accumulation
of hits in one muon station. Despite the fact, that this proposal 
was made in the context of detection of decaying in flight neutralinos,
it could be used also for decaying staus, but to measure long lifetimes
these staus must stop inside the yoke. Unfortunately only small fraction
of staus will do that. Larger fraction will be stopped in the concrete and
rocks of the cavern. 

There were proposals to drill out the cavern walls to
recuperate the part with stopped CHAMP or to install water tanks for CHAMPs
capture, but if CHAMPs are staus than there is also much simpler solution.
17\% of stau decays produce muons. These muons could be detected not only
for staus stopped inside detector but also for staus stooped in the
cavern walls if muon is released backward. The first possibility is rather
hopeless for the ATLAS detector because of air muon system
but the second could be more efficient for ATLAS than for CMS because ATLAS 
is bigger and is closer to the cavern wall and so, it will have 
larger angular size when seen from the CHAMP decay point.  

Although it is straight forward to estimate for a~given model a~probability 
for a muon from the stopped stau decay to cross again the detector full 
simulation is needed to know detector answer to such muon. In principle
these muons are very similar to cosmic muons. The only difference is
a~homogeneous distribution of incoming directions in contrast to top
bottom directions expected for cosmics. 

Without full detector simulation (which already started, 
but no official results
of it have been released) it is difficult to know if the sensitivity
could be sufficient for measuring long lifetimes of stau. However, no
such measurement is possible without cosmic trigger. Another problem
is tracking of low momentum staus, which will go outside 25\,ns window.
Although triggering on such particles is not possible its offline
recovery is not hopeless because drift tubes, thanks to their operation mode,
remember data form many bunch crossings.

It is important to note, that cosmic trigger, even if possibly harmful 
for normal LHC operation, should be studied in detail, to be switched on
 if long lived CHAMPs will discovered at LHC.

\subsection{Conclusions}

It was underlined that lithium BBN problem could be solved by
long lived CHAMPs, next, that such particles, if not to heavy, could be
discovered at the LHC and that the most efficient way of
measuring their very long lifetime is to design cosmic trigger for
LHC detectors.



\clearpage
\newpage

\setcounter{figure}{0}
\setcounter{table}{0}
\setcounter{equation}{0}
\setcounter{footnote}{0}



\section{
Precision measurements of the stop mass at the ILC
\protect\footnote{A.~Sopczak, A.~Freitas, C.~Milst\'ene and M.~Schmitt}}


\subsection{Introduction}
Supersymmetric particles are likely to be produced
and observed in high-energy proton-proton collisions at the LHC.  
However, it will be difficult to confirm their identity as
superpartners of the known Standard Model particles and to
measure their properties precisely. For this, one needs 
experiments at a linear $\rm e^+e^-$ collider such as the proposed 
ILC at $\sqrt{s}=500$~GeV.
The importance of scalar top studies has been emphazised in the
'2005 Les Houches' proceedings~\cite{leshouches05_stop}.
This work extends these studies.

An experiment at the ILC will be able to make many precise 
measurements from which particle properties, and ultimately,
the outlines of a particle physics model may be inferred.
Due to the high statistical precision expected at the ILC, the optimization 
of the systematic errors is of particular importance.
We have studied one specific example, the extraction
of the mass of a scalar top quark from cross-section
measurements near threshold.  We have devised a method which
reduces most systematic uncertainties and leads to a potentially
very accurate measurement of the stop quark mass.  This method
is general and could be applied to other particles
that are pair-produced in an $\rm e^+e^-$ collider.

The method relies on the comparison of production rates at two different
center-of-mass energies, and knowledge of how the cross-section
varies as a function of $\sqrt{s}$ and the mass of the particle.

We have chosen the case of a light scalar top with
a mass not much higher than the mass of the lightest neutralino
since production of this particle was already extensively 
studied in an ILC context.
It was concluded that a conventional approach to the measurement of the
stop quark mass culminated in an uncertainty of about 1~GeV~\cite{Carena:2005gc}.
The new method improves substantially on this result.
The presented results are preliminary and being finalized~\cite{Freitas:2007zr}.

For this analysis, we have performed realistic simulations of the 
signal and backgrounds, and used two techniques to separate the 
signal from the background. The first technique is based on conventional 
sequential cuts, while the second employs an Iterative 
Discriminant Analysis (IDA). Furthermore, the hadronization followed by
fragmentation of the 
stop has been included and we have carefully studied the systematic 
uncertainties arising from this and other sources.

There are theoretical motivations for studying a light 
stop quark with a small mass difference.  Specifically, we
evoke a scenario within the Minimal Supersymmetric
extension of the Standard Model (MSSM) which is able to explain the
dark matter density of the universe as well as the baryon
asymmetry through the mechanism of electroweak baryogenesis~\cite{Carena:2005gc}.

A small mass difference between the stop and the lightest
neutralino can help to bring the dark matter relic density into the observed
region~\cite{Spergel:2006hy,Tegmark:2003uf}
 due to co-annihilation between the stop and the neutralino. For this
mechanism to be effective, the typical mass difference is rather small,
$\Delta m = m_{\rm \tilde{t}_1} - m_{\tilde{\chi}^0_1}
\,\raisebox{-.1ex}{$_{\textstyle <}\atop^{\textstyle\sim}$}\, 30$ GeV~\cite{Balazs:2004bu}.
The dominant decay mode of the stop is 
$\mbox{$\rm \tilde{t}_1$} \to c\,\tilde{\chi}^0_1$, resulting 
in a final state with two soft charm jets and missing energy.

\begin{wraptable}{r}{8cm}
\vspace*{-5mm}
\centering
\footnotesize
\begin{tabular}{lcc}\hline
Method & $\Delta m_{\rm \tilde{t}_1}$ (GeV) & luminosity \\
\hline
Polarization & 0.57 & $2 \times 500 \mathrm{\, fb}^{-1}$ \\
Threshold scan & 1.2 & $300 \mathrm{\, fb}^{-1}$ \\
End point & 1.7 & $500\mathrm{\, fb}^{-1}$\\
Minimum mass & 1.5 & $500 \mathrm{\, fb}^{-1}$ \\ \hline
\end{tabular}
\caption{Comparison of precision for scalar top mass determination
         for the SPS-5 benchmark ($m_{\rm \tilde{t}_1}=220.7$~GeV).} 
\label{tab:stop_previous}
\end{wraptable}

Previous methods to determine the scalar top quark mass 
were discussed for the SPS-5 benchmark ($m_{\rm \tilde{t}_1}=220.7$~GeV)~\cite{Sopczak:2006xj}
and results are summarized in Table~\ref{tab:stop_previous}.
For the cosmology motivated benchmark with $m_{\rm \tilde{t}_1}=122.5$~GeV
and $m_{\tilde{\chi}^0_1} = 107.2$~GeV, 
an experimental precision of $\Delta m_{\rm \tilde{t}_1} = \pm 1.0$~GeV was obtained~\cite{Carena:2005gc},
and about $\pm1.2$~GeV including theoretical uncertainties.
The following study investigates the same signal scenario and
it is based on the same background reactions and event preselection.

\subsection{Mass determination method}

This method proposes to derive the stop mass from measurements at two center-of-mass
energies, one measuring the stop production cross-section near the threshold (th), and 
the other measuring it at a center-of-mass energy where the cross-section has approximately a 
peak (pk).
Using both measurements leads to a cancellation of systematic uncertainties in the mass
determination. A parameter $Y$ is defined as 
\begin{equation}
Y= \frac{N_{\rm th}-B_{\rm th}}{N_{\rm pk}-B_{\rm pk}}
=\frac{\sigma_{\rm th}}{\sigma_{\rm pk}}
\cdot\frac{\epsilon_{\rm th}}{\epsilon_{\rm pk}}
\cdot\frac{{\cal L}_{\rm th}}{{\cal L}_{\rm pk}},
\end{equation}
where $N$ is the total number of expected events after event selection and $B$ the number
of corresponding background events, $\sigma$ is the stop production cross-section,
$\epsilon$ the selection efficiency, and $\cal L$ the lumi\-no\-sity. The center-of-mass
energies 260 and 500~GeV have been chosen. Near the threshold, the production cross-section
is very sensitive to the stop mass.

\begin{wrapfigure}{r}{8cm}
\vspace*{-6mm}
\includegraphics[width=0.5\textwidth]{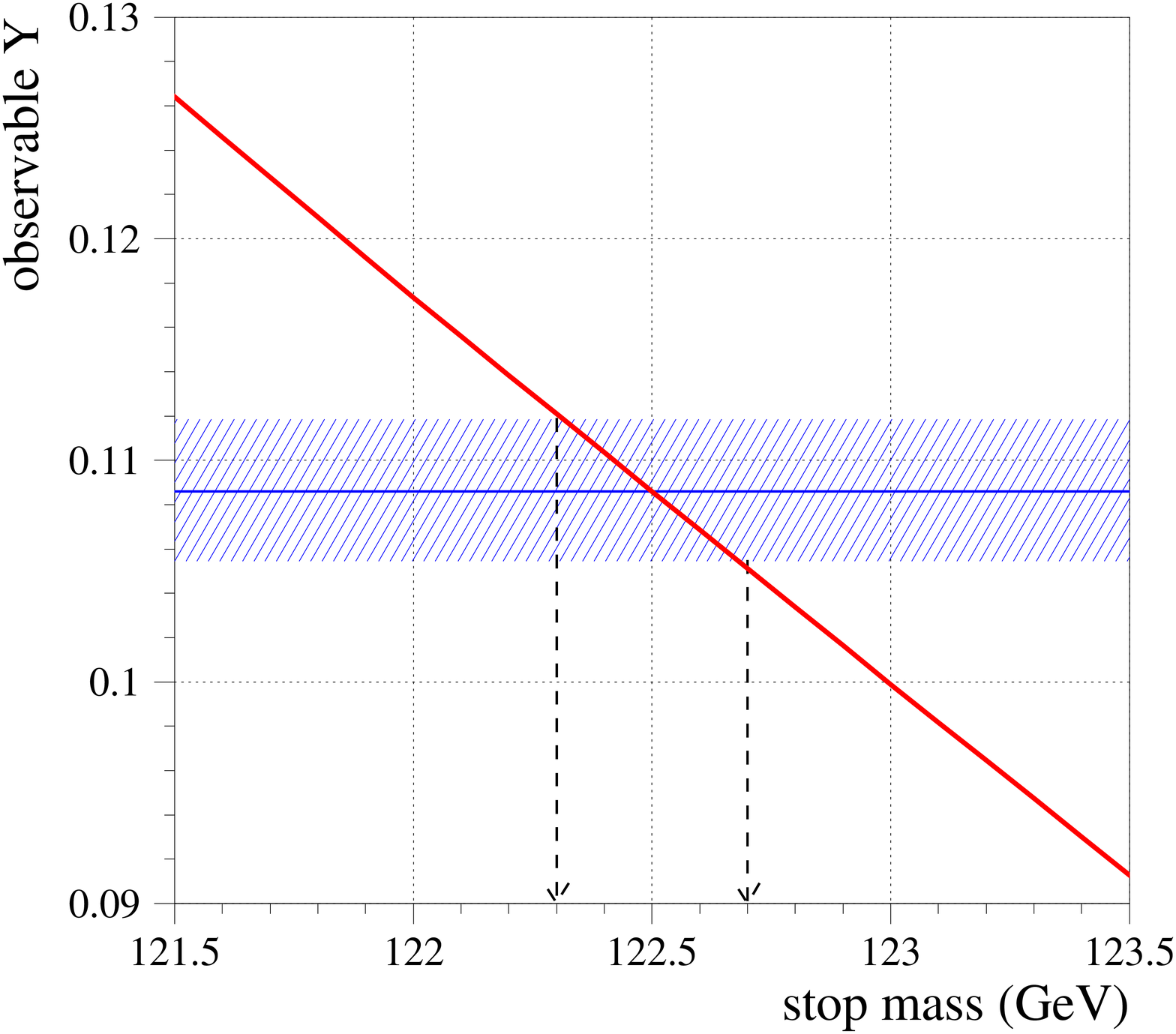}
\vspace*{-10mm}
\caption{Example of mass uncertainty derivation from the 
uncertainty of the observable $Y$.}
\vspace*{-8mm}
\label{fig:stop_example}  
\end{wrapfigure}

In this study we assume that the ILC will operate primarily 
at $\sqrt{s}=500$~GeV with a total luminosity of ${\cal L} = 500$~fb$^{-1}$, 
and a small luminosity of 
${\cal L} = 50$~fb$^{-1}$ will be collected at $\sqrt{s}=260$~GeV. 
Table~\ref{tab:stop_xsec} summarizes the expected production cross-sections.
The detector response was modeled with the SIMDET package~\cite{Pohl:2002vk} 
including the LCFI~\cite{LCFI} vertex detector concept.

The relation of the observable $Y$ and the stop mass is used to determine the stop mass
with precision. For example a variation of $Y$ by 3\% in a realistic scenario would
lead to an uncertainty of the stop mass $\Delta m_{\rm \tilde{t}_1} = 0.2$~GeV as illustrated in 
Fig.~\ref{fig:stop_example}. 

\subsection{Sequential-cut analysis}

In order to cancel the systematic uncertainties 
to a large extent with the described method,
the same sequential cuts are applied for the $\sqrt{s}=260$ and 500~GeV analyses. 
Details of the event selection are given in Table~\ref{tab:stop_cuts} and the results are
given in Table~\ref{tab:stop_nev}.

\begin{table}[htbp]
\begin{minipage}{0.3\textwidth}
\caption{Cross-sections for the stop signal~\cite{Eberl:1996wa} 
and Standard Model background processes
for $\sqrt{s} = 260$~GeV and $\sqrt{s} = 500 $~GeV and different
polarization
combinations. The signal is given for a right-chiral stop of $m_{\tilde{t}} =
122.5$~GeV. 
Negative polarization values refer to
left-handed polarization and positive values to right-handed polarization.
\label{tab:stop_xsec}}
\end{minipage}\hfill
\begin{minipage}{0.65\textwidth}
\footnotesize
\begin{tabular}{lrrrrrr}
\hline
Process &  \multicolumn{3}{c}{$\sigma$ (pb) at $\sqrt{s} = 260 $~GeV} 
        &  \multicolumn{3}{c}{$\sigma$ (pb) at $\sqrt{s} = 500 $~GeV} \\
\hline
$P(e^-) / P(e^+)$ \hspace*{-5mm} & 0/0 & -.8/+.6 & +.8/-.6
                  & 0/0 & -.8/+.6 & +.8/-.6 \\
\hline
$\tilde{t}_1 \tilde{t}_1^*$ & 0.032 & 0.017 & 0.077 & 0.118 & 0.072 & 0.276 \\
\hline
$W^+W^-$ & 16.9\phantom{0} & 48.6\phantom{0} & 1.77 
         & 8.6\phantom{0} & 24.5\phantom{0} & 0.77 \\
$ZZ$    & 1.12 & 2.28 & 0.99 & 0.49 & 1.02 & 0.44 \\
$W e\nu$ & 1.73 & 3.04 & 0.50 & 6.14 & 10.6\phantom{0} & 1.82 \\
$e e Z$  & 5.1\phantom{0} & 6.0\phantom{0} 
         & 4.3\phantom{0} & 7.5\phantom{0} & 8.5\phantom{0} & 6.2\phantom{0} \\
$q \bar{q}$, $q \neq t$ & 49.5\phantom{0} & 92.7\phantom{0} & 53.1\phantom{0} 
                        & 13.1\phantom{0} & 25.4\phantom{0} & 14.9\phantom{0} \\
$t \bar{t}$ & 0.0\phantom{0} & 0.0\phantom{0} & 0.0\phantom{0} & 0.55 & 1.13 &
0.50 \\
2-photon & 786\phantom{.00}&&&  \hspace*{-5mm} 936\phantom{.00}&& \\[-1ex]
$p_T > 5$ GeV\hspace*{-5mm} &&&&&& \\
\hline
\end{tabular}
\end{minipage}
\end{table}

\begin{table}[htbp]
\centering
\begin{minipage}{0.3\textwidth}
\caption{Selection cuts for $\sqrt{s}=260$ and 500~GeV.
Also listed are the selection efficiencies optimized for right-chiral stop quarks.
}
\label{tab:stop_cuts}
\end{minipage}\hfill
\begin{minipage}{0.65\textwidth}
\footnotesize
\begin{tabular}{lcc}
\hline
Variable &  $\sqrt{s} = 260 $~GeV 
         &  $\sqrt{s} = 500 $~GeV \\
\hline
number of charged tracks      
   &  $5 \le N_{\mathrm{tracks}} \le 25$ 
   &  $5 \le N_{\mathrm{tracks}} \le 20$ 
\\
visible energy $E_{\mathrm{vis}}$
   &  $0.1 < E_{\mathrm{vis}}/\sqrt{s} < 0.3$
   &  $0.1 < E_{\mathrm{vis}}/\sqrt{s} < 0.3$
\\
event long. momentum
   &  $ |p_L / p_{\mathrm{tot}}| < 0.85$
   &  $ |p_L / p_{\mathrm{tot}}| < 0.85$
\\
event transv. momentum $p_T$
   &  $15 < p_T < 45$~GeV
   &  $22 < p_T < 50$~GeV
\\
thrust $T$
   &  $0.77 < T < 0.97$
   &  $0.55 < T < 0.90$
\\
Number of jets $N_{\mathrm{jets}}$ 
   &  $N_{\mathrm{jets}} \ge 2$ 
   &  $N_{\mathrm{jets}} \ge 2$
\\
extra-jet veto
   &  $E_{\mathrm{jet}} < 25$~GeV
   &  $E_{\mathrm{jet}} < 25$~GeV
\\
charm tagging likelihood $P_{\mathrm{c}}$ 
   & $P_{\mathrm{c}}> 0.6$ 
   & $P_{\mathrm{c}}> 0.6$
\\
di-jet invariant mass $m_{\mathrm{jj}}$ 
   &\hspace*{-6mm} $m_{\mathrm{jj}}^2 < 5500~\mathrm{GeV}^2$ or 
   &\hspace*{-3mm} $m_{\mathrm{jj}}^2 < 5500~\mathrm{GeV}^2$ or 
\\

   & $m_{\mathrm{jj}}^2 > 8000~\mathrm{GeV}^2$
   & $m_{\mathrm{jj}}^2 > 10000~\mathrm{GeV}^2$
\\
\hline
signal efficiency & 0.340 & 0.212 \\
\hline
\end{tabular}
\end{minipage}
\end{table}

\clearpage

\begin{table}[htbp]
\vspace*{2mm}
\centering
\begin{minipage}{0.3\textwidth}
\caption{\label{tab:stop_nev} 
Numbers of generated events, and expected events
for the sequential-cut analysis at $\sqrt{s} = 260$ and 500~GeV
for total luminosities of 50~fb$^{-1}$ and 200~fb$^{-1}$
with unpolarized and polarized beams.
}
\end{minipage}\hfill
\begin{minipage}{0.65\textwidth}
\footnotesize
\centering
\begin{tabular}{lrrrrrr}
\hline
 &  \multicolumn{3}{c}{$\sqrt{s} = 260$~GeV} 
 &  \multicolumn{3}{c}{$\sqrt{s} = 500$~GeV}\\
\hline
                  & generated & \multicolumn{2}{c}{${\cal{L}} = 50~\mathrm{fb}^{-1}$}
                  & generated & \multicolumn{2}{c}{${\cal{L}} = 200~\mathrm{fb}^{-1}$} \\
\hline
$P(e^-) / P(e^+)$ &   & 0/0 & {.8/-.6}
                  &   & 0/0 & {.8/-.6} \\
\hline
$\tilde{t}_1 \tilde{t}_1^*$ & 50,000 & 544 & 1309 
                            & 50,000 & 5170 & 12093  \\
\hline
$W^+W^-$ & 180,000 &   $38$ &   $4$ &   $210,000$ &   $16$ &  $2$  \\
$ZZ$     &  30,000 &    $8$ &   $7$ &    $30,000$ &    $36$ &   $32$ \\
$W e\nu$ & 210,000 &  $208$ &  $60$ &   $210,000$ & $7416$ & $2198$ \\
$e e Z$  & 210,000 &    $2$ &   $2$ &   $210,000$ &   $<7$ &  $<6$ \\ 
$q \bar{q}$, $q \neq t$ 
         & 350,000 &   $42$ &  $45$ &   $350,000$ &    $15$ &   $17$ \\
$t \bar{t}$ & ---  &    $0$ &   $0$ &   $180,000$ &    $7$ &   $7$ \\
2-photon & $1.6\times 10^6$ 
         & $53$ & $53$ & $8.5\times 10^6$ & $12$ & $12$  \\
\hline
\hspace*{-2mm} total background \hspace*{-2mm}& --- & $351$ & $171$ & --- & $7509$ & $2274$ \\
\hline
\end{tabular}
\end{minipage}
\end{table}

\begin{wraptable}{r}{9.8cm}
\vspace*{-5mm}
\centering
\footnotesize
\begin{tabular}{lrrrrrr}
\hline
 &  \multicolumn{3}{c}{$\sqrt{s} = 260$~GeV} 
 &  \multicolumn{3}{c}{$\sqrt{s} = 500$~GeV}\\
\hline
                  & generated & \multicolumn{2}{c}{${\cal{L}} = 50~\mathrm{fb}^{-1}$}
                  & generated & \multicolumn{2}{c}{${\cal{L}} = 200~\mathrm{fb}^{-1}$} \\
\hline
$P(e^-) / P(e^+)$\hspace*{-6mm} &   & 0/0 & {.8/-.6}
                  &   & 0/0 & {.8/-.6} \\
\hline
$\tilde{t}_1 \tilde{t}_1^*$ & 50,000 &   $619$  & $1489$
                            & 50,000 & $9815$  & $22958$  \\
\hline
$W^+W^-$ & 180,000 &   $11$ &   $1$ &   $210,000$ &   $<8$ &   $<1$ \\
$ZZ$     &  30,000 &   $<2$ &  $<2$ &    $30,000$ &    $20$ &   $18$ \\
$W e\nu$ & 210,000 &   $68$ &  $20$ &   $210,000$ &  $1719$ & $510$ \\
$e e Z$  & 210,000 &    $3$ &   $2$ &   $210,000$ &   $<7$ &  $<6$ \\ 
$q \bar{q}$, $q \neq t$ 
         & 350,000 &   $16$ &  $17$ &   $350,000$ &    $18$ &   $21$ \\
$t \bar{t}$ & ---  &    $0$ &   $0$ &   $180,000$ &     $1$ &    $1$ \\
2-photon & $1.6\times 10^6$ 
         & $27$ & $27$ & $8.5\times 10^6$ & $294$ & $294$  \\
\hline
\hspace*{-2mm} total background \hspace*{-6mm}& --- & $125$ & $67$ & --- & $2067$ & $851$ \\
\hline
\end{tabular}
\caption{\label{tab:stop_ida2}
Numbers of generated events, and expected events
for the IDA at $\sqrt{s} = 260$ and 500~GeV
for total luminosities of 50~fb$^{-1}$ and 200~fb$^{-1}$
with unpolarized and polarized beams.
}
\end{wraptable}

\subsection{Iterative discriminant analysis}
\begin{figure}[thp]
\vspace*{3mm}
\begin{center}
\includegraphics[width=0.49\textwidth]{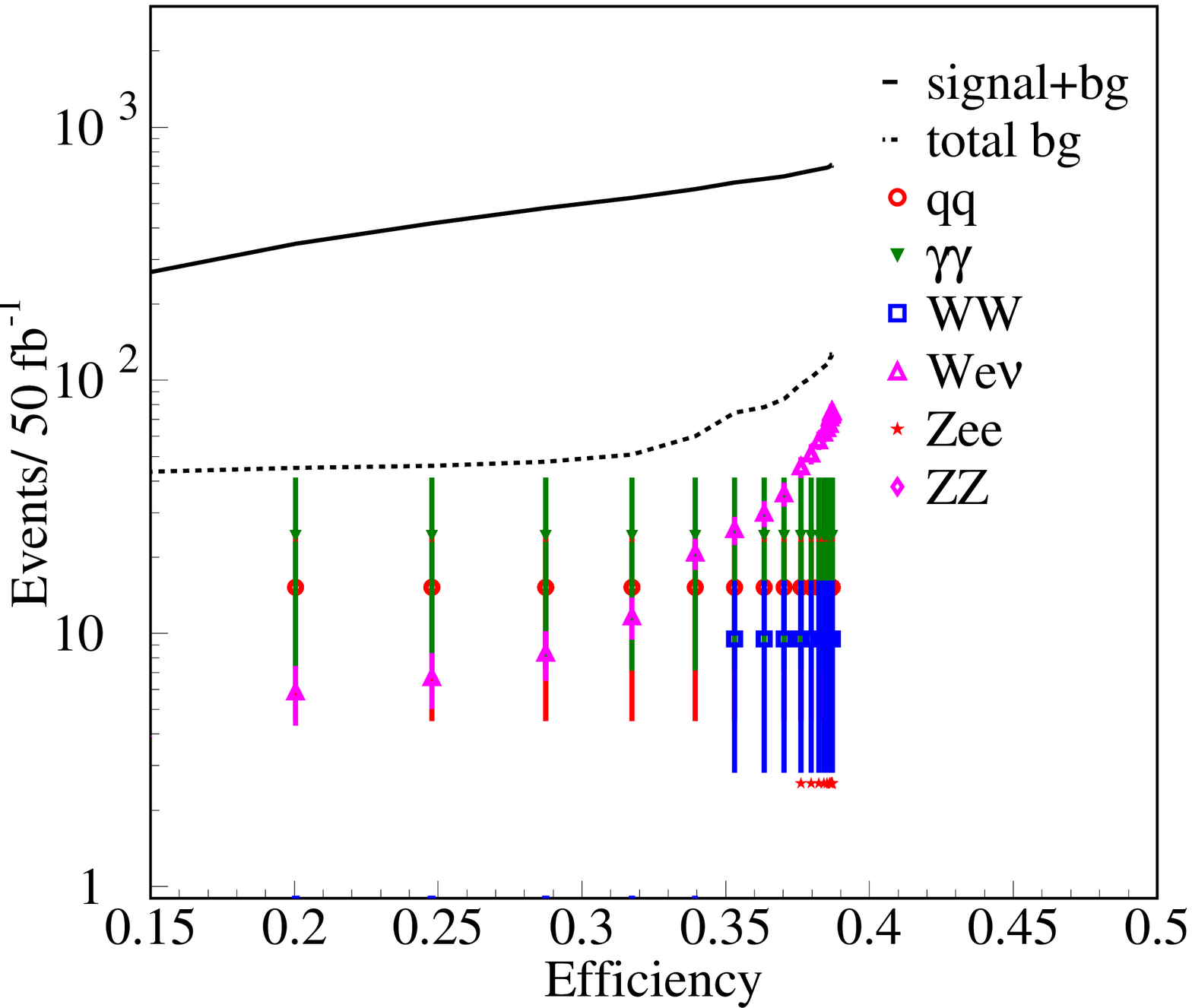} \hfill
\includegraphics[width=0.49\textwidth]{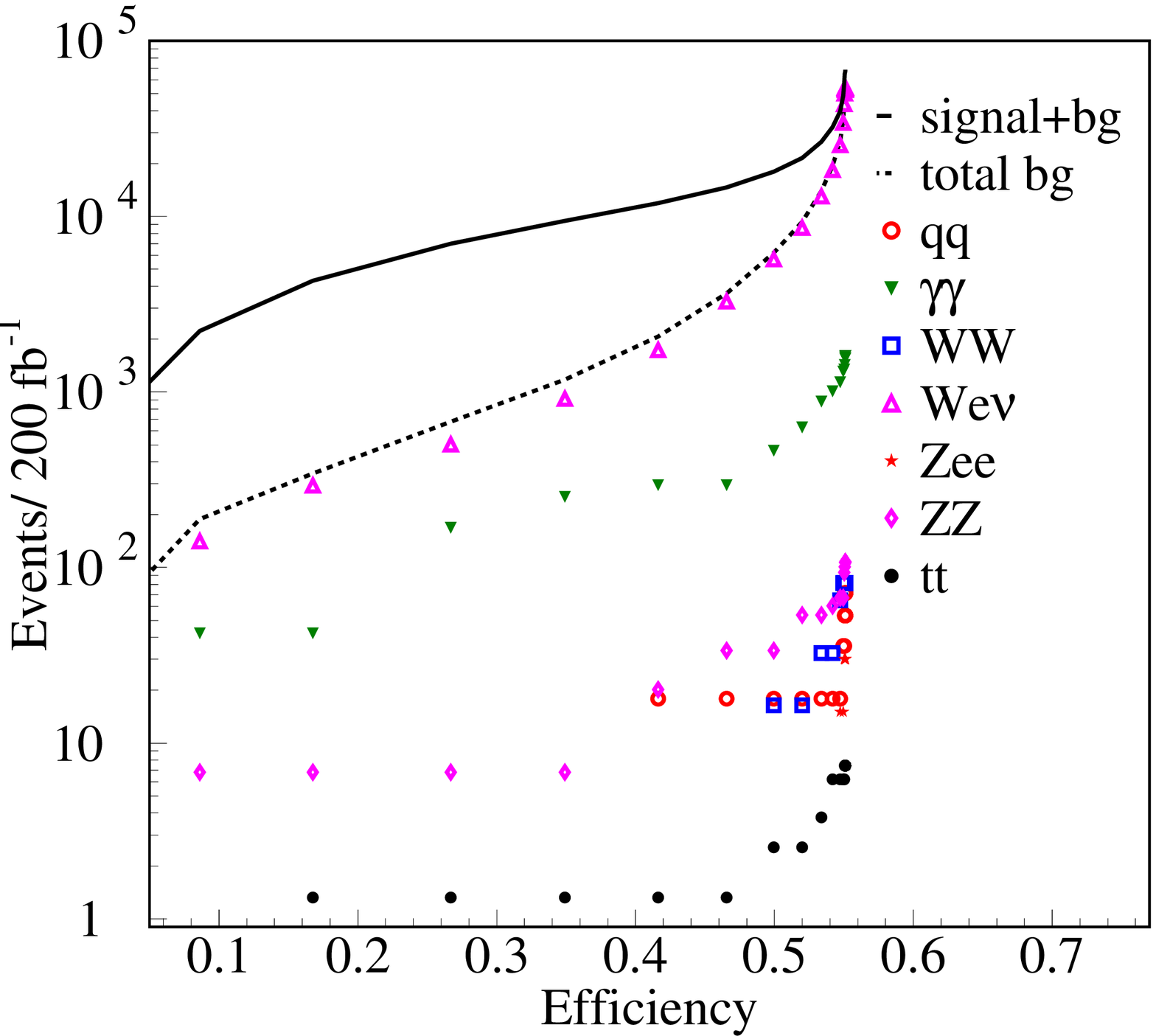}
\end{center}
\vspace*{-6mm}
\caption{IDA: Expected background events as a function of the signal efficiency. 
         Left: ${\cal L} = 50$~fb$^{-1}$ at $\sqrt{s} = 260$ GeV.
         Right: ${\cal L} = 200$~fb$^{-1}$ at $\sqrt{s} = 500$ GeV.
}
\label{fig:stop_perf260}
\end{figure}

The Iterative Discriminant Analysis (IDA)
\cite{Malmgren:1998xr} is applied to increase
the discriminant power between signal and background
compared to the sequential-cut-based analysis, and thus reduce the 
statistical uncertainty in the stop mass measurement. Figure~\ref{fig:stop_perf260}
gives the results of expected number of background events
as a function of the signal efficiency. The chosen working points have efficiencies
of 38.7\% and 41.6\% for the $\sqrt{s}=260$ and 500~GeV analyses, respectively.
Table~\ref{tab:stop_ida2} lists the corresponding expected background events.

\clearpage
\subsection{Systematic uncertainties}
Both the sequential-cut-based analysis and the IDA method lead to a small 
statistical uncertainty resulting in \mbox{$\Delta m_{\rm \tilde{t}_1} < 0.2$}~GeV and thus
systematic uncertainties are particularly important to evaluate.
Three classes of systematic uncertainties are distinguished:
\begin{itemize}
\item instrumental uncertainties related to the detector and accelerator:
      detector calibration (energy scale),
      track reconstruction efficiency,
      charm-quark tagging efficiency, and
      integrated luminosity.
\item Monte Carlo modeling uncertainty of the signal: charm and stop fragmentation effects.
      The Peterson fragmentation function~\cite{Peterson:1982ak} was used with
      $\epsilon_{\rm c}=-0.031\pm 0.011$ (OPAL)~\cite{Alexander:1995bk}. For 
      $\epsilon_{\rm b}=-0.0041\pm 0.0004$ (OPAL)~\cite{Abbiendi:2002vt} and 
      $\epsilon_{\rm b}=-0.0031\pm 0.0006$ (ALEPH)~\cite{Heister:2001jg} 
      an average uncertainty of 15\% was taken, 
      and a factor 2 improvement at the ILC has been assumed, 
      leading to $\Delta\epsilon_{\rm \tilde{t}_1}=0.6\times 10^{-6}$ where 
      $\epsilon_{\rm \tilde{t}_1}=
      \epsilon_b(m_{\rm b}/m_{\rm m_{\rm \tilde{t}_1}})^2$~\cite{Peterson:1982ak,Ackerstaff:1998eb}.
      Fragmentation effects and gluon radiation increase the number of jets significantly and
      the importance of c-quark tagging is stressed in order to resolve the combinatorics.
\item neutralino mass $108.2\pm 0.3$~GeV~\cite{Carena:2006gb}.
\item theoretical uncertainties on the signal and background. Some improvement compared
      to the current loop calculation techniques is assumed, and an even larger reduction of this uncertainty
      is anticipated before the start of the ILC operation.
\end{itemize}

Tables~\ref{tab:stop_sys} and~\ref{tab:stop_sysida} list the systematic uncertainties 
for the sequential-cut analysis and the IDA. The systematic uncertainty
using the IDA method from detector calibration (energy scale) is larger. 
This is because the sequential-cut analysis pays particular attention to cancellation 
of this uncertainty between the two \mbox{analyses} at the different center-of-mass energies. 

\begin{table}[hp]
\begin{minipage}{0.4\textwidth}
\caption{Sequential-cut analysis experimental systematic uncertainties
on the signal efficiency. The first column indicates the variable that is cut
on. The second column contains the expected systematic uncertainty for this variable 
based on experience from LEP.
The third column shows by how much the signal efficiency for
$\sqrt{s} = 260$~GeV varies as a result of varying the cut value by this uncertainty. 
The fourth column shows the same for $\sqrt{s} = 500$~GeV, and
the fifth column lists the resulting error on the observable $Y$.}
\label{tab:stop_sys}
\end{minipage}\hfill
\begin{minipage}{0.58\textwidth}
\footnotesize
\begin{tabular}{lllll}
\hline
 & error on &  \multicolumn{2}{c}{rel. shift on signal eff.} & \\
variable & variable & 260~GeV & 500~GeV &\hspace*{-3mm} error on $Y$\hspace*{-3mm} \\
\hline
energy scale    & $1\%$ &  $3.7\%$ &  $3.1\%$ & $0.6\%$ 
\\
$N_{\mathrm{tracks}}$      & $0.5\%$ &  \multicolumn{3}{c}{negligible}
\\
charm tagging   & $0.5\%$ &  \multicolumn{3}{c}{taken to be $0.5\%$}
\\
luminosity      & -- & $0.4\%$ & $0.2\%$ & $0.4\%$ 
\\
charm fragmentation & $0.011$ & $0.3\%$ & $0.8\%$ & $0.6\%$
\\
stop fragmentation & $0.6\times 10^{-6}$\hspace*{-3mm} & $0.6\%$ & $0.2\%$  & $0.7\%$ 
\\
neutralino mass & 0.3~GeV & 3.8\% & 3.0\% & 0.8\%
\\
background estimate & -- & 0.8\% & 0.1\% & 0.8\% 
\\
\hline
\end{tabular}
\end{minipage}
\end{table}

\begin{table}[bph]
\centering
\begin{minipage}{0.4\textwidth}
\caption{
IDA experimental systematic uncertainties.
\label{tab:stop_sysida}
}
\vspace*{-1mm}
\end{minipage}\hfill
\begin{minipage}{0.58\textwidth}
\footnotesize
\begin{tabular}{lllll}
\hline
 & error on &  \multicolumn{2}{c}{rel. shift on signal eff.} & \\
variable & variable & 260~GeV & 500~GeV & \hspace*{-3mm}error on $Y$\hspace*{-3mm} \\
\hline
energy scale    & $1\%$ &  $3.4\%$ &  $1.3\%$ & $2.3\%$ 
\\
$N_{\mathrm{tracks}}$      & $0.5\%$ &  \multicolumn{3}{c}{negligible}
\\
charm tagging   & $0.5\%$ &  \multicolumn{3}{c}{taken to be $0.5\%$}
\\
luminosity      & -- & $0.4\%$ & $0.2\%$ & $0.4\%$ 
\\
charm fragmentation & $0.011$ & $0.1\%$ & $0.6\%$ & $0.5\%$
\\
stop fragmentation & $0.6\times 10^{-6}$\hspace*{-3mm} & $0.1\%$ & $0.8\%$  & $0.7\%$ 
\\
neutralino mass & 0.3~GeV & 3.7\% & 1.6\% & 2.2\%
\\
background estimate & -- & 0.3\% & 0.2\% & 0.1\% 
\\
\hline
\end{tabular}
\end{minipage}
\end{table}

\clearpage

\subsection{Mass determination}

The assessment of the achievable stop mass precision is based on the
statistical and systematic uncertainties on the observable $Y$ (eq. (1))
as summarized in Table~\ref{tab:stop_sum}. 
The IDA method has a smaller statistical uncertainty, and also a smaller background 
uncertainty due to a smaller number of expected background events.
The expected stop mass uncertainty 
is inferred from the uncertainty on $Y$ as given in Table~\ref{tab:stop_mstoperr}.

\begin{table}[hp]
\centering
\begin{minipage}{0.4\textwidth}
\caption{Summary of statistical and systematic uncertainties on
the observable~$Y$.
}
\label{tab:stop_sum}
\end{minipage}\hfill
\begin{minipage}{0.58\textwidth}
\footnotesize
\begin{tabular}{lcc}
\hline
error source for $Y$ & sequential cuts & IDA method \\
\hline
statistical                           & 3.1\% &   2.7\% \\
detector effects                      & 0.9\% &   2.4\% \\
charm  fragmentation                  & 0.6\% &   0.5\% \\
stop fragmentation                    & 0.7\% &   0.7\% \\
neutralino mass                       & 0.8\% &   2.2\% \\
background estimate                   & 0.8\% &   0.1\% \\
\hline
sum of experimental systematics       & 1.7\% &   3.4\% \\
sum of experimental errors            & 3.5\% &   4.3\% \\
\hline
theory for signal cross-section       & 5.5\% &   5.5\% \\
\hline
total error $\Delta Y$                & 6.5\% &   7.0\% \\
\hline
\end{tabular}
\end{minipage}
\end{table}

\begin{table}[hp]
\begin{minipage}{0.4\textwidth}
\caption{Estimated measurement errors (in~GeV) on the stop quark mass.
\label{tab:stop_mstoperr}}
\end{minipage} \hfill
\begin{minipage}{0.58\textwidth}
\footnotesize
\begin{tabular}{lcc}
\hline
 & \multicolumn{2}{c}{measurement error $\Delta m_{\rm \tilde{t}_1}$ (GeV)} \\
error category & sequential cuts & IDA method \\
\hline
statistical                           & $0.19$   & $0.17$ \\
sum of experimental systematics       & $0.10$   & $0.21$ \\
beam spectrum and calibration         & $0.1\phantom{0}$    & $0.1\phantom{0}$  \\
sum of experimental errors            & $0.24$   & $0.28$ \\
sum of all exp. and th. errors
                                      & $0.42$   & $0.44$ \\
\hline
\end{tabular}
\end{minipage}
\end{table}

\subsection{Cold dark matter interpretation}
\begin{figure}[hbp]
\begin{minipage}{0.59\textwidth}
\begin{center}
\includegraphics[width=\textwidth]{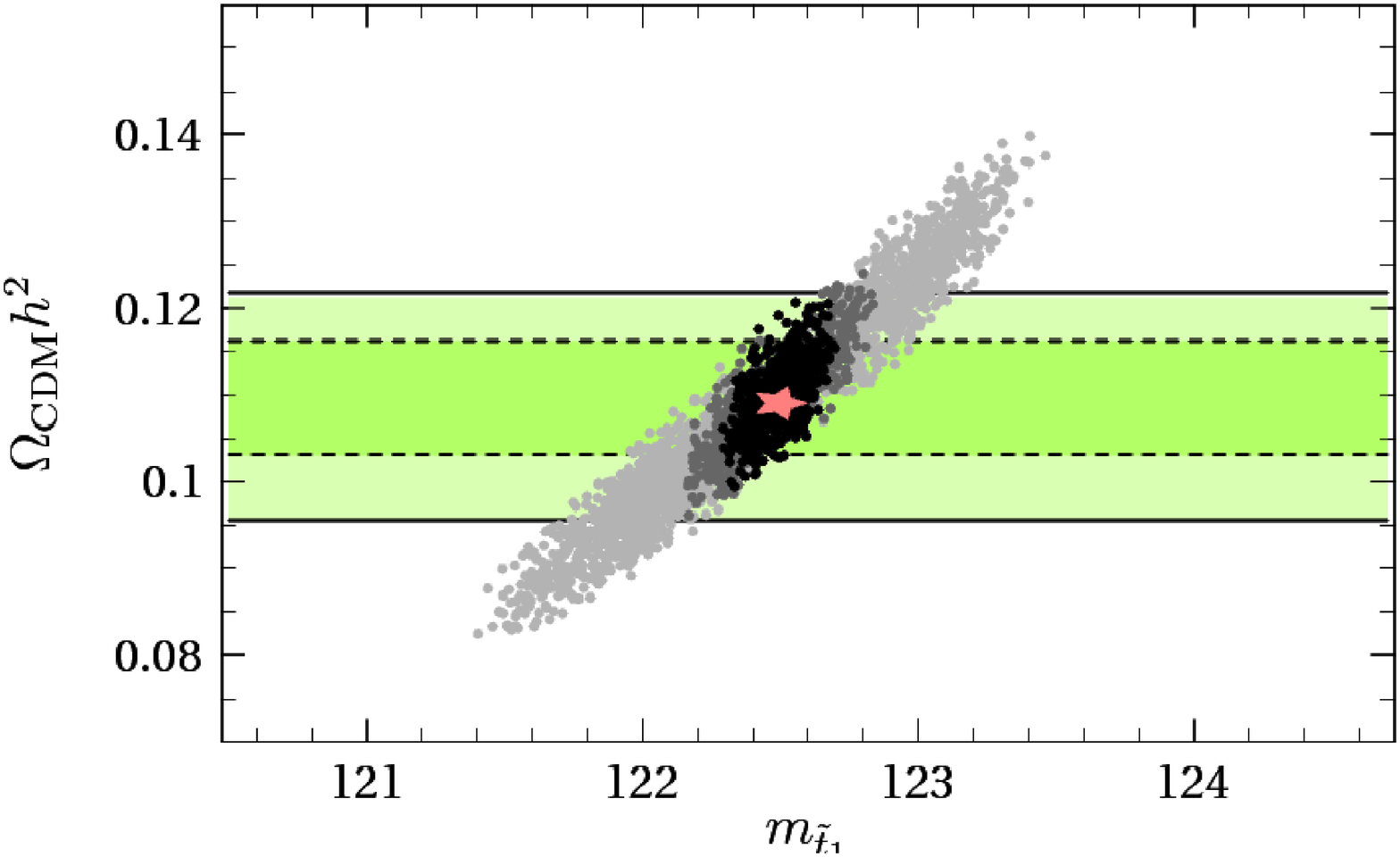}
\end{center}
\end{minipage}\hfill
\begin{minipage}{0.39\textwidth}
\caption{
Computation of dark matter relic abundance $\Omega_{\rm CDM} h^2$
taking into account estimated experimental errors for stop, chargino, neutralino
sector measurements at the future ILC. The black dots correspond to
a scan over the 1$\sigma$ ($\Delta \chi^2 \leq 1$) region including the 
total expected experimental uncertainties (detector and simulation),
the grey-dotted region includes also the theory uncertainty,
and the light grey-dotted area are the previous results~\cite{Carena:2005gc}.
The red star indicates the best-fit point. 
The horizontal shaded bands show the
1$\sigma$ and 2$\sigma$ constraints on the relic
density measured by WMAP.}
\label{fig:stop_stoppar}
\end{minipage}
\end{figure}

The chosen benchmark parameters are compatible with the mechanism of electroweak
baryogenesis~\cite{Carena:2005gc}.
They correspond to a value for the dark matter relic abundance
within the WMAP bounds, $\Omega_{\rm CDM} h^2 = 0.109$~\cite{Spergel:2006hy}.
The relic dark matter density has been computed as in Ref.~\cite{Carena:2005gc}\footnote{
The assumed benchmark parameters changed slighty (larger slepton masses assumed) and
$\Omega_{\rm CDM} h^2$ changed from 0.1122~\cite{Carena:2005gc} to 0.109.}.
In the investigated scenario, the stop and lightest neutralino masses are 
$m_{\rm \tilde{t}_1} = 122.5~$GeV and $m_{\tilde{\chi}^0_1} = 107.2$~GeV, 
and the stop mixing angle is 
$\cos \theta_{\rm \tilde{\rm t}} = 0.0105$,
i.e. the light stop is almost completely right-chiral. 
The improvement compared to Ref.~\cite{Carena:2005gc} regarding the CDM precision 
determination is shown in Fig.~\ref{fig:stop_stoppar} and summarized in Table~\ref{tab:stop_errors}.

\begin{table}[tb]
\begin{minipage}{0.45\textwidth}
\caption{Estimated precision for the determination of stop mass and dark
matter relic density for different assumptions about the 
systematic errors.
}
\label{tab:stop_errors}
\end{minipage}\hfill
\begin{minipage}{0.50\textwidth}
\footnotesize
\renewcommand{\arraystretch}{1.2}
\begin{tabular}{lcl}
\hline
 & $\Delta m_{\rm \tilde{t}_1}$ (GeV) & $\Omega_{\rm CDM} h^2$ \\
\hline
exp. and th. errors   &
 0.42 & $0.109^{+0.015}_{-0.013}$ \\
stat. and exp. errors only && \\ 
\rule{0mm}{0mm}\hspace{1em} sequential-cut analysis &
 0.24 & $0.109^{+0.012}_{-0.010}$ \\
\rule{0mm}{0mm}\hspace{1em} IDA &
 0.28 & $0.109^{+0.012}_{-0.010}$ \\
\hline
\end{tabular}
\renewcommand{\arraystretch}{1.}
\end{minipage}
\end{table}

\subsection*{Conclusions}

Scalar top quarks could be studied with precision at a future 
International Linear Collider (ILC).
The simulations for small stop-neutralino mass difference are motivated by cosmology.
The precision mass determination at the future ILC is possible with a method 
using two center-of-mass energies, e.g. $\sqrt{s}=260$ and 500~GeV.
This method can also be applied to other analyses to improve the mass resolution
in searches for new particles.
The precision of two independent analysis methods, one with a sequential-cuts and the
other with an Interative Discriminant Analysis (IDA) lead to very similar results.
The new proposed method increases the mass precision by about a factor of three due 
to the error cancellation using two center-of-mass energies with one near the production 
threshold.
Including experimental and theoretical uncertainties, the mass of a 122.5~GeV scalar top could be 
determined with a precision of 0.42~GeV. The interpretation of this benchmark scenario leads to a 
uncertainty on $\Omega_{\rm CDM} h^2$ of $-0.013$ and $+0.015$ which is about a factor two
better compared to previous results, and comparable to current cosmological (WMAP) 
measurement uncertainties.
With the new stop mass determination, the stop mass uncertainty is no longer the 
dominant uncertainty in the $\Omega_{\rm CDM} h^2$ calculation.

\subsection*{Acknowledgements}
We would like to thank the organizers of the 2005 and 2007 editions of the
Les Houches workshops ``Physics at TeV Colliders', where part of this work
was carried out.



\clearpage
\newpage

\setcounter{figure}{0}
\setcounter{table}{0}
\setcounter{equation}{0}
\setcounter{footnote}{0}
\section{Comparison of SUSY spectrum codes: the NUHM case
\protect\footnote{S.~Kraml and S.~Sekmen}}


\subsection{Introduction}

Recent analyses of uncertainties in SUSY spectrum calculations~\cite{Allanach:2003jw,Belanger:2005jk} 
have triggered important improvements in the various spectrum codes. This concerns in particular the 
treatment of the top and bottom Yukawa couplings, and the finite corrections in translating 
$\rm \overline{DR}$ parameters to on-shell masses and mixing angles, see e.g.~\cite{Baer:2005pv}.
Moreover, all public codes now apply full two-loop renormalization group (RG) evolution 
of the SUSY-breaking  parameters, plus one-loop self-energy corrections for sparticle masses.

So far, comparisons of spectrum computations have concentrated on the constrained MSSM
or mSUGRA models. 
In this contribution, we extend these studies to models with non-universal Higgs masses (NUHM).  
We use the most recent versions of the four public spectrum codes,
ISAJET\,7.75~\cite{Paige:2003mg,Baer:2005pv}, 
SOFTSUSY\,2.0.14~\cite{Allanach:2001kg},
SPHENO\,2.2.3~\cite{Porod:2003um} and 
SUSPECT\,2.3~\cite{Djouadi:2002ze}.
We first compare the results from the four public codes for the NUHM benchmark points 
of  \cite{DeRoeck:2005bw}, hence also comparing with the private code SSARD.  
It turns out that there is good agreement once the different sign convention 
in SSARD as compared to the other codes is taken into account. 
Then we discuss the case of gaugino mediation, for which $~10\%$ differences in 
left-chiral slepton masses can occur for very large $m_{H_{d}}^2-m_{H_{u}}^2$.

\subsection{NUHM benchmarks}

NUHM models have recently become very popular because they lead to very interesting 
phenomenological effects beyond the well-studied CMSSM case, 
see e.g.~\cite{Nath:1997qm,Ellis:2002iu,Baer:2005bu}. 
In general, the non-universality of the Higgs doublets can be specified either through GUT-scale 
masses $m_{H_{u}}^2$ and $m_{H_{d}}^2$, or though $\mu$ and $m_A$ at the weak scale.\footnote{We 
take $\mu\equiv\mu(M_{\rm EWSB})$, where $M_{\rm EWSB}$ is the scale of electroweak 
symmetry breaking, and $m_A\equiv m_A({\rm pole})$.} 
In particular, the group around Baer uses $m_{H_{u,d}}^2$, while the group around Ellis and Olive follows 
the second approach. From the four public spectrum codes, only two currently have the $(\mu,\,m_A)$ 
approach implemented. We therefore start our discussion by reproducing the NUHM benchmark points 
proposed in~\cite{DeRoeck:2005bw} with a scan in the $m_{H_u}^2$\,--\,$m_{H_d}^2$ plane.
 
\begin{figure}
\begin{center}
\includegraphics[width=0.46\textwidth]{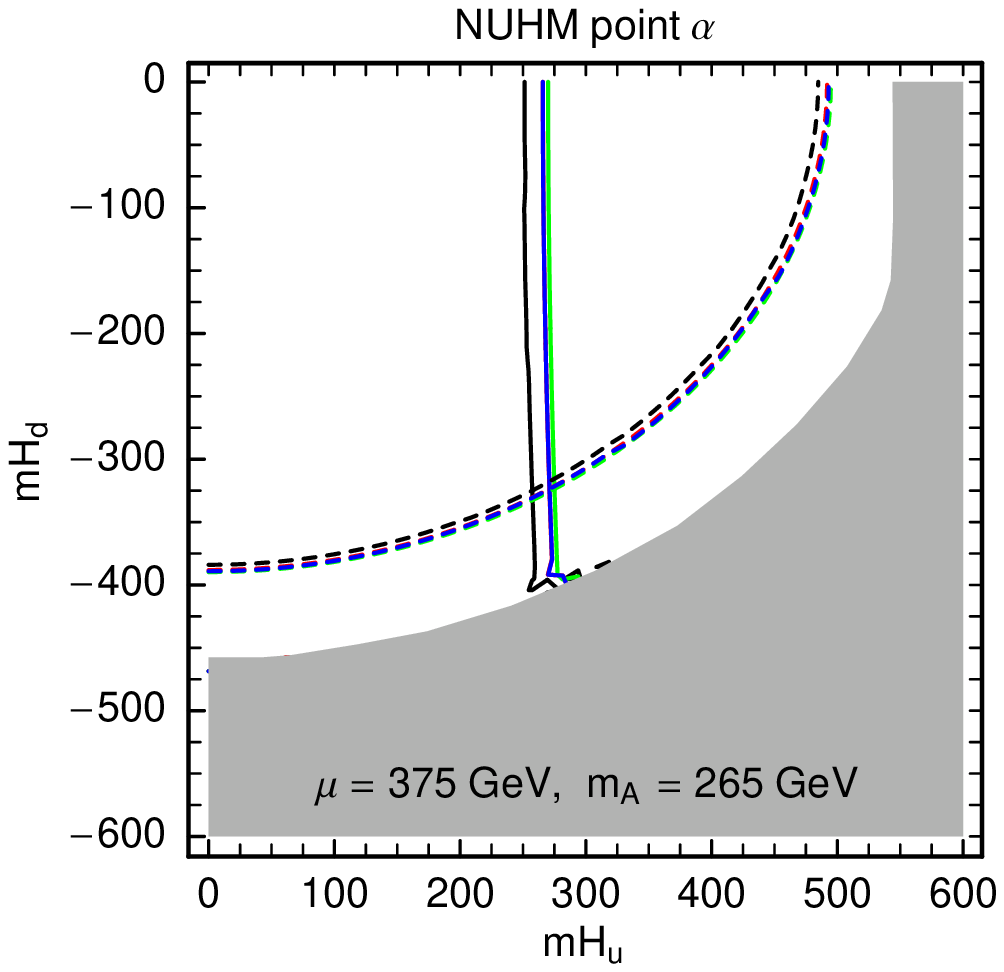}\qquad
\includegraphics[width=0.46\textwidth]{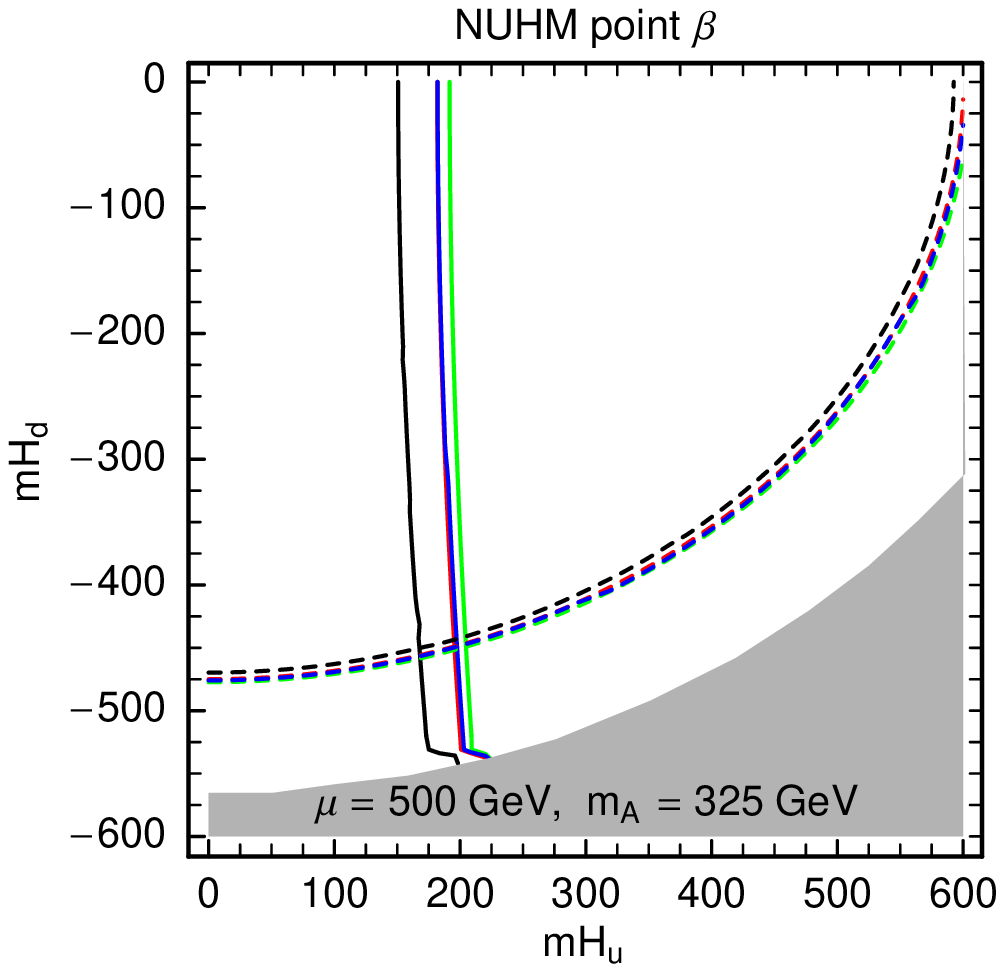}\\[5mm]
\includegraphics[width=0.46\textwidth]{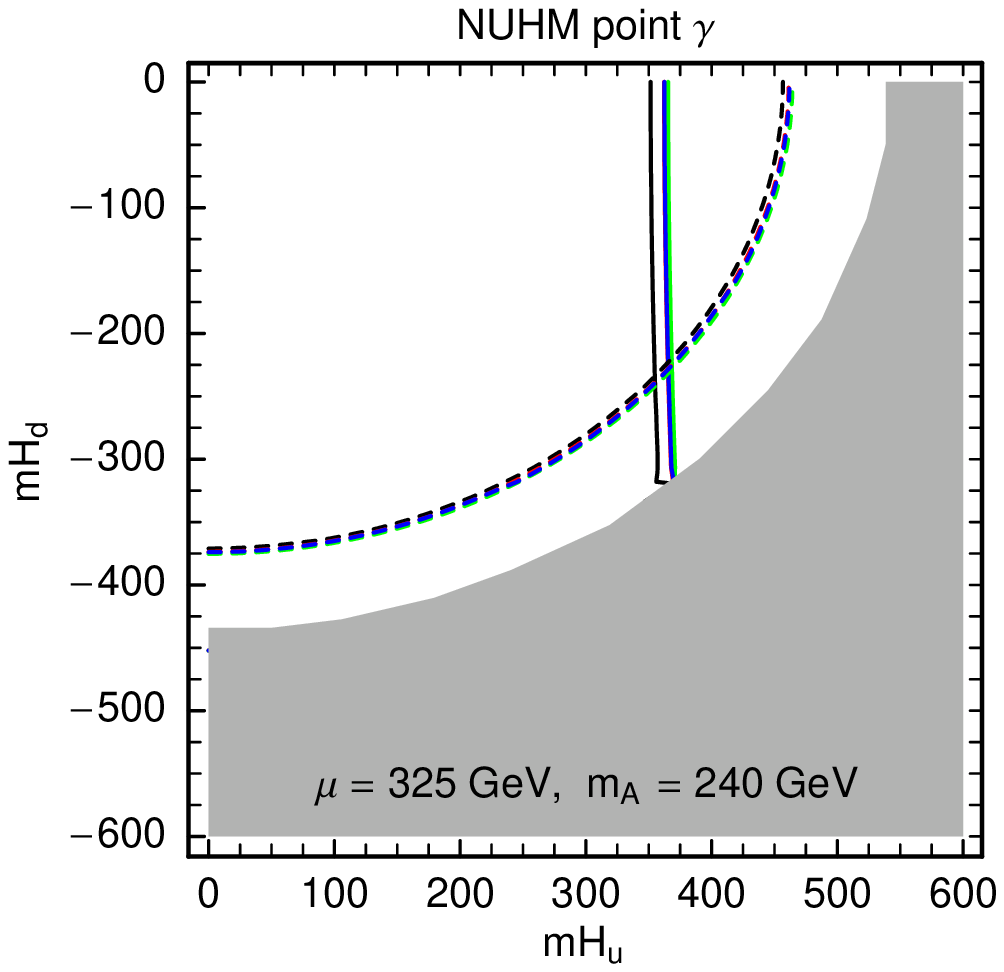}
\caption{Contours of constant $\mu$ (solid lines) and $m_A$ (dashed lines) from the four public spectrum codes 
in the $m_{H_u}$ versus $m_{H_d}$ plane for the NUHM benchmark scenarios $\alpha$, $\beta$, $\gamma$ 
of \cite{DeRoeck:2005bw}. 
The black, red, green and blue lines are for ISAJET, SOFTSUSY, SPHENO and SUSPECT, respectively. 
Each benchmark point is approximately reproduced where the respective contours of $\mu$ and $m_A$ intersect. 
In the grey areas no solution is obtained.}
\label{fig:nuhmBM}
\end{center}
\end{figure}

Figure~\ref{fig:nuhmBM} shows contours of constant $\mu$ and $m_A$ from the four public 
spectrum codes in the plane of $m_{H_u}$ versus $m_{H_d}$, 
$m_{H_{u,d}}\equiv {\rm sign}(m_{H_{u,d}}^2)\sqrt{|m_{H_{u,d}}^2}|$,  
for the NUHM benchmark scenarios $\alpha$, $\beta$, $\gamma$ of \cite{DeRoeck:2005bw}. 
Each benchmark point is approximately reproduced where the respective 
contours of $\mu$ and $m_A$ intersect. As can be seen, the solutions for the various codes lie close 
to each other. One can therefore also expect good agreement on the resulting physical spectrum. 

That this is indeed the case is examplified in Table~\ref{tab:alpha} for benchmark point $\alpha$. 
We find agreement of about 1--2\% on the sparticle and Higgs masses, and about 10\% 
on the relic density, which we compute using MICROMEGAS~\cite{Belanger:2004yn,Belanger:2006is} 
for the ISAJET, SOFTSUSY, SPHENO and SUSPECT spectra.
Note also that the agreement within the public codes themselves is better than the overall agreement 
including SSARD, cf.~Table~\ref{tab:alpha}.
For point $\beta$, there is also 1--2\% agreement on the masses. For the relic density, however, 
the spectra of the public codes interfaced to MICROMEGAS give $\Omega h^2\sim 0.07$ with 15\% variation, 
while SSARD gives $\Omega h^2=0.1$. The source of the difference is the A-funnel annihilation 
cross section. 

In SOFTSUSY, there is also the option to input $\mu$ and $m_A$ 
instead of $m_{H_{u,d}}^2$ at $M_{\rm GUT}$. To this aim, the progam makes an initial 
guess of the GUT-scale $m_{H_{u,d}}^2$ for the first iteration. For later iterations, only the 
EWSB-scale boundary conditions are used for $m_{H_{u,d}}^2$. The procedure works 
extremely well, giving exactly the same results irrespective of whether one uses 
GUT-scale $m_{H_{u,d}}^2$ or EWSB-scale $\mu$, $m_A$ as inputs 
---at least for the cases we have tried.

There is an analogous option in ISAJET, through the {\tt NUSUG} keywords. However, 
the procedure applied is more complicated, using boundary conditions for $m_{H_{u,d}}^2$ 
at both the GUT and the EWSB scales. The results slightly depend on what kind of input 
is used. For example, for point $\alpha$, input of $\mu=375$~GeV and $m_A=265$~GeV 
gives $m_{H_{u}}^2=(250)^2$~GeV$^2$ and $m_{H_{d}}^2=-(320.2)^2$~GeV$^2$ at $M_{\rm GUT}$.
On the other hand, GUT-scale input of $m_{H_{u}}^2=(250)^2$~GeV$^2$ and 
$m_{H_{d}}^2=-(320.2)^2$~GeV$^2$ gives $\mu=378$~GeV and $m_A=274.6$~GeV 
at the weak scale.

SPHENO and SUSPECT do not yet have $\mu$ and $m_A$ input for SUGRA scenarios.  

\begin{table}\centering
\begin{tabular}{|c||c|c|c|c|c||c|}
\hline
     & SSARD\,\cite{DeRoeck:2005bw} & ISAJET & SOFTSUSY & SPHENO & SUSPECT & $\Delta$ [\%] \\
\hline
   $m_{H_u}^2(M_{\rm GUT})$ & $-(333)^2$  & $+(257.4)^2$ & $+(271.0)^2$ & $+(275.0)^2$ & $+(271.1)^2$ & \\
   $m_{H_d}^2(M_{\rm GUT})$ & $+(294)^2$  & $-(325.0)^2$ & $-(323.6)^2$ & $-(323.7)^2$ & $-(323.7)^2$ & \\
\hline
    $h^0$ & 115 & 112.0 & 112.5 & 112.7 & 112.5 & 2.66 \\
    $H^0$ & 266 & 265.0 & 265.0 & 265.7 & 265.7 & 0.38 \\
    $A^0$ & 265 & 265.0 & 265.0 & 265.0 & 265.0 & --- \\
    $H^\pm$ & 277 & 280.0 & 277.0 & 277.1 & 277.1 & 1.08 \\
\hline
    $\tilde u_R$ & 637 & 647.2 & 644.9 & 646.3 & 643.9 & 1.58 \\
    $\tilde u_L$ & 648 & 659.7 & 653.1 & 659.3 & 656.1 & 1.79 \\
    $\tilde d_R$ & 630 & 639.3 & 633.2 & 638.1 & 635.8 & 1.46 \\
    $\tilde d_L$ & 653 & 664.8 & 659.3 & 663.9 & 660.8 & 1.79 \\
    $\tilde t_1$ & 471 & 475.1 & 475.5 & 477.1 & 476.9 & 1.28 \\
    $\tilde t_2$ & 652 & 655.3 & 652.7 & 654.9 & 655.1 & 0.50 \\
    $\tilde b_1$ & 590 & 599.7 & 591.9 & 594.4 & 597.8 & 1.63 \\
    $\tilde b_2$ & 629 & 637.4 & 630.5 & 637.0 & 635.4 & 1.33 \\
\hline
    $\tilde e_R$ & 216 & 218.8 & 219.3 & 218.9 & 218.1 & 1.51 \\
    $\tilde e_L$ & 296 & 296.9 & 296.7 & 296.7 & 295.1 & 0.61 \\
    $\tilde\nu_{e}$ & 285 & 285.1 & 285.2 & 285.7 & 284.6 & 0.39 \\
    $\tilde\tau_1$ & 212 & 216.3 & 215.9 & 215.4 & 214.8 & 2.00 \\
    $\tilde\tau_2$ & 298 & 297.9 & 298.3 & 298.3 & 296.8 & 0.50 \\
    $\tilde\nu_{\tau}$ & 285 & 284.0 & 284.9 & 285.4 & 284.3 & 0.49 \\
\hline
    $\tilde\chi^0_1$ & 115 & 112.4 & 111.7 & 111.9 & 112.2 & 1.16 \\
    $\tilde\chi^0_2$ & 212 & 208.2 & 207.8 & 208.1 & 208.2 & 2.01 \\
    $\tilde\chi^0_3$ & 388 & 380.3 & 383.8 & 381.7 & 380.6 & 2.01 \\
    $\tilde\chi^0_4$ & 406 & 401.3 & 401.5 & 400.7 & 400.9 & 1.32 \\
\hline
    $\tilde\chi^\pm_1$ & 212 & 208.3 & 209.0 & 207.7 & 207.6 & 2.11 \\
    $\tilde\chi^\pm_2$ & 408 & 400.4 & 398.1 & 401.5 & 400.9 & 2.45 \\
\hline
    $\tilde g$ & 674 & 691.2 & 687.7 & 685.6 & 688.7 & 2.51 \\
\hline
    $\Omega h^2$ & 0.12 & 0.109 & 0.114 & 0.111 & 0.108 & 10.68 \\
\hline     
\end{tabular}
\caption{Comparison of results for NUHM benchmark point $\alpha$ given by 
$m_0=210$~GeV, $m_{1/2}=285$~GeV, $A_0=0$, $\tan\beta=10$, $m_t=178$~GeV, $\mu=375$~GeV and 
$m_A=265$~GeV.}
\label{tab:alpha}
\end{table}

\subsection{Slepton masses in gaugino mediation}

In general, in SUSY-breaking models  with universal scalar and gaugino masses, 
the right-chiral charged sleptons, $\tilde\ell_R$, are lighter than the 
left-chiral ones and the sneutrinos, $\tilde\ell_L$ and $\tilde\nu_\ell$  ($\ell = e,\mu,\tau$).  
However, owing to $S$-term contributions in the RG evolution, for large enough  
$m_{H_d}^2-m_{H_u}^2>0$, $\tilde\ell_L$ and/or $\tilde\nu_\ell$ can become lighter than $\tilde\ell_R$, 
and even lighter than the $\tilde\chi^0_1$~\cite{Nath:1997qm,Ellis:2002iu,Baer:2005bu}.
In such a setup, if R parity is conserved, the lightest SUSY particle
(LSP) should be a gravitino or axino, and the next-to-lightest one (NLSP)
a $\tilde\tau_1$ or $\tilde\nu_\tau$. This has recently attracted quite some interest 
in the context of gaugino mediation~\cite{Buchmuller:2005ma,Buchmuller:2006nx,Covi:2007xj,Kraml:2007sx}.

In gaugino mediation~\cite{Kaplan:1999ac,Chacko:1999mi}, gauge couplings and gaugino masses 
each unify at the compactification scale $M_C$, while there are no-scale boundary conditions for 
sfermion masses and trilinear couplings, i.e.\ $m_0=A_0=0$) \cite{Chacko:1999mi}.
The free parameters of the model are hence $m_{1/2}$, $m^2_{H_u}$, $m^2_{H_d}$, $\tan\beta$, 
and the sign of $\mu$; $|\mu|$ being determined by radiative electroweak symmetry
breaking. Following~\cite{Buchmuller:2005ma,Buchmuller:2006nx,Covi:2007xj,Kraml:2007sx}, we take 
$M_C=M_{\rm GUT}$ and $m_t=172.5$~GeV. 

\begin{figure}
\begin{center}
\includegraphics[width=0.46\textwidth]{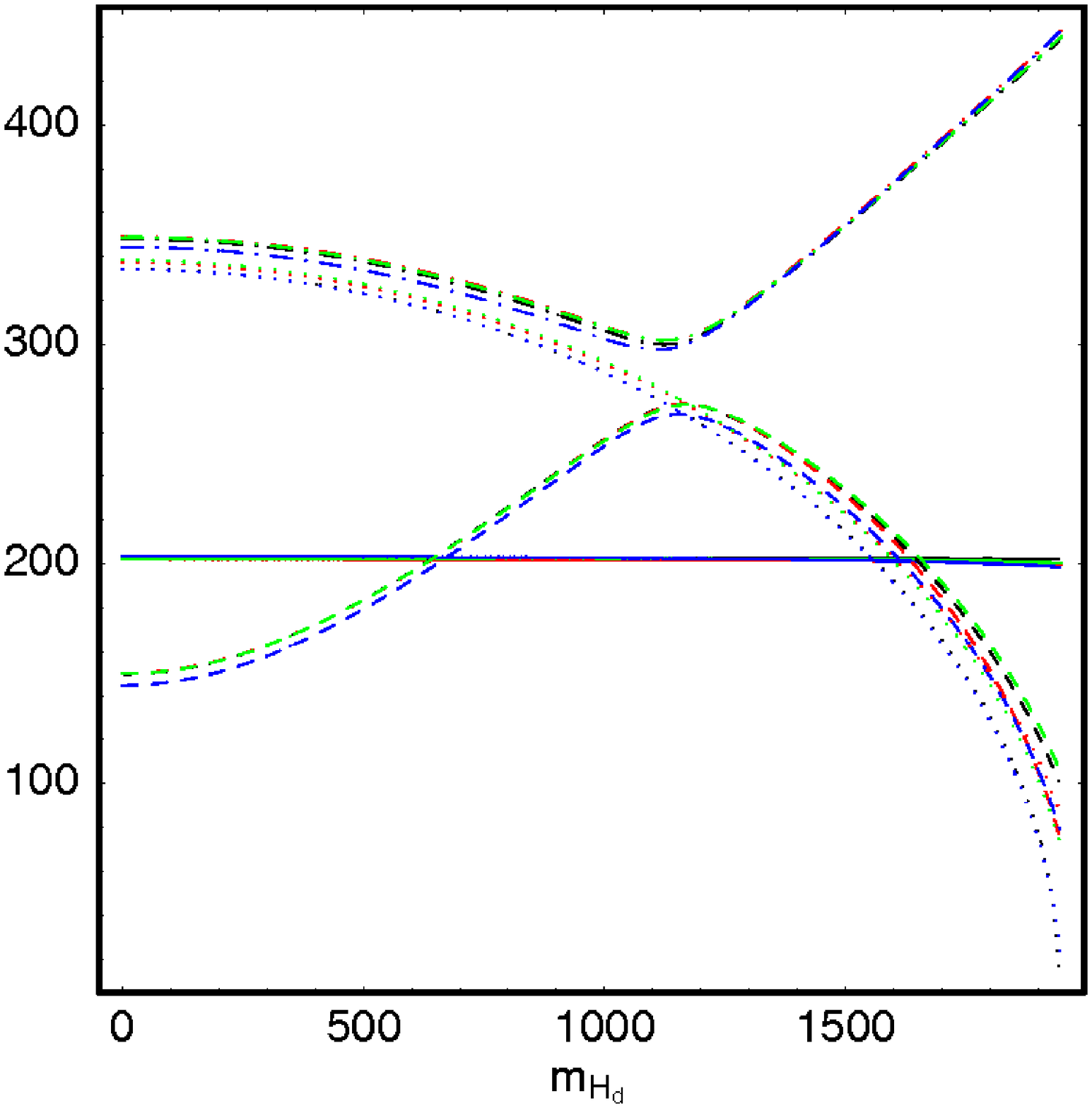}\\
  \begin{picture}(100,0)\setlength{\unitlength}{1mm}
    \put(-6,44){$\tilde\chi^0_1$}
    \put(4,32){$\tilde\tau_1$}
    \put(35,66){$\tilde\tau_2$}
    \put(0,55){$\tilde\nu_\tau$}
  \end{picture}\\[-6mm]
\caption{Neutralino, stau and tau-sneutrino masses (in GeV) from the four public spectrum codes as 
a function of $m_{H_d}$, for $m_{1/2}=500$~GeV, $\tan\beta=10$, $m0=A_0=m_{H_u}=500$. 
The black, red, green and blue lines are for ISAJET, SOFTSUSY, SPHENO and SUSPECT, respectively.}
\label{fig:gauginomed}
\end{center}
\end{figure}

Figure~\ref{fig:gauginomed} shows the $\tilde\chi_1^0$, $\tilde\tau_{1,2}^{}$ and $\tilde\nu_\tau$ masses 
as a function of $m_{H_d}$, for $m_{1/2}=500$~GeV, $\tan\beta=10$, $m_{H_u}=500$.\footnote{In 
SUSPECT we use $m_0=10^{-3}$, since this code does not give a spectrum for $m_0\equiv0$.} 
As can be seen, for $m_{H_d}\sim1.2$~TeV, the left-ciral sleptons become lighter than the 
right-chiral ones; for $m_{H_d}\sim1.6$~TeV, they are lighter than the lightest neutralino. 
Moreover, there is an overall good agreement between the codes. Only for very large 
$m_{H_d}^2-m_{H_u}^2$, when there are large $S$-term corrections, the differences in 
the left-chiral slepton masses reach $\sim 10\%$. This is also the region where the stau or 
sneutrino is the (pseudo)\,LSP.
Explicit numbers are given in Tables~\ref{tab:gm1} and \ref{tab:gm2} 
for $m_{H_d}=900$~GeV and $1.8$~TeV, respectively. 
Note also that there is $\sim3$--$4\%$ difference in $\mu$. The masses of coloured sparticles 
and Higgs bosons are not shown due to lack of space; they agree to $\sim1\%$ or better. 

\begin{table}[t]\centering
\begin{tabular}{|c||c|c|c|c||c|}
\hline
     & ISAJET & SOFTSUSY & SPHENO & SUSPECT & $\Delta m$ [\%] \\
\hline
    $\tilde e_R$ & 253.6 & 256.7 & 256.7 & 254.1 & 1.25 \\
    $\tilde e_L$ & 318.8 & 317.4 & 316.7 & 311.4 & 2.34 \\
    $\tilde\nu_{e}$ & 304.8 & 305.2 & 306.4 & 301.5 & 1.61 \\
    $\tilde\tau_1$ & 241.3 & 241.1 & 240.9 & 238.2 & 1.29 \\
    $\tilde\tau_2$ & 313.7 & 315.7 & 315.2 & 310.0 & 1.82 \\
    $\tilde\nu_{\tau}$ & 296.6 & 300.4 & 301.6 & 296.7 & 1.67 \\
\hline
    $\tilde\chi^0_1$ & 203.0 & 201.8 & 202.6 & 202.8 & 0.59 \\
    $\tilde\chi^0_2$ & 364.4 & 365.6 & 367.8 & 367.2 & 0.93 \\
    $\tilde\chi^0_3$ & 463.5 & 477.3 & 481.2 & 477.7 & 3.73 \\
    $\tilde\chi^0_4$ & 500.2 & 508.4 & 513.4 & 511.1 & 2.6 \\
\hline
    $\tilde\chi^\pm_1$ & 365.0 & 367.3 & 367.3 & 366.3 & 0.63 \\
    $\tilde\chi^\pm_2$ & 499.1 & 504.9 & 513.1 & 510.2 & 2.76 \\
\hline
\end{tabular}
\caption{Comparison of neutralino, chargino, and slepton masses for a gaugino-mediation 
scenario with $m_0=A_0=0$, $m_{1/2}=500$~GeV, $\tan\beta=10$, $m_{H_u}=500$ 
and $m_{H_d}=900$~GeV.}
\label{tab:gm1}
\end{table}

\begin{table}[t]\centering
\begin{tabular}{|c||c|c|c|c||c|}
\hline
     & ISAJET & SOFTSUSY & SPHENO & SUSPECT & $\Delta m$ [\%] \\
\hline
    $\tilde e_R$ & 432.1 & 436.2 & 433.8 & 436.0 & 0.94 \\
    $\tilde e_L$ & 199.4 & 182.7 & 192.3 & 180.1 & 10.23 \\
    $\tilde\nu_{e}$ & 170.3 & 179.9 & 175.0 & 162.5 & 10.12 \\
    $\tilde\tau_1$ & 159.9 & 151.1 & 163.4 & 148.6 & 9.5 \\
    $\tilde\tau_2$ & 410.5 & 413.3 & 411.1 & 413.0 & 0.68 \\
    $\tilde\nu_{\tau}$ & 130.6 & 149.5 & 144.5 & 148.6 & 13.19 \\
\hline
    $\tilde\chi^0_1$ & 202.5 & 200.9 & 201.3 & 200.1 & 1.19 \\
    $\tilde\chi^0_2$ & 351.7 & 353.6 & 357.2 & 356.4 & 1.55 \\
    $\tilde\chi^0_3$ & 425.3 & 439.4 & 444.5 & 440.9 & 4.39 \\
    $\tilde\chi^0_4$ & 474.9 & 481.1 & 486.7 & 484.6 & 2.45 \\
\hline
    $\tilde\chi^\pm_1$ & 351.8 & 355.6 & 356.2 & 355.0 & 1.24 \\
    $\tilde\chi^\pm_2$ & 473.0 & 478.9 & 486.3 & 483.6 & 2.77 \\
\hline     
\end{tabular}
\caption{Same as Table~\ref{tab:gm1} but for $m_{H_d}=1.8$~TeV.}
\label{tab:gm2}
\end{table}

\subsection{Conclusions}

We have compared results of the latest versions of the public SUSY spectrum codes 
ISAJET, SOFTSUSY, SPHENO and SUSPECT for models with non-universal Higgs 
masses. We find very good agreement for the mass spectra and the resulting $\Omega h^2$.
Only for edges of parameter space, the differences in the most sensitive masses become large. 
For instance, we have found differences in slepton masses of the order of 10\% for 
large $m_{H_{d}}^2-m_{H_{u}}^2$.

\subsection*{Acknowledgements}
SS acknowledges the support by T\"UB\'ITAK grant 106T457 (TBAG-HD 190).

\clearpage
\newpage

\setcounter{figure}{0}
\setcounter{table}{0}
\setcounter{equation}{0}
\setcounter{footnote}{0}
\section{The SLHA2 Conventions
\protect\footnote{participating/corresponding authors: B.C.~Allanach, P.~Skands and P.~Slavich}}

\subsection{Introduction}

The states and couplings appearing in the general minimal
supersymmetric standard model (MSSM) can be defined in a number of
ways. In principle, this is not a problem, as 
translations between different conventions
can usually be carried out without ambiguity. From the point of
view of practical application, however, such translations
are, at best, tedious, and at worst they introduce an unnecessary
possibility for error. 
To deal with this problem, and to create a more transparent situation
for non-experts, the original SUSY Les Houches Accord (SLHA1) was proposed
\cite{Skands:2003cj}. However, SLHA1 was designed exclusively with the
MSSM with real parameters and R-parity conservation in mind. 
We here summarize conventions \cite{Allanach:2008qg}
relevant for R-parity violation (RPV), flavour violation, 
and CP-violation (CPV) in the minimal supersymmetric standard
model. We also consider next-to-minimal models which we shall
collectively label NMSSM. 
For simplicity, we still limit the scope of the SLHA2 in two regards: 
for the MSSM, we restrict our attention to \emph{either} CPV or
RPV, but not both. 
For the NMSSM, we define one catch-all model and extend the SLHA1 mixing only 
to include the new states, with CP, R-parity, and flavour still
assumed conserved. 
For brevity, this document only describes our  
convention choices and not the full ASCII data structures that go with
them (these are the focus of a complementary summary \cite{Skands:2007zz}). 
The complete SLHA2 is described in detail in \cite{Allanach:2008qg}.

\subsection{Extensions of SLHA1 \label{sec:slha1}}
Firstly, we allow for using either the $A^0$ or $H^+$ pole masses,
respectively, as input instead of the parameter
$m_A^2(\ensuremath{M_{\mathrm{input}}})$ defined in
\cite{Skands:2003cj}.  Secondly, we also optionally allow for
different parameters to be defined at different scales (e.g., $\mu$
defined at $M_{\mathrm{EWSB}}$, the remaining parameters defined at
$\ensuremath{M_{\mathrm{input}}}$).

To define the general properties of the model, we introduce new global
switches for field content (either MSSM or NMSSM), RPV (either off or
on), CPV (either no CPV, just the CKM phase, or general CPV), and
flavour violation (either no flavour violation or quark and/or lepton
flavour violation).

Also note the recent proposal \cite{Alwall:2007mw} for a joint
SLHA+LHEF (Les Houches Event File \cite{Alwall:2006yp}) format for BSM
event generation.

\subsection{Flavour violation \label{sec:flv}}
\label{flavchange}
In the Super-CKM basis of the squarks \cite{Hall:1985dx}, 
the quark mass matrix is diagonal, and the
squarks are rotated in parallel to their superpartners. This implies
that only physically measurable parameters are present. 
Actually, once  the electroweak symmetry is broken, a rotation in 
flavour space 
\begin{equation}
  D^{\,o}     \  = \ V_d \,D\,,           \hspace*{0.8truecm}
  U^{\,o}     \  = \ V_u \,U\,,           \hspace*{0.8truecm}
  \bar{D}^{o}     = \ U_d^\ast \,\bar{D}\,, \hspace*{0.8truecm}
  \bar{U}^{o}       = \ U_u^\ast \,\bar{U}\,, 
\label{superotation}
\end{equation}
of all matter superfields in the (s)quark superpotential 
\begin{equation}
 W_Q  \ = \ \epsilon_{ab}  
  \left[ \left(Y_D\right)_{ij}  H_1^a Q_i^{b\,o} \bar{D}_j^{\,o}
  +    \left(Y_U\right)_{ij} H_2^b  Q_i^{a\,o}  \bar{U}_j^{\,o}
 \, \right]
\label{superpot}
\end{equation}
brings fermions from the interaction eigenstate basis 
$\{d_L^o,u_L^o,d_R^o,u_R^o\}$ to their mass eigenstate basis 
$\{d_L,u_L,d_R,u_R\}$:  
$d_L^o = V_d d_L$, $u_L^o = V_u u_L$, $d_R^o = U_d d_R$, $u_R^o = U_u
u_R$, and the scalar superpartners to the basis 
$\{ \tilde{d}_L, \tilde{u}_L, 
    \tilde{d}_R, \tilde{u}_R \}$.
Through this rotation, the Yukawa matrices $Y_D$ and $Y_U$ are 
reduced to their diagonal form $\hat{Y}_D$ and $\hat{Y}_U$: 
\begin{equation}
(\hat{Y}_D)_{ii} = (U_d^\dagger Y_D^T V_d)_{ii}= \sqrt{2}\frac{m_{d\,i}}{v_1}\,,
\hspace*{1.0truecm} 
(\hat{Y}_U)_{ii} = (U_u^\dagger Y_U^T V_u)_{ii}=
\sqrt{2}\frac{m_{u\,i}}{v_2}\,. 
\label{diagyukawa}
\end{equation}
Tree-level mixing terms among quarks of different generations are due 
to the misalignment of $V_d$ and $V_u$, expressed via the CKM matrix 
\cite{Cabibbo:1963yz,Kobayashi:1973fv}, $V_{\mathrm{CKM}}=V_u^\dagger V_d$,
which is proportional to the tree-level
$\bar{u}_{Li}d_{Lj}W^+ $, 
$\bar{u}_{Li}d_{Rj}H^+ $, and 
$\bar{u}_{Ri}d_{Lj}H^+$ couplings ($i,j=1,2,3$). 

In the super-CKM  basis the $6\times 6$ mass matrices for the up- and
down-type squarks are defined as
\begin{equation}
{\cal L}^{\rm mass}_{\tilde q} ~=~ 
- \Phi_u^{\dagger}\,
{\cal M}_{\tilde u}^2\, 
\Phi_u
- \Phi_d^{\dagger}\,
{\cal M}_{\tilde d}^2\, 
\Phi_d~,
\end{equation}
where $\Phi_u = (\tilde u_L,\tilde c_L, \tilde t_L,
\tilde u_R,\tilde c_R, \tilde t_R)^T$ and 
$\Phi_d = (\tilde d_L,\tilde s_L, \tilde b_L,
\tilde d_R,\tilde s_R, \tilde b_R)^T$.
We diagonalise the squark mass matrices via $6\times6$ unitary
matrices $R_{u,d}$, such that $R_{u,d} \,{\cal M}_{{\tilde
    u},{\tilde d}}^2\,R_{u,d}^\dagger$ are diagonal matrices with increasing
mass squared values. 
The flavour-mixed mass matrices read:
\begin{eqnarray}
{\cal M}_{\tilde u}^2 &=&  \left( \begin{array}{cc}
  V_{\mathrm{CKM}} \,{\hat{m}_{\tilde Q}}^2\, V_{\mathrm{CKM}}^\dagger + m^2_{u}
  + D_{u\,LL} 
  &   
  \frac{v_2}{\sqrt{2}} {\hat T}_U^\dagger  - \mu m_u \cot\beta      \\[1.01ex]
  \frac{v_2}{\sqrt{2}} {\hat T}_U  - \mu^* m_u \cot\beta  &
       {\hat m}^2_{\tilde u} + m^2_{u} + D_{u\,RR} \\                 
 \end{array} \right) \,\, ,
\label{massmatrixu}  \\
\nonumber\\
{\cal M}_{\tilde d}^2 &=&  \left( \begin{array}{cc}
   {\hat m_{\tilde Q}}^2 + m^2_{d} + D_{d\,LL}           
& 
   \frac{v_1}{\sqrt{2}} {\hat T}_D^\dagger  - \mu m_d \tan\beta      \\[1.01ex]
   \frac{v_1}{\sqrt{2}} {\hat T}_D  - \mu^* m_d \tan\beta  &
   {\hat m}^2_{\tilde d} + m^2_{d} + D_{d\,RR}
             \\                 
 \end{array} \right) \,\, .
\label{massmatrixd}  
\end{eqnarray}
In the equations above we introduced the $3\times 3$ matrices
\begin{equation}
{\hat m_{\tilde Q}}^2 \equiv V^\dagger_d \,m^2_{\tilde Q}\, V_d\,,~~~
{\hat m_{\tilde u,\tilde d}}^2 
  \equiv U^\dagger_{u,d} \,{m^2_{\tilde u, \tilde{d}}}^T\, U_{u,d}\,,~~~
{\hat T_{U}} \equiv U^\dagger_u \,T_{U}^T\, V_u\,,~~~
{\hat T_{D}} \equiv U^\dagger_d \,T_{D}^T\, V_d\,,~~~
\label{eq:that}
\end{equation}
where the un-hatted mass matrices $m^2_{Q,u,d}$ 
and trilinear interaction matrices $T_{U,D}$ 
are given in the electroweak basis of \cite{Skands:2003cj}, in which
the soft SUSY-breaking potentials $V_3$ and
$V_2$ have the following forms:
\begin{eqnarray}
V_3 & = & \epsilon_{ab} \sum_{ij}
\left[
(T_E)_{ij} H_1^a \tilde{L}_{i_L}^{b} \tilde{e}_{j_R}^* +
(T_D)_{ij} H_1^a               \tilde{Q}_{i_L}^{b}  \tilde{d}_{j_R}^* +
(T_U)_{ij}  H_2^b \tilde{Q}_{i_L}^{a} \tilde{u}_{j_R}^*
\right]
+ \mathrm{h.c.}~, \label{eq:slha1v3soft}\\[2mm]
V_2 &=& m_{H_1}^2 {{H^*_1}_a} {H_1^a} + m_{H_2}^2 {{H^*_2}_a} {H_2^a} +
{\tilde{Q}^*}_{i_La} (m_{\tilde Q}^2)_{ij} \tilde{Q}_{j_L}^{a} +
{\tilde{L}^*}_{i_La} (m_{\tilde L}^2)_{ij} \tilde{L}_{j_L}^{a}  
+ \nonumber \\ &&
\tilde{u}_{i_R} (m_{\tilde u}^2)_{ij} {\tilde{u}^*}_{j_R} +
\tilde{d}_{i_R} (m_{\tilde d}^2)_{ij} {\tilde{d}^*}_{j_R} +
\tilde{e}_{i_R} (m_{\tilde e}^2)_{ij} {\tilde{e}^*}_{j_R} -
(m_3^2 \epsilon_{ab} H_1^a H_2^b + \mathrm{h.c.})~. \label{eq:slha1v2soft}
\end{eqnarray} 
 
The
matrices $m_{u,d}$ are the diagonal up-type and down-type quark masses
and $D_{f\,LL,RR}$ are the D-terms given by:
 \begin{equation}
 D_{f\,LL,RR} =  \cos 2\beta \, m_Z^2 
   \left(T_f^3 - Q_f \sin^2\theta_W \right) 
{{\mathchoice {\rm 1\mskip-4mu l} {\rm 1\mskip-4mu l}
{\rm 1\mskip-4.5mu l} {\rm 1\mskip-5mu l}}}_3\,,
\label{dterm}
\end{equation}
which are also flavour diagonal, and $Q_f$ is the electric
charge of the left-handed chiral supermultiplet to which the squark
belongs, i.e., it is $2/3$ for $U$ and $-2/3$ for $U^c$.
Note that the up-type and down-type
squark mass matrices in eqs.~(\ref{massmatrixu}) and
(\ref{massmatrixd}) cannot be simultaneously flavour-diagonal unless
${\hat m_{\tilde Q}}^2$ is flavour-universal. 

For the lepton sector, we adopt a super-PMNS basis and 
cover all cases that lead to a low energy effective field theory
with Majorana neutrino masses and one sneutrino per family. 
In terms of this low energy effective theory, the lepton mixing phenomenon is
analogous to the quark mixing case. After electroweak symmetry breaking, the
neutrino sector of the MSSM contains the Lagrangian pieces
\begin{equation}
{\mathcal L} = -\frac12 {\nu}^{oT} (m_\nu) \nu^o
+ {\rm h.c.},
\end{equation}
where $m_\nu$ is a $3 \times 3$ symmetric matrix. The 
interaction eigenstate basis neutrino fields $\nu^o$ are related to the mass
eigenstate ones $\nu$ by
$\nu^o = V_\nu \nu$, reducing the mass matrix $m_\nu$ 
to its diagonal form $\hat{m}_\nu$ 
\begin{equation}
  (\hat{m}_\nu)_{ii} = (V_\nu^T m_\nu V_\nu)_{ii} =
  m_{\nu_i}. \label{eq:diagmL} 
\end{equation}
The charged
lepton fields have a $3 \times 3$ Yukawa coupling matrix defined in the
superpotential piece \cite{Skands:2003cj}
\begin{equation}
W_E = \epsilon_{ab}  (Y_E)_{ij} H_1^a L_i^{bo} \bar{E}_j^o,
\end{equation}
where the charged lepton interaction eigenstates 
$\{ e_L^o, e_R^o \}$ are related to the mass eigenstates
by 
$e_L^o = V_e e_L$ and $e_R^o = U_e e_R$.
The equivalent diagonalised charged lepton Yukawa matrix is 
\begin{equation}
(\hat{Y}_E)_{ii} = (U_e^\dag Y_E^T V_e)_{ii} = \sqrt{2}
  \frac{m_{ei}}{v_1}~~~. \label{eq:diagYL}
\end{equation}

Lepton mixing in the charged current interaction can then be characterised by
the PMNS matrix \cite{Maki:1962mu,Pontecorvo:1967fh},
$U_{PMNS}=V_e^\dag V_\nu$, 
which is proportional to the tree-level
$\bar{e}_{Li}{\nu_j}W^-$ and $\bar{e}_{Ri}{\nu_j}H^-$
couplings.

Rotating the interaction eigenstates of the sleptons identically to
their leptonic counterparts, we obtain the super-PMNS basis for the charged
sleptons and the sneutrinos, described by the Lagrangian (neglecting 
  the possible term $\Phi_\nu^T {\hat {\mathcal M}}_{\tilde \nu}^2
  \Phi_\nu$) 
\begin{equation}
{\mathcal L}_{\tilde l}^{mass} = - \Phi_e^\dag {\mathcal  M}_{\tilde
  e}^2 \Phi_e - 
\Phi_\nu^\dag {\mathcal M}_{\tilde \nu}^2 \Phi_\nu,
\end{equation}
where $\Phi_\nu=( {\tilde \nu}_e, {\tilde \nu}_\mu, {\tilde \nu}_\tau )^T$ and
$\Phi_e = ( {\tilde e}_L, {\tilde \mu}_L, {\tilde \tau}_L, {\tilde e}_R,
{\tilde \mu}_R, {\tilde \tau}_R )^T$. ${\mathcal M}_{\tilde e}^2$ is the $6 \times 6$ matrix
\begin{equation}
{\mathcal M}^2_{\tilde e} = \left( 
\begin{array}{cc} 
 {\hat{m}}^2_{\tilde L} + m_e^2 + {D_e}_{LL} & 
\frac{v_1}{\sqrt{2}} \hat{T}_E^\dag - \mu m_e \tan \beta \\[1.01ex]
\frac{v_1}{\sqrt{2}} \hat{T}_E - \mu^* m_e \tan \beta &
\hat{m}_{\tilde e}^2 + m_e^2 + {D_e}_{RR}
\end{array}
\right)~,
\end{equation}
and ${\mathcal M}_{\tilde \nu}^2$ is the $3\times 3$ matrix
\begin{equation}
{\mathcal M}_{\tilde \nu}^2 = U_{PMNS}^\dag \  \hat{m}_{\tilde L}^2 \ U_{PMNS}
+ {D_\nu}_{LL}, \label{numass}
\end{equation}
where  ${D_{e}}_{LL}$
and ${D_\nu}_{LL}$ are given in eq.~(\ref{dterm}).
In the equations above we introduced the $3 \times 3$ matrices
\begin{equation}
{\hat{m}_{\tilde L}}^2 \equiv V^\dagger_e \,m^2_{\tilde L}\, V_e\,,~~~
{\hat{m}_{\tilde e}}^2 \equiv U^\dagger_e \,{m^2_{\tilde e}}^T\, U_e\,,~~~
{\hat{T}_{E}} \equiv U^\dagger_e \,T_{E}^T\, V_e\,,~~~
\label{eq:tlhat}
\end{equation}
where the un-hatted matrices $m_{L,e}^2$ and 
 $T_E$ 
are given in the interaction basis of ref.~\cite{Skands:2003cj}. We  
diagonalise the charged slepton and sneutrino mass matrices via the
unitary 6$\times$6 and $3\times 3$
matrices $R_{e,\nu}$ respectively. Thus, $R_{e,\nu} {\mathcal M}^2_{\tilde e,
  \tilde \nu} R_{e,\nu}^\dagger$ are diagonal with increasing entries toward
the bottom right of each matrix. 

\subsection{R-parity violation \label{sec:rpv}}

We write the R-parity violating superpotential in the interaction
basis as
\begin{eqnarray}
W_{\mathrm{RPV}} &=& \epsilon_{ab} \left[
\frac{1}{2} \lambda_{ijk} L_i^a L_j^b \bar{E}_k +
\lambda'_{ijk} L_i^a Q_j^{bx} \bar{D}_{kx}
- \kappa_i L_i^a H_2^b \right] + \frac{1}{2} \lambda''_{ijk} \epsilon_{xyz} \bar{U}_{i}^x
\bar{D}_{j}^y \bar{D}_{k}^z 
, \label{eq:Wrpv}
\end{eqnarray}
where $x,y,z=1,\ldots,3$ are fundamental SU(3)$_C$ indices and
$\epsilon^{xyz}$ is the antisymmetric tensor in 3 dimensions with
$\epsilon^{123}=+1$. 
In eq.~(\ref{eq:Wrpv}), $\lambda_{ijk}, \lambda'_{ijk}$ and $\kappa_i$ break
lepton number, whereas $\lambda''_{ijk}$ violate baryon
number.  
To ensure proton stability, either lepton number
conservation or baryon number conservation is usually still assumed,
resulting in either 
$\lambda_{ijk}=\lambda'_{ijk}=\kappa_{i}=0$ or $\lambda''_{ijk}=0$ for all
$i,j,k=1,2,3$. 

The trilinear R-parity violating terms in the soft SUSY-breaking potential
are 
\begin{eqnarray}
V_{3,\mathrm{RPV}} &=& \epsilon_{ab} \left[
\frac12(T)_{ijk} {\tilde L}_{iL}^a {\tilde L}_{jL}^b {\tilde e}^*_{kR} +
(T')_{ijk} {\tilde L}_{iL}^a {\tilde Q}_{jL}^b {\tilde d}^*_{kR}
\right] 
+\frac12 (T'')_{ijk}\epsilon_{xyz} {\tilde u}_{iR}^{x*} {\tilde d}_{jR}^{y*} {\tilde d}_{kR}^{z*}
+ \mathrm{h.c.} \label{eq:trilinear}~~~.
\end{eqnarray}
Note that we do not factor 
out the $\lambda$ couplings (e.g.\ as in 
${T_{ijk}}/{\lambda_{ijk}}\equiv A_{\lambda,ijk}$) in order to avoid
potential problems with $\lambda_{ijk}=0$ but $T_{ijk}\ne 0$.
This usage is 
consistent with the convention for the R-conserving sector elsewhere in
this  report.

The bilinear R-parity violating soft terms (all lepton number
violating) are 
\begin{equation}
V_{2,\mathrm{RPV}} = -\epsilon_{ab}D_i {\tilde L}_{iL}^a H_2^b  + {\tilde
L}_{iaL}^\dag m_{{\tilde L}_i H_1}^2 H_1^a + \mathrm{h.c.}~.
\label{eq:bilinear}
\end{equation}

When lepton number is not conserved the sneutrinos may 
acquire vacuum expectation values (VEVs)
$\langle {\tilde \nu}_{e,\mu,\tau} \rangle \equiv v_{e, \mu,\tau}/\sqrt{2}$.
We generalise the SLHA1 parameter $v$ to:
\begin{equation}
v=\sqrt{v_1^2+v_2^2+v_{e}^2+v_{\mu}^2+v_{\tau}^2}~=246~\mathrm{GeV}.
\end{equation}
The addition of sneutrino VEVs allows for
various different definitions of $\tan
\beta$, but we here choose to keep the SLHA1 definition $\tan
\beta=v_2/v_1$. 

We 
use the super-CKM/PMNS basis throughout, as defined in
subsection~\ref{sec:flv}, with the following considerations specific to
the R-parity violating case. Firstly, the $d$-quark mass matrices
are given by
\begin{eqnarray}
(m_d)_{ij} = (Y_D)_{ij} v_1 + \lambda'_{kij} v_k  \,\,.
\label{eq:dquarkmassrp}
\end{eqnarray}
where $v_k$ are the sneutrino VEVs. 
Secondly, in the lepton number violating case, we 
restrict our attention to scenarios in which 
there are no right-handed neutrinos and, thus, neutrino
masses are generated solely by the RPV couplings.
In this case, the PMNS matrix is not an independent input but an
output. 

We define the super-CKM basis as
the one where the Yukawa couplings $Y_D$ and $Y_U$ are 
diagonal.
The PMNS basis is defined as the basis where $Y_E$ is diagonal and the 
loop-induced neutrino mass matrix is diagonalised. 
In this way one obtains a uniquely defined set of parameters:
\begin{eqnarray}
{\hat \lambda}_{ijk} &\equiv& \lambda_{rst} V_{\nu,ri} V_{e,sj} 
 U^\dagger_{e,tk}~,
 \label{eq:rotlam} \\
{\hat \lambda'}_{ijk} &\equiv& \lambda'_{rst}  V_{\nu,ri} V_{d,sj} 
U^\dagger_{d,tk}~, \\
{\hat \kappa}_i &\equiv&   \kappa_r V_{e,ri}~, \label{eq:rotbi} \\
{\hat \lambda''}_{ijk} &\equiv& \lambda''_{rst}  U^\dagger_{u,ri}
 U^\dagger_{d,sj} U^\dagger_{d,tk}~,
\label{eq:rotlampp}
\end{eqnarray}
where  the fermion mixing matrices are defined in subsection~\ref{sec:flv}.
The Lagrangian for the quark-slepton interactions then takes the following
form:
\begin{eqnarray}
{\cal L} = - {\hat \lambda}'_{ijk}  \tilde \nu_i \bar{d}_{Rk} d_{Lj}
 + {\hat \lambda'}_{rsk} U_{PMNS,ri}^\dagger V_{CKM,sj}^\dagger 
 \tilde l_{L,i} \bar{d}_{Rk} u_{Lj} \, + \mathrm{h.c.}~.
\end{eqnarray}
Similarly one obtains the soft SUSY breaking couplings in this basis
by replacing 
the superpotential quantities in
eqs.~(\ref{eq:rotlam})--(\ref{eq:rotlampp}) by the 
corresponding soft SUSY breaking couplings. In addition we define:
\begin{eqnarray}
\hat m^2_{\tilde L_i H_1} \equiv V^\dagger_{e,ir}m^2_{\tilde L_r H_1}~.
\end{eqnarray}

As for the R-conserving MSSM, the bilinear terms (both SUSY-breaking
and SUSY-respect\-ing ones, including $\mu$) and the VEVs are not
independent parameters. Specifically, out of the 12 parameters 
$\kappa_i$, $D_i$, sneutrino vevs, and $m^2_{\tilde{L}_iH_1}$, only 9
are independent. 

\subsubsection*{Particle mixing}
In general, the
neutrinos mix with the neutralinos. This requires a change in the definition of
the $4\times 4$ neutralino mixing matrix $N$ to a $7\times 7$ matrix. 
The Lagrangian contains the (symmetric) neutrino/neutralino mass matrix as 
\begin{equation}
\mathcal{L}^{\mathrm{mass}}_{{\tilde \chi}^0} =
-\frac12{\tilde\psi^0}{}^T{\mathcal M}_{\tilde\psi^0}\tilde\psi^0 +
\mathrm{h.c.}~, 
\end{equation}
in the basis of 2--component spinors $\tilde\psi^0 =$
$( \nu_e, \nu_\mu, \nu_\tau, -i\tilde b, -i\tilde w^3,  
\tilde h_1, \tilde h_2)^T$. We define the
unitary $7 \times 7$ 
neutrino/neutralino mixing matrix $N$, such that:
\begin{equation}
-\frac12{\tilde\psi^0}{}^T{\mathcal M}_{\tilde\psi^0}\tilde\psi^0
= -\frac12\underbrace{{\tilde\psi^0}{}^TN^T}_{{{\tilde \chi}^0}{}^T}
\underbrace{N^*{\mathcal
    M}_{\tilde\psi^0}N^\dagger}_{\mathrm{diag}(m_{{\tilde \chi}^0})}
\underbrace{N\tilde\psi^0}_{{\tilde \chi}^0}~,  \label{eq:neutmass}
\end{equation}
where the 7 (2--component) generalised neutrinos ${\tilde
  \chi}^0=(\nu_1,...,\nu_7)^T$ are
defined strictly mass-ordered.

In the limit of CP conservation, the default convention is that $N$ be
a real matrix and one or more of the mass eigenstates may
have an apparent negative mass. The minus sign may be removed by phase
transformations on $\tilde \chi^0_i\equiv\nu_i$ as explained in
SLHA1~\cite{Skands:2003cj}.

Charginos and charged leptons may also mix in the case of $L$-violation. 
In a similar spirit to the neutralino mixing, we define
\begin{equation}
\mathcal{L}^{\mathrm{mass}}_{{\tilde \chi}^+} =
-\frac12{\tilde\psi^-}{}^T{\mathcal M}_{\tilde\psi^+}\tilde\psi^+ +
\mathrm{h.c.}~, 
\end{equation}
in the basis of 2--component spinors $\tilde\psi^+ =$
$({e}^+,{\mu}^+,{\tau}^+,-i\tilde w^+, \tilde h_2^+)^T$,
$\tilde\psi^- =$ $({e}^-,{\mu}^-,{\tau}^-,-i\tilde w^-,\tilde h_1^-)^T$  
where $\tilde w^\pm = (\tilde w^1 \mp \tilde w^2) / \sqrt{2}$. Note
that, in the limit of no RPV the lepton fields are mass eigenstates.

 We define the
unitary $5 \times 5$
charged fermion mixing matrices $U,V$
such that:
\begin{equation}
-\frac12{\tilde\psi^-}{}^T{\mathcal M}_{\tilde\psi^+}\tilde\psi^+
= -\frac12\underbrace{{\tilde\psi^-}{}^TU^T}_{{{\tilde \chi}^-}{}^T}
\underbrace{U^*{\mathcal
    M}_{\tilde\psi^+}V^\dagger}_{\mathrm{diag}(m_{{\tilde \chi}^+})}
\underbrace{V\tilde\psi^+}_{{\tilde \chi}^+}~,  \label{eq:chargmass}
\end{equation}
where the generalised charged leptons $\tilde
\chi^+\equiv(e^+_1,e^+_2,e^+_3,e^+_4,e^+_5)$ 
are defined as strictly mass ordered.
In the limit of CP
conservation, $U$ and $V$ are chosen to be real by default. 

R-parity violation via lepton number violation implies that the
 sneutrinos can mix with the Higgs bosons. In the limit of 
 CP conservation the CP-even (-odd) Higgs bosons mix with real (imaginary)
 parts of the sneutrinos. We write the neutral scalars as $\phi^0
 \equiv \sqrt{2} \Re{(H_1^0, H_2^0, {\tilde \nu}_e, {\tilde \nu}_\mu, {\tilde
\nu}_\tau)^T}$, with the mass term
\begin{equation}
{\mathcal L} = - \frac12 {\phi^0}^T {\mathcal M}_{\phi^0}^2 \phi^0~,
\end{equation}
where ${\mathcal M}_{\phi^0}^2$ is a $5 \times 5$ symmetric mass matrix. 
We define the orthogonal $5 \times 5$ 
mixing matrix $\aleph$  by
\begin{equation}
-{\phi^0}{}^T{\mathcal M}_{\phi^0}^2
\phi^0
= -\underbrace{{\phi^0}{}^T {\mathbf \aleph}^T}_{{{
      \Phi}^0}{}^T} 
\underbrace{{\mathbf \aleph}{\mathcal
    M}_{\phi^0}^2\aleph^T}_{\mathrm{diag}(m_{{ \Phi}^0}^2)}
\underbrace{{\mathbf \aleph}\phi^0}_{{ \Phi}^0}~,  
\label{eq:sneutmass}
\end{equation}
where $\Phi^0 \equiv (h^0_1, h^0_2, h^0_3, h^0_4, h^0_5)$ are the
neutral scalar mass eigenstates in strictly increasing mass orderx. 

We write the neutral pseudo-scalars 
as $\bar\phi^0 \equiv \sqrt{2} \Im{(H_1^0, H_2^0,
  {\tilde \nu}_e, {\tilde 
  \nu}_\mu, {\tilde \nu}_\tau)^T}$,
with the mass term
\begin{equation}
{\mathcal L} = - \frac12 {{\bar \phi}^0}{}^T {\mathcal M}_{{\bar
  \phi}^0}^2 {\bar \phi}^0~,
\end{equation}
where ${\mathcal M}_{\bar \phi^0}^2$ is a $5 \times 5$ symmetric mass
matrix. We define 
the $4 \times 5$ mixing matrix $\bar \aleph$ by
\begin{equation}
-{\bar \phi^0}{}^T{\mathcal M}_{\bar \phi^0}^2
\bar \phi^0
= -\underbrace{{\bar \phi^0}{}^T {\bar \aleph}^T}_{{{
      \bar \Phi}^0}{}^T} 
\underbrace{{\bar \aleph}{\mathcal
    M}_{\bar \phi^0}^2{\bar \aleph}^T}_{\mathrm{diag}(m_{{\bar \Phi}^0}^2)}
\underbrace{{\bar \aleph}\bar \phi^0}_{{\bar \Phi}^0}~,  
\label{eq:sneutmass2}
\end{equation}
where $\bar\Phi^0 \equiv (A^0_1, A^0_2, A^0_3, A^0_4)$ 
are the pseudoscalar mass eigenstates, again in increasing
mass order. 

The charged sleptons and charged Higgs bosons also mix in the $8 \times 8$ mass
squared matrix ${\mathcal M}^2_{\phi^\pm}$ which we diagonalize 
by a $7 \times 8$ matrix $C$:
\begin{equation}
{\mathcal L}=- 
\underbrace{({H_1^-}^*, {H^+_2},\tilde{e}_{L_i}^*,\tilde{e}_{R_j}^*) C^\dagger}_{{\Phi^+}} 
\underbrace{C {\mathcal M}^2_{\phi^\pm} C^\dagger}_{ \mathrm{diag}({\mathcal M}^2_{\Phi^\pm})} 
C \left (
\begin{array}{c} {H_1^-}\\[1mm] {H_2^+}^* \\  \tilde{e}_{L_k} \\
\tilde{e}_{R_l} \end{array} \right )\, \label{eq:higgsslep}~,
\end{equation}
where $i,j,k,l\in \{1,2,3\}$ and
$\Phi^+=\Phi^-{}^\dagger\equiv(h^+_1,h^+_2,h^+_3,h^+_4,h^+_5,h^+_6,h^+_7)$.
  
R-parity violation may also generate contributions to down-squark
mixing via additional left-right mixing terms, 
\begin{eqnarray}
\frac{1}{\sqrt{2}} v_1 {\hat T}^\dagger_{D,ij} - \mu m_{d,i}
 \tan\beta \delta_{ij}
+ \frac{v_k}{\sqrt{2}} {\hat T}^\dagger_{\lambda',kij} 
\end{eqnarray}
where $v_k$ are the sneutrino vevs.
  However, this only mixes the six down-type squarks amongst
themselves and so is identical to the effects of flavour mixing.  This
is covered in subsection~\ref{sec:flv}.

\subsection{CP violation \label{sec:cpv}}

In general, we write complex parameters in terms of their real and
imaginary parts, rather than in terms of phase and modulus. (The SLHA1
data structures are then understood to refer to the real parts, and 
the imaginary parts are provided in data blocks of the same name 
but prefaced by {\tt IM}.) The defaults for all imaginary parameters 
will be zero.

One special case is the $\mu$ parameter. When $|\mu|$ 
is determined by the conditions for  electroweak symmetry breaking,
only the phase $\varphi_\mu$ is taken as an input parameter, see
\cite{Allanach:2008qg}.

When CP symmetry is broken, quantum corrections cause mixing between
the CP-even and CP-odd Higgs states. 
Writing the neutral scalar interaction eigenstates as 
$\phi^0 \equiv \sqrt{2} (\Re{H_1^0},$
$\Re{H_2^0},$ $\Im{H_1^0},$ $\Im{H_2^0})^T$ 
we define the $3 \times 4$  mixing matrix $S$ by 
\begin{equation}
-{\phi^0}{}^T{\mathcal M}_{\phi^0}^2
 \phi^0
= -\underbrace{{ \phi^0}{}^T {S}^T}_{{{
       \Phi}^0}{}^T} 
\underbrace{{S}^*{\mathcal
    M}_{\phi^0}^2{S}^\dagger}_{\mathrm{diag}(m_{{ \Phi}^0}^2)}
\underbrace{{S}\phi^0}_{{ \Phi}^0}~,  
\label{eq:cvhmass}
\end{equation}
where $\Phi^0 \equiv (h_1^0,h_2^0,h_3^0)^T$ are the mass eigenstates.

For the neutralino and chargino mixing matrices, the default convention 
in SLHA1 is that they be real. One or more mass eigenvalues may then
have an apparent negative sign, which can be removed by a
phase transformation on $\tilde \chi_i$ as explained in
SLHA1~\cite{Skands:2003cj}. When going to CPV, the reason for
introducing the negative-mass convention in the first place, namely
maintaining the mixing matrices strictly real, disappears. 
In the CP violating case, we therefore take
all masses real and positive, with $N$, $U$, and $V$ complex. 

\subsection{The next-to-minimal supersymmetric standard model\label{sec:nmssm}}
We write the most general CP conserving NMSSM superpotential as:
\begin{equation}\label{eq:nmssmsup}
W_{NMSSM} = W_{MSSM} - \epsilon_{ab}\lambda {S} {H}^a_1 {H}^b_2 + \frac{1}{3}
\kappa {S}^3 + \mu' S^2 +\xi_F S \ , \end{equation} 
where $W_{MSSM}$ is the MSSM superpotential,
in the conventions of ref.~\cite[eq.~(3)]{Skands:2003cj}. 
A non-zero $\lambda$ in combination with a VEV $\left< S
\right>$ of the singlet generates a contribution to the effective 
$\mu$ term $\mu_\mathrm{eff}= \lambda \left< S
\right> + \mu$, where the MSSM $\mu$ term is normally assumed to be
zero in NMSSM constructions, 
yielding $\mu_{\mathrm{eff}}=\lambda \left< S \right>$. 
The remaining terms represent a general cubic 
potential for the singlet; $\kappa$ is dimensionless, $\mu'$ has
dimension of mass, and $\xi_F$ has dimension of mass squared. 
The soft SUSY-breaking terms relevant to the NMSSM are
\begin{equation}\label{eq:nmssmsoft}
V_\mathrm{soft} = V_{2,MSSM} + V_{3,MSSM} + m_\mathrm{S}^2 | S |^2 +
(-\epsilon_{ab}\lambda A_\lambda {S} {H}^a_1 {H}^b_2 + 
\frac{1}{3} \kappa A_\kappa {S}^3  
+ m_{S}'^2 S^2 +\xi_S S
+ \mathrm{h.c.}) \ , \end{equation}
where $V_{i,MSSM}$ are the MSSM soft terms defined in
eqs.~(\ref{eq:slha1v3soft}) and (\ref{eq:slha1v2soft}), and we have
introduced the notation $m_{S}'^2 \equiv B'\mu'$. 

At tree level, there are thus 15 parameters (in addition to $m_Z$
which fixes the sum of the squared Higgs VEVs) 
that are 
relevant for the Higgs sector of the R-parity and CP-conserving NMSSM: 
\begin{equation} 
\tan\!\beta,\ \mu,\ m_{H_1}^2,\ m_{H_2}^2,\ m_3^2,\
\lambda,\  \kappa,\ A_{\lambda},\  A_{\kappa},\ \mu',\ m_{S}'^2,\ \xi_F,\
\xi_S,\ \lambda \left< S \right>,\ m_S^2~.\label{eq:nmssmpar}
\end{equation}
The minimisation of the effective 
potential imposes 3 conditions on these
parameters, such that only 12 of them can be considered
independent. 

Note that we write the soft parameter $m_3^2$ in the form
$m_3^2/(\cos\beta\sin\beta)$, see \cite{Skands:2003cj}. The notation
$m_A^2$ that was used for that parameter in the SLHA1 is no longer
relevant in the NMSSM context, but by keeping the definition in terms
of $m_3^2$ and $\cos\beta\sin\beta$ unchanged, we maintain an
economical and straightforward correspondence between the two cases.

\subsubsection*{Particle mixing}

In the CP-conserving NMSSM, the CP-even interaction eigenstates are 
$\phi^0 \equiv\sqrt{2} \Re{(H_{1}^0, H_{2}^0, S)^T}$. 
We define the orthogonal $3 \times 3$ mixing matrix $S$ by 
\begin{equation}
-{\phi^0}{}^T{\mathcal M}_{\phi^0}^2 \phi^0
= -\underbrace{{\phi^0}{}^T {S}^T}_{{{
      \Phi}^0}{}^T} 
\underbrace{{S}{\mathcal
    M}_{\phi^0}^2{S}^T}_{\mathrm{diag}(m_{{\Phi}^0}^2)}
\underbrace{{S} \phi^0}_{{\Phi}^0}~,  
\end{equation}
where $\Phi^0 \equiv (h^0_1, h^0_2, h^0_3)$ are the mass eigenstates
ordered in mass. 

In the CP-odd sector the interaction eigenstates are 
$\bar\phi^0 \equiv\sqrt{2} \Im{(H_{1}^0, H_{2}^0, S)^T}$. 
We define the $2 \times 3$ mixing matrix $P$ by
\begin{equation}
-{\bar \phi^0}{}^T{\mathcal M}_{\bar \phi^0}^2
\bar \phi^0
= -\underbrace{{\bar \phi^0}{}^T {P}^T}_{{{
      \bar \Phi}^0}{}^T} 
\underbrace{{P}{\mathcal
    M}_{\bar \phi^0}^2{P}^T}_{\mathrm{diag}(m_{{\bar \Phi}^0}^2)}
\underbrace{{P}\bar \phi^0}_{{\bar \Phi}^0}~,  
\end{equation}
where $\bar\Phi^0 \equiv (A^0_1, A^0_2)$ are the mass eigenstates
ordered in mass. 

The neutralino sector of the NMSSM requires a change in the definition
of the $4 \times 4$ neutralino mixing matrix $N$ to a $5 \times 5$ matrix.  The
Lagrangian contains the (symmetric) neutralino mass matrix as 
\begin{equation}
\mathcal{L}^{\mathrm{mass}}_{{\tilde \chi}^0} =
-\frac12{\tilde\psi^0}{}^T{\mathcal M}_{\tilde\psi^0}\tilde\psi^0 +
\mathrm{h.c.}~, 
\end{equation}
in the basis of 2--component spinors $\tilde\psi^0 =$ $(-i\tilde b,$
$-i\tilde w^3,$  $\tilde h_1,$ $\tilde h_2,$ $\tilde s)^T$. 
We define the unitary $5 \times 5$ neutralino mixing matrix $N$ such that:
\begin{equation}\label{eq:nmssmneutmass}
-\frac12{\tilde\psi^0}{}^T{\mathcal M}_{\tilde\psi^0}\tilde\psi^0
= -\frac12\underbrace{{\tilde\psi^0}{}^TN^T}_{{{\tilde \chi}^0}{}^T}
\underbrace{N^*{\mathcal
    M}_{\tilde\psi^0}N^\dagger}_{\mathrm{diag}(m_{{\tilde \chi}^0})}
\underbrace{N\tilde\psi^0}_{{\tilde \chi}^0}~,  
\end{equation}
where the 5 (2--component) neutralinos ${\tilde \chi}_i$ are defined
such that the absolute value of their masses increase with $i$. As in
SLHA1, our convention is that $N$ be a real matrix. One or more mass
eigenvalues may then have an apparent negative sign, which can be
removed by a phase transformation on $\tilde \chi_i$. 

\subsection{Summary}
The Supersymmetry Les Houches Accord (SLHA)~\cite{Skands:2003cj} 
provides a universal set of conventions
for supersymmetry 
analysis problems in high energy physics. 
Here, we summarise extensions of the conventions of the first SLHA 
to include various generalisations \cite{Allanach:2008qg}: 
the minimal supersymmetric
standard model with flavour violation, RPV, and CPV, 
as well as the simplest next-to-minimal model. For updates
and examples, see\\ 
\texttt{http://home.fnal.gov/$\sim$skands/slha/} 


\clearpage
\newpage

\bibliography{slha2houches,gmtwosfitter_main,splitsfitter_main,benAndFawz_bib,ofsdec,blind_susy_search,lh07_KramlSekmen,diracgauginos,KramlRaklev,barles,hybrid,lh07_stop,li6,tgbsfitter_main}

\end{document}